\documentclass{emulateapj}
\usepackage{lscape}
\usepackage{apjfonts}

\slugcomment{Published in the Astrophysical Journal Supplement Series,
2005 Dec: ApJS, 161, 304-360}

\shorttitle{Massive Star Clusters}
\shortauthors{McLaughlin \& van der Marel}

\begin{document}

\title{Resolved Massive Star Clusters in the Milky Way and its Satellites: \\
Brightness Profiles and a Catalogue of Fundamental Parameters}

\author{Dean E. McLaughlin}
\affil{University of Leicester, Dept. of Physics and Astronomy,
University Road, \\ Leicester, UK LE1 7RH}
\email{dean.mclaughlin@astro.le.ac.uk}
\and
\author{Roeland P. van der Marel}
\affil{Space Telescope Science Institute, 3700 San Martin Drive,
Baltimore, MD  21218}
\email{marel@stsci.edu}

\begin{abstract}

We present a database of structural and dynamical properties for 153
spatially resolved star clusters in the Milky Way, the Large and Small
Magellanic Clouds, and the Fornax dwarf spheroidal. This database complements
and extends others in the literature, such as those of \citet{har96}
and \citet{mg03a,mg03b,mg03c}. Our cluster sample
comprises 50 ``young massive clusters'' in the LMC and SMC, and 103 old
globular clusters between the four galaxies. The parameters we list include
central and half-light averaged surface brightnesses and mass densities; core
and effective radii; central potentials, concentration parameters, and tidal
radii; predicted central velocity dispersions and escape velocities; total
luminosities, masses, and binding energies; central phase-space densities;
half-mass relaxation times; and ``$\kappa$-space'' parameters. We use
publicly available population-synthesis models to compute stellar-population
properties (intrinsic $B-V$ colors, reddenings, and $V$-band mass-to-light
ratios) for the same 153 clusters plus another 63 globulars in the Milky Way.
We also take velocity-dispersion measurements from the literature for a subset
of 57 (mostly old) clusters to derive dynamical mass-to-light ratios for them,
showing that these compare very well to the population-synthesis predictions.
The combined dataset is intended to serve as the basis for future
investigations of structural correlations and the fundamental plane of massive
star clusters, including especially comparisons between the systemic
properties of young and old clusters.

The structural and dynamical parameters are derived from fitting three
different models---the modified isothermal sphere of \citet{king66};
an alternate modified isothermal sphere based on the ad hoc stellar
distribution function of \citet{wil75}; and asymptotic power-law models with
constant-density cores---to the surface-brightness profile of each cluster.
Surface-brightness data for the LMC, SMC, and Fornax clusters are based in
large part on the work of \citet{mg03a,mg03b,mg03c}, but include significant
supplementary data culled from the literature and important corrections to
Mackey \& Gilmore's $V$-band magnitude scale. The profiles of Galactic
globular clusters are taken from \citet{tkd95}. We address the question of
which model fits each cluster best, finding in the majority of cases that the
Wilson models---which are spatially more extended than King models but still
include a finite, ``tidal'' cut-off in density---fit clusters of any age, in
any galaxy, as well as or better than King models. Untruncated, asymptotic
power laws often fit about as well as Wilson models but can be significantly
worse. We argue that the extended halos known to characterize many Magellanic
Cloud clusters may be examples of the {\it generic} envelope structure of
self-gravitating star clusters, not just transient features associated
strictly with young age.

\end{abstract}


\section{Introduction}
\label{sec:intro}

Next to elliptical galaxies, globular clusters are the most thoroughly modeled
and best understood class of ``hot'' stellar system. The simple models of
single-mass, isotropic, lowered isothermal spheres developed by \citet{king66}
have been fit to a large majority of the $\sim150$ Galactic globulars currently
known, yielding comprehensive catalogues of cluster structural parameters
and derived physical properties \citep{djo93,pry93,tkd95,har96}.
These have been used to explore a multitude of scaling relations and
interdependences between the various properties \citep{djo94}, leading to the
definition of a fundamental plane for globular clusters which is analogous to
but physically distinct from that for early-type galaxies and bulges
\citep[e.g.,][]{djo95,bur97,bel98,mcl00}. Spectroscopic and HST imaging
data have also been collected, and fit with \citet{king66} models, for a large
number of globulars in M31 \citep[e.g.,][]{djo97,dub97,bar02},
M33 \citep{lar02}, and NGC 5128 = Centaurus A \citep{hol99,har02,mar04}.
These clusters appear to follow essentially the same scaling relations and
lie on the same fundamental plane as Galactic globulars.

That these data contain important information on the formation and evolution
of globular clusters (GCs) is clear. But cleanly separating formative from
evolutionary influences on the present-day form of GC structural correlations
and the fundamental plane is difficult in the absence of any rigorous,
first-principles predictions for the systemic properties of newly born star
clusters. This issue only gains in importance with the emergent consensus that
the ``super'' star clusters now forming in nearby starbursts and galaxy
mergers, and even massive young clusters in more quiescent disks, may well be
close analogues to what globular clusters were a Hubble time ago. What,
then, can the properties of GCs tell us about how star clusters form in
general? Conversely, what can a comparison between the properties of young
clusters and GCs add to our understanding of the globulars?

Important steps towards addressing these questions have been made
by studies of the mass-radius relations for young massive clusters in
merging galaxies \citep[e.g.,][]{zep99} and relatively nearby spiral disks
\citep{lar04}, and, on a somewhat different scale, for the (young) nuclear
clusters in very late-type spirals \citep{boe04,wal05}. However, the data on
these clusters---all of which are at distances of many Mpc---are necessarily
much less complete, and in some respects more uncertain, than those for
globular clusters in the Milky Way and even out to NGC 5128. A comprehensive
study of young cluster properties vs.~GC properties requires a catalogue for
the former which includes the full suite of physical parameters routinely
calculated for Galactic GCs, preferably obtained within the same modeling
framework and to comparable precision and accuracy.

Our main goal in this paper is to contribute to such a database, focusing
specifically on the nearby populations of massive clusters in the Large and
Small Magellanic Clouds. In particular, we fit a variety of models to obtain
derived parameters for 53 LMC clusters and 10 SMC clusters with
high-quality, HST-based surface-density profiles of their inner parts measured
by \citet{mg03a,mg03b}. Thirteen of these objects are globular clusters older
than 10 Gyr; the other 50 range in age from several Myr to a few Gyr. At
the same time, \citet{mg03c} have published new surface-brightness profiles
for the 5 old GCs in the Fornax dwarf spheroidal, so we include these in our
catalogue as well. As we describe below, some aspects of our analyses of these
LMC, SMC, and Fornax clusters go beyond what is normally done for GCs. Thus,
we also take surface-brightness profiles for globulars in the Milky Way from
the collection of \citet{tkd95}, and some supplementary data from the
catalogue of \citet{har96}, in order to model as many Galactic GCs as is
practical in precisely the same way that we treat the Mackey \& Gilmore
datasets. The exact number of Galactic GCs that we include depends on the
context of our various calculations.

Much in the existing catalogues of Galactic GC properties is based on the
same surface-brightness data from \citet{tkd95} that we employ here; and, as
we said above, the catalogues already contain the results of reasonably
uniform modeling of the clusters. However, all of this is within the context
of the theoretical \citet{king66} isotropic, modified isothermal spheres.
\citet{mg03a,mg03b,mg03c}, for their part, tabulate estimates of cluster
central surface brightnesses, core radii, total luminosities, and total masses
derived from fits of power-law models for their surface-density profiles:
$I(R)\propto [1+(R/r_0)^2]^{-(\gamma-1)/2}$. (Mackey \& Gilmore's cluster
masses further depend on their application of population-synthesis models.)
One of our aims here is to homogenize the analysis of the basic data on young
and globular clusters alike, and thus we completely re-fit all of the LMC,
SMC, Fornax, and Milky Way cluster profiles with {\it both} \citet{king66}
{\it and} power-law models. It is then important to recognize Mackey \&
Gilmore's motivation for fitting asymptotic power laws to their cluster sample
in the first place.

In their pioneering study of the structure of young clusters in the LMC,
\citet{eff87} found that \citet{king66} models are not always capable of
describing the outer-envelope structures of these objects, which are very
extended spatially and may be better fit by power laws that do not include the
sharp tidal cut-offs built into King models. Some subsequent studies of other
young massive clusters have come to the same conclusion
\citep[e.g.,][]{lar04,sch04}. It has been argued that this reflects
the presence of rather massive halos of unbound stars around the clusters,
which (it is presumed) will eventually be stripped away by tides to leave
behind a more familiar, King-like body. However, so far as we are aware,
there have not been any systematic attempts made to fit large samples of young
clusters with structural models that are more extended than \citet{king66}
but still spatially truncated rather than formally infinite like power laws.
Nor, for that matter, has any such test been performed on old GCs.
Thus, in this paper we also fit our combined LMC/SMC/Fornax/Milky Way cluster
sample with a third type of model which is intermediate to King and power-law
models. For this we have chosen a spherical and isotropic version of the model
originally developed by \citet{wil75} for application to elliptical galaxies.
Wilson's model is essentially a single-mass, modified isothermal sphere like a
\citet{king66} model in its core but with an ad hoc change in the
``lowering'' term in the stellar distribution function to give more extended
(though still finite) halos for otherwise similar clusters.

In \S\ref{sec:data} below, we discuss in detail the surface-brightness
data that we use to fit models to the clusters in our sample. In the
LMC and SMC, we supplement many of the HST-derived profiles in
\citet{mg03a,mg03b} with older (ground-based) starcount densities for the
same clusters. In many cases this allows the density profiles to be defined
out to substantially larger projected radii. For the Fornax clusters, we rely
exclusively on the \citet{mg03c} data. We work in the $V$ band, and in all
cases we check Mackey \& Gilmore's magnitude scales against ground-based
aperture magnitudes from the literature. We find it necessary to recalibrate
Mackey \& Gilmore's surface brightnesses, since as published they lead to
integrated magnitudes that are always too faint---often by several tenths of
a mag---relative to the ground-based numbers. The zeropoint correction for
every cluster is tabulated. Our discussion of the Milky Way GC
surface-brightness data is very brief, as we simply take them (also in the
$V$ band) from \citet{tkd95} with minor modifications.

Following that, in \S\ref{sec:popsyn} we explore the use of
population-synthesis models to infer an extinction $A_V$ and mass-to-light
ratio $\Upsilon_V$ for each of the 68 LMC, SMC, and Fornax clusters from
\citet{mg03a,mg03b,mg03c}---who have also provided an age and metallicity
for every cluster, either compiled from CMD studies in the literature or
estimated themselves---and for 148 Galactic GCs with metallicities listed in
the catalogue of \cite{har96}. We resort to the use of extinctions based on
population-synthesis model colors to derive intrinsic luminosities, etc.,
from the observations
of LMC and SMC clusters, because direct measurements of reddening are not
available for the majority of that sample. (Measured extinctions are available
for the Fornax and Milky Way GCs, and we use those as a check on our
population-synthesis model values.) We need theoretical mass-to-light ratios
for most of the clusters in all four galaxies in order to convert from
luminosity to mass when determining a number of dynamical cluster parameters
(velocity dispersions have been observed for only a fraction of the clusters
under consideration here). We tabulate $A_V$ and $\Upsilon_V$ for every
cluster as determined using each of two publicly available
population-synthesis
codes \citep{frv97,bru03} under a number of different assumptions on the
form of the stellar IMF. For our subsequent modeling, however, we only
use the values predicted by the code of \citet{bru03} with the disk-star
IMF of \citet{cha03}. \citet{mg03a,mg03b,mg03c} also used population-synthesis
modeling to estimate mass-to-light ratios for their LMC/SMC/Fornax clusters;
but they assumed a much steeper IMF than we do and obtained systematically
different numbers for $\Upsilon_V$.

Section \ref{sec:modeling} first describes (\S\ref{subsec:mods}) some salient
aspects of King, power-law, and Wilson models. Then \S\ref{subsec:fits}
shows in detail how each compares to the observed
surface-brightness and internal velocity-dispersion profiles of
the well studied GC $\omega$ Centauri \citep[see also][]{mcl03}, before
presenting the bulk of our results, the basic parameters of each model fit
to each of the 68 LMC, SMC, and Fornax clusters and 85 Galactic globulars.
These are accompanied by a number of derived cluster properties, all evaluated
within each of the three fitted models: central and
half-light averaged surface brightnesses and mass densities; core and
effective radii; concentrations and tidal radii (or infinite power-law slope);
predicted central velocity dispersions and escape velocities; total
luminosities, masses, and binding energies; central phase-space densities; and
half-mass relaxation times. In \S\ref{subsec:fitcomp} we compare our fits
to those in existing catalogues (when the latter overlap with our work),
and we check that our derived cluster properties are generally well
defined regardless of which model is fit to the data.

We also investigate in \S\ref{subsec:fitcomp} the question of which model
tends to fit these clusters best. We find that, for $\approx90\%$ of our
full sample of young massive clusters {\it and} old globular clusters, the
more extended \citet{wil75} models provide equally good or significantly
better fits than \citet{king66} models. Untruncated power laws generally
describe the young clusters about as well as the spatially limited Wilson
models; only occasionally are they slightly better. Thus,
we conclude that (1) unlimited power laws are not the only description
possible for the halos of young LMC/SMC clusters in particular, and
(2) structure extending beyond what is predicted by \citet{king66} models is
a fairly generic feature of any star cluster, rather than something
present only at young ages. {\it It is not obvious that all such extended
halos must necessarily be unbound}, but it is a subtle problem to measure
accurately the tidal radius of a cluster and to compare it properly to the
theoretical Roche lobe defined by a galaxy's tidal field. We do not tackle
the issue in any detail in this paper.

In \S\ref{sec:dynml} we collect velocity-dispersion data from the literature
for a subset of 19 Magellanic Cloud and Fornax clusters and 38 Galactic GCs,
and use these in conjunction with our surface-brightness fits to derive
dynamical mass-to-light ratios for them. We compare these directly to the
population-synthesis model values which we used to derive all mass-dependent
cluster properties. The agreement is very good on average.

Finally, \S\ref{sec:kappa} presents the ``$\kappa$-space'' parameters of
\citet{bbf92} and \citet{bur97} (or, more precisely, mass-equivalent versions
of these) for the 153 young and globular clusters to which we fit all three of
our structural models. We also tabulate in this Section the galactocentric
radii of all the objects. These quantities complete what is needed to
construct the fundamental plane of star clusters in any of the equivalent
formulations that can be found in the current literature. However, we leave
the actual delineation and interpretation of all cluster correlations,
including discussion of young vs.~old populations, for future work.

\section{Starcount Data, Aperture Photometry, and Surface-Brightness Profiles}
\label{sec:data}

In this Section we collect and combine, in a uniform fashion,
available data defining the run of surface brightness as a function
of radius in a significant number of nearby and well-resolved globular and
young, massive star clusters.

In the case of the Milky Way globular cluster (GC) system, this task has in
fact already been completed by \citet{tkd95}, who gathered 
inhomogeneous surface-brightness (SB) and starcount data from the
literature for 124 GCs and combined them objectively into a single,
zeropointed $V$-band SB profile for each cluster.\footnotemark
\footnotetext{We note in passing that Trager et al.~give the number of
clusters in their catalogue as 125. However, according to
\citet{har96}, the GCs Terzan 5 and Terzan 11, which appear separately in
\citet{tkd95}, are in fact the same object.}
This database has served as the raw material for the standard catalogues
\citep[updated 2003 February\footnotemark]{djo93,har96}
of \citet{king66} model parameters for Galactic GCs.
\footnotetext{Available online at \hfill\break
{\tt http://physwww.mcmaster.ca/\%7Eharris/mwgc.dat}\ .}
Here we similarly work from the \citet{tkd95} GC profile data
for our modeling of Milky Way globulars. Although
some of these profiles have been superseded by more recent work on a few
individual clusters, we have not attempted to incorporate any such updates
into the Trager et al.~catalogue, preferring instead to draw on data that have
been processed in a homogeneous way by a single set of authors. A few further
details on our handling of the Trager et al.~data are given in
\S\ref{subsec:trager} below.

The situation for well-resolved massive clusters in the Magellanic Clouds
and the Fornax dwarf spheroidal is not quite as simple, although in these
cases the recent studies of \citet{mg03a,mg03b,mg03c} have gone a long way
towards providing a close analogue to the \citet{tkd95}
dataset for Milky Way globulars. Mackey \& Gilmore performed starcounts on the
inner $\approx100\arcsec$ of 68 massive clusters (including 18 old globulars)
from archival HST (WFPC2) images, providing the most accurate definitions of
the {\it core} structures of these objects. (In the LMC, at a distance of
$D=50.1$ kpc, 100\arcsec\ corresponds to 24.2 pc; in the SMC, with $D=60.0$
kpc, $100\arcsec=29.1$ pc; for the Fornax dwarf, $D=137$ kpc and
$100\arcsec=66.4$ pc.) Mackey \& Gilmore further converted their starcount
densities to $V$-band fluxes and published their results as standard
surface-brightness profiles.

Our modeling of the LMC/SMC/Fornax clusters is therefore based in largest part
on the work of \citet{mg03a,mg03b,mg03c}. Again in the interest of
confining our attention to as uniform a raw dataset as possible, we have
not brought into the sample any other Magellanic Cloud clusters studied by
other authors. (Note however, that Mackey \& Gilmore's clusters do include
many previously observed from the ground and modeled by other authors; see
their papers for details of the overlap and comparisons with earlier work.)

We did, however, find it useful to go back through the literature to supplement
Mackey \& Gilmore's HST data with (1) any available ground-based starcount
data at projected radii $R\ga100\arcsec$, and (2) all available
concentric-aperture photometry for these 68 clusters. The detailed comparison
of the HST data against ground-based aperture photometry and large-radius
starcounts, and our combination of these into a single, properly zeropointed 
$V$-band SB profile for each cluster, are what we describe now.

\subsection{LMC, SMC, and Fornax Clusters}
\label{subsec:mackey}

Table \ref{tab:starcount} lists all sources of star counts that we have taken
from the literature for the 53 massive clusters (including 12 old globulars)
studied by \citet{mg03a} in the Large Magellanic Cloud; the 10
(including one GC) studied by \citet{mg03b} in the SMC; and the
5 globular clusters of the Fornax dwarf spheroidal \citep{mg03c}.
We note again that, although Mackey \& Gilmore have published their data as
SB profiles, they are in fact number densities converted to net
mag arcsec$^{-2}$; the other papers listed in Table \ref{tab:starcount}
publish their results directly as number densities $N$ per unit area on the
sky, either already corrected for or accompanied by an estimate of background
contamination. The older data can be combined with those of
Mackey \& Gilmore by transforming the ground-based $N$ as
$\mu=C-2.5\,\log\,N$, with the constant $C$ chosen to provide the best
average agreement with the HST surface brightnesses at clustercentric radii
where the observations overlap.

For each of their 68 clusters in these three Galactic satellites,
\citet[hereafter referred to collectively as MG03]{mg03a,mg03b,mg03c} have
produced a ``primary'' surface brightness profile in the $V$
band and a ``secondary'' profile in one of $B$ or $I$. We have
chosen to work on the $V$ magnitude scale, but to make maximal use of the new
HST data, we have combined the data in the secondary bands with the main
$V$-band data by applying a constant color term to the former. These
instrumental colors are determined simply from integrating to find the total
magnitude of every MG03 cluster in both $V$ and $B$ or $I$, and then adding the
difference $(V-B)$ or $(V-I)$ to every point in the secondary SB profile. This
produces a single profile with two independent estimates of $\mu_V$---both used
in our model fitting below---at every clustercentric radius. The instrumental
colors we applied are listed for reference in Table \ref{tab:colors} and are
in agreement with the total cluster magnitudes already calculated by MG03.

\begin{deluxetable*}{ll}
\tabletypesize{\scriptsize}
\tablecaption{Published Starcount Data for Magellanic Cloud and Fornax
Clusters \label{tab:starcount}}
\tablewidth{0pt}
\tablecolumns{2}
\tablehead{
\colhead{Source} & \colhead{Clusters}
}
\startdata

\sidehead{LMC:}

\citet{kck87}   &   NGC 1711, 1786, 1835, 1847,
                        1850, 1856, 2019, 2100   \\
\citet{khk87}   &   NGC 1777, 1868; Hodge 14;
                    SL 842                        \\
\citet{ckk89}   &   NGC 1754, 1805, 1898, 2031,
                        2121, 2136, 2173, 2210,   \\
       ~~       &   \phantom{NGC }2213, 2231      \\
\citet{eff87}   &   NGC 1818, 1831, 1866, 2004,
                        2156, 2157, 2159, 2164,   \\
       ~~       &   \phantom{NGC }2172, 2214      \\
\citet{mg03a}   &   all of the above, plus:       \\
       ~~       &   NGC 1466, 1651, 1718, 1841,
                        1860, 1916, 1984, 2005,   \\
       ~~       &   \phantom{NGC }2011, 2153, 2155,
                    2162, 2193, 2209, 2249, 2257; \\
        ~~      &   Hodge 4, 11; R136 (30 Dor);
                    SL 663, 855                   \\

\sidehead{SMC:}

\citet{kdk82}   & Kron 3; NGC 152, 176, 361, 458  \\
\citet{kk83}    & NGC 121, 330, 339, 411, 416     \\
\citet{mg03b}   & all of the above                \\

\sidehead{Fornax Dwarf Spheroidal:}

\citet{mg03c}   & FORNAX 1, 2, 3, 4, 5            \\

\enddata
\end{deluxetable*}

\begin{deluxetable*}{lcrclcr}
\tabletypesize{\scriptsize}
\tablewidth{0pt}
\tablecaption{Instrumental Colors of Magellanic Cloud and Fornax Clusters
\label{tab:colors}}
\tablecolumns{7}
\tablehead{
\colhead{Cluster}  &  \colhead{Color}  &  \colhead{Value}  &
\colhead{~~}       &
\colhead{Cluster}  &  \colhead{Color}  &  \colhead{Value}  \\
\colhead{(1)} & \colhead{(2)} & \colhead{(3)} & &
\colhead{(1)} & \colhead{(2)} & \colhead{(3)}
}
\startdata
 LMC-HODGE11 & $(B-V)$ &   $0.35\pm0.09$ & &
 LMC-NGC2156 & $(V-I)$ &   $0.04\pm0.14$ \\
 LMC-HODGE14 & $(B-V)$ &   $0.41\pm0.26$ & &
 LMC-NGC2157 & $(B-V)$ &  $-0.12\pm0.08$ \\
  LMC-HODGE4 & $(B-V)$ &   $0.42\pm0.11$ & &
 LMC-NGC2159 & $(V-I)$ &   $0.10\pm0.11$ \\
 LMC-NGC1466 & $(V-I)$ &   $0.77\pm0.06$ & &
 LMC-NGC2162 & $(B-V)$ &   $0.46\pm0.11$ \\
 LMC-NGC1651 & $(B-V)$ &   $0.49\pm0.11$ & &
 LMC-NGC2164 & $(V-I)$ &   $0.13\pm0.06$ \\
 LMC-NGC1711 & $(V-I)$ &   $0.21\pm0.06$ & &
 LMC-NGC2172 & $(V-I)$ &   $0.12\pm0.12$ \\
 LMC-NGC1718 & $(B-V)$ &   $0.50\pm0.10$ & &
 LMC-NGC2173 & $(B-V)$ &   $0.49\pm0.10$ \\
 LMC-NGC1754 & $(V-I)$ &   $1.02\pm0.07$ & &
 LMC-NGC2193 & $(B-V)$ &   $0.50\pm0.14$ \\
 LMC-NGC1777 & $(B-V)$ &   $0.39\pm0.10$ & &
 LMC-NGC2209 & $(B-V)$ &   $0.09\pm0.20$ \\
 LMC-NGC1786 & $(V-I)$ &   $0.87\pm0.05$ & &
 LMC-NGC2210 & $(V-I)$ &   $0.82\pm0.06$ \\
 LMC-NGC1805 & $(V-I)$ &   $0.08\pm0.10$ & &
 LMC-NGC2213 & $(B-V)$ &   $0.47\pm0.12$ \\
 LMC-NGC1818 & $(V-I)$ &   $0.05\pm0.06$ & &
 LMC-NGC2214 & $(B-V)$ &   $0.00\pm0.09$ \\
 LMC-NGC1831 & $(B-V)$ &   $0.19\pm0.05$ & &
 LMC-NGC2231 & $(B-V)$ &   $0.42\pm0.13$ \\
 LMC-NGC1835 & $(V-I)$ &   $1.02\pm0.06$ & &
 LMC-NGC2249 & $(B-V)$ &   $0.27\pm0.08$ \\
 LMC-NGC1841 & $(V-I)$ &   $0.88\pm0.11$ & &
 LMC-NGC2257 & $(B-V)$ &   $0.37\pm0.10$ \\
 LMC-NGC1847 & $(B-V)$ &  $-0.05\pm0.11$ & &
    LMC-R136 & $(V-I)$ &   $0.36\pm0.08$ \\
 LMC-NGC1850 & $(B-V)$ &   $0.18\pm0.06$ & &
   LMC-SL663 & $(B-V)$ &   $0.29\pm0.21$ \\
 LMC-NGC1856 & $(B-V)$ &   $0.26\pm0.04$ & &
   LMC-SL842 & $(B-V)$ &   $0.40\pm0.19$ \\
 LMC-NGC1860 & $(B-V)$ &  $-0.07\pm0.14$ & &
   LMC-SL855 & $(B-V)$ &   $0.17\pm0.27$ \\
 LMC-NGC1866 & $(V-I)$ &   $0.18\pm0.05$ & &
   SMC-KRON3 & $(B-V)$ &   $0.42\pm0.09$ \\
 LMC-NGC1868 & $(B-V)$ &   $0.29\pm0.07$ & &
  SMC-NGC121 & $(B-V)$ &   $0.49\pm0.06$ \\
 LMC-NGC1898 & $(V-I)$ &   $0.86\pm0.07$ & &
  SMC-NGC152 & $(B-V)$ &   $0.39\pm0.11$ \\
 LMC-NGC1916 & $(V-I)$ &   $1.12\pm0.04$ & &
  SMC-NGC176 & $(B-V)$ &  $-0.01\pm0.18$ \\
 LMC-NGC1984 & $(V-I)$ &   $0.56\pm0.12$ & &
  SMC-NGC330 & $(B-V)$ &  $-0.06\pm0.07$ \\
 LMC-NGC2004 & $(B-V)$ &  $-0.13\pm0.11$ & &
  SMC-NGC339 & $(B-V)$ &   $0.44\pm0.11$ \\
 LMC-NGC2005 & $(V-I)$ &   $0.92\pm0.08$ & &
  SMC-NGC361 & $(B-V)$ &   $0.37\pm0.11$ \\
 LMC-NGC2011 & $(V-I)$ &   $0.29\pm0.16$ & &
  SMC-NGC411 & $(B-V)$ &   $0.44\pm0.11$ \\
 LMC-NGC2019 & $(V-I)$ &   $0.94\pm0.06$ & &
  SMC-NGC416 & $(B-V)$ &   $0.36\pm0.08$ \\
 LMC-NGC2031 & $(V-I)$ &   $0.28\pm0.06$ & &
  SMC-NGC458 & $(B-V)$ &   $0.07\pm0.11$ \\
 LMC-NGC2100 & $(B-V)$ &   $0.02\pm0.11$ & &
     FORNAX1 & $(V-I)$ &   $0.64\pm0.12$ \\
 LMC-NGC2121 & $(B-V)$ &   $0.39\pm0.10$ & &
     FORNAX2 & $(V-I)$ &   $0.78\pm0.09$ \\
 LMC-NGC2136 & $(B-V)$ &   $0.17\pm0.11$ & &
     FORNAX3 & $(V-I)$ &   $0.89\pm0.08$ \\
 LMC-NGC2153 & $(B-V)$ &   $0.30\pm0.15$ & &
     FORNAX4 & $(V-I)$ &   $1.06\pm0.13$ \\
 LMC-NGC2155 & $(B-V)$ &   $0.51\pm0.13$ & &
     FORNAX5 & $(V-I)$ &   $0.90\pm0.09$ \\
\enddata

\tablecomments{Cluster colors are those implied by the integrated magnitudes
derived from surface-brightness profiles out to $R\approx100\arcsec$ as
published by \citet{mg03a,mg03b,mg03c}. Colors are strictly
instrumental, in that the published surface brightnesses have not been
corrected for any zeropoint changes in any bandpass. For clusters with
instrumental $(B-V)$ colors given here, compare with the {\it true} colors
listed in Column (4) of Table \ref{tab:poptable}.}

\end{deluxetable*}

\begin{deluxetable*}{ll}[!t]
\tabletypesize{\scriptsize}
\tablecaption{Published Aperture Photometry for Magellanic Cloud and Fornax
Clusters \label{tab:apphot}}
\tablewidth{0pt}
\tablecolumns{2}
\tablehead{
\colhead{Source} & \colhead{Clusters}
}
\startdata

\sidehead{LMC:}

\citet{ber74}          & NGC 1835, 2019                         \\
\citet{ber75}          & NGC 2173                               \\
\citet{vdb81}          & \citet{mg03a} sample
                         excluding SL 663, 855                  \\
\citet{gor83}          & NGC 1835, 1841, 1856, 1916, 2121       \\
\citet{eff87}          & NGC 1818, 1831, 1866, 2004, 2156,   
                             2157, 2159, 2164,                  \\
                 ~~    & \phantom{NGC }2172, 2214               \\
\citet{bic96}          & \citet{mg03a} sample
                         excluding SL 663, 855                  \\

\sidehead{SMC:}

\citet{vdb81}          & full \citet{mg03b} sample              \\
\citet{gor83}          & full \citet{mg03b} sample              \\

\sidehead{Fornax Dwarf Spheroidal:}

\citet{hod65,hod69}; \citet{web85} & FORNAX 1                               \\
\citet{dev70}                      & FORNAX 2, 3, 4, 5                      \\
\citet{gor83}                      & FORNAX 3, 4                            \\

\enddata
\end{deluxetable*}

After combining the primary $V$ and secondary $B/I$ SB data for each cluster,
we parsed them to be left only with strictly independent surface brightnesses.
MG03 performed starcounts in four different series of concentric apertures
covering each cluster, each series using a different width for the annuli.
Within a given series, the annuli are all of the same width and not
overlapping, but there is annulus overlap between the series. Thus, the
SB profile tabulations from MG03 include interdependent data points. To
avoid such correlations during our model fitting, we simply adopted only the
$(V+B/I)$ surface brightnesses derived by MG03 in their narrowest annuli
($\Delta R=1\farcs5$) for the inner regions of every cluster, and switched
over to take only the widest-annulus results ($\Delta R=4\arcsec$) in the
outer parts. We discarded altogether the SB points estimated from the
intermediate annulus widths. Additionally, to each annulus that we kept we
assigned an average (median) radius determined by the local slope of the
surface-brightness distribution, as described in \citet{king88} (rather than
associating each SB point with the straight mean of the inner and outer radii
of its annulus, as MG03 do).

At this point we were able to match the ground-based starcounts to the
combined HST SB profiles, on the $V$-band magnitude scale defined by MG03.
Background-corrected stellar number densities $N$ and uncertainties are
tabulated for 32 LMC
clusters in the Mackey \& Gilmore sample, and for all 10 of the SMC clusters,
in the papers listed in Table \ref{tab:starcount}. We first converted all of
these to uncalibrated ``surface brightnesses,'' $-2.5\,\log\,N$. In many cases,
multiple plates with different exposure times and limiting magnitudes were
used by the original authors to derive independent estimates of the density
throughout overlapping ranges of radius $R$ within a single cluster. We
shifted such partial $N(R)$ profiles by constants chosen to make the median
difference of the various $\log\,N(R)$, taken over all radii where any count
sets overlap, vanish. The single, ground-based
profile in $-2.5\,\log\,N(R)$ that resulted always extended inward to at least
$R=100\arcsec$ in each cluster, thus overlapping with the MG03 HST $\mu_V$
data. The difference $[\mu_V(R_i)+2.5\log\,N(R_i)]$ averaged over the radii
where ground- and space-based data overlap then defined a constant $C$ which,
when added to $-2.5\,\log\,N(R)$, brings the older starcount data onto
the $V$-band surface brightness scale of Mackey \& Gilmore. In determining
these ``calibrations,'' we never included the ground-based density at the
innermost or outermost radius of any count set, as these are the radii
potentially most susceptible, respectively, to crowding and background errors.

\begin{deluxetable*}{lccccrclccccr}
\tabletypesize{\scriptsize}
\tablewidth{0pt}
\tablecaption{Zeropoint Offsets Applied to Mackey \& Gilmore $V$-band
Surface Brightness \label{tab:offset}}
\tablecolumns{13}
\tablehead{
\colhead{Cluster}         & \colhead{$R_{\rm ap}$}      &
\colhead{$V_{\rm ap}$}    & \colhead{Ref.}              &
\colhead{$V_{\rm ap}$}    & \colhead{$\Delta\mu_V$}     &
\colhead{~~}              &
\colhead{Cluster}         & \colhead{$R_{\rm ap}$}      &
\colhead{$V_{\rm ap}$}    & \colhead{Ref.}              &
\colhead{$V_{\rm ap}$}    & \colhead{$\Delta\mu_V$}     \\
\colhead{}       & \colhead{[sec]}  & \colhead{[lit.]}  &
\colhead{}       & \colhead{[MG03]} & \colhead{}        &
\colhead{~~}     &
\colhead{}       & \colhead{[sec]}  & \colhead{[lit.]}  &
\colhead{}       & \colhead{[MG03]} & \colhead{}        \\
\colhead{(1)} & \colhead{(2)}  & \colhead{(3)}  & \colhead{(4)}  &
\colhead{(5)} & \colhead{(6)}  & &
\colhead{(1)} & \colhead{(2)}  &
\colhead{(3)} & \colhead{(4)} & \colhead{(5)} & \colhead{(6)}
}
\startdata
 LMC-HODGE11 &  30.5 & 11.93 &  1 & $12.158$ & $-0.228\pm0.046$ & &
 LMC-NGC2156 &  27.6 & 11.38 &  3 & $12.670$ & $-1.290\pm0.104$ \\
 LMC-HODGE14 &  31.0 & 13.42 &  2 & $13.571$ & $-0.151\pm0.084$ & &
 LMC-NGC2157 &  30.0 & 10.16 &  2 & $10.920$ & $-0.760\pm0.074$ \\
  LMC-HODGE4 &  19.0 & 13.33 &  1 & $13.489$ & $-0.159\pm0.080$ & &
 LMC-NGC2159 &  36.0 & 11.38 &  2 & $12.569$ & $-1.189\pm0.065$ \\
 LMC-NGC1466 &  30.0 & 11.59 &  2 & $11.963$ & $-0.373\pm0.040$ & &
 LMC-NGC2162 &  31.0 & 12.70 &  2 & $12.814$ & $-0.114\pm0.079$ \\
 LMC-NGC1651 &  50.0 & 12.28 &  1 & $12.254$ &  $0.026\pm0.063$ & &
 LMC-NGC2164 &  30.0 & 10.34 &  2 & $11.603$ & $-1.263\pm0.040$ \\
 LMC-NGC1711 &  30.0 & 10.11 &  2 & $11.431$ & $-1.321\pm0.041$ & &
 LMC-NGC2172 &  36.0 & 11.75 &  2 & $12.976$ & $-1.226\pm0.069$ \\
 LMC-NGC1718 &  31.0 & 12.25 &  2 & $12.219$ &  $0.031\pm0.059$ & &
 LMC-NGC2173 &  75.0 & 11.88 &  1 & $11.901$ & $-0.021\pm0.059$ \\
 LMC-NGC1754 &  50.0 & 11.57 &  1 & $11.907$ & $-0.337\pm0.052$ & &
 LMC-NGC2193 &  19.0 & 13.42 &  1 & $13.582$ & $-0.162\pm0.089$ \\
 LMC-NGC1777 &  19.0 & 12.80 &  1 & $12.997$ & $-0.197\pm0.048$ & &
 LMC-NGC2209 &  34.0 & 13.15 &  2 & $13.940$ & $-0.790\pm0.147$ \\
 LMC-NGC1786 &  30.0 & 10.88 &  2 & $11.453$ & $-0.573\pm0.029$ & &
 LMC-NGC2210 &  34.0 & 10.94 &  2 & $11.492$ & $-0.552\pm0.033$ \\
 LMC-NGC1805 &  30.0 & 10.63 &  2 & $11.945$ & $-1.315\pm0.073$ & &
 LMC-NGC2213 &  31.0 & 12.38 &  2 & $12.466$ & $-0.086\pm0.076$ \\
 LMC-NGC1818 &  36.0 &  9.70 &  2 & $11.181$ & $-1.481\pm0.047$ & &
 LMC-NGC2214 &  30.0 & 10.93 &  2 & $11.408$ & $-0.478\pm0.088$ \\
 LMC-NGC1831 &  30.0 & 11.18 &  2 & $11.219$ & $-0.039\pm0.040$ & &
 LMC-NGC2231 &  22.0 & 13.20 &  2 & $13.516$ & $-0.316\pm0.105$ \\
 LMC-NGC1835 &  31.0 & 10.17 &  2 & $10.652$ & $-0.482\pm0.037$ & &
 LMC-NGC2249 &  75.0 & 11.94 &  1 & $12.035$ & $-0.095\pm0.060$ \\
 LMC-NGC1841 &  93.5 & 11.43 &  1 & $12.164$ & $-0.734\pm0.087$ & &
 LMC-NGC2257 &  30.5 & 12.62 &  1 & $12.747$ & $-0.127\pm0.066$ \\
 LMC-NGC1847 &  36.0 & 11.06 &  2 & $11.512$ & $-0.452\pm0.076$ & &
    LMC-R136 &  30.0 &  8.27 &  2 &  $8.985$ & $-0.715\pm0.078$ \\
 LMC-NGC1850 &  25.0 &  9.57 &  1 &  $9.721$ & $-0.151\pm0.060$ & &
   LMC-SL663 & \nodata & \nodata & -- &  \nodata &  $0$         \\
 LMC-NGC1856 &  31.0 & 10.06 &  2 & $10.042$ &  $0.018\pm0.032$ & &
   LMC-SL842 &  19.0 & 14.15 &  1 & $14.196$ & $-0.046\pm0.160$ \\
 LMC-NGC1860 &  36.0 & 11.04 &  2 & $12.852$ & $-1.812\pm0.134$ & &
   LMC-SL855 & \nodata & \nodata & -- &  \nodata &  $0$         \\
 LMC-NGC1866 &  36.0 &  9.73 &  2 & $10.838$ & $-1.108\pm0.034$ & &
   SMC-KRON3 &  31.0 & 12.05 &  2 & $12.198$ & $-0.148\pm0.046$ \\
 LMC-NGC1868 &  31.0 & 11.57 &  2 & $11.633$ & $-0.063\pm0.044$ & &
  SMC-NGC121 &  31.0 & 11.24 &  2 & $11.548$ & $-0.308\pm0.034$ \\
 LMC-NGC1898 &  20.0 & 11.86 &  1 & $12.463$ & $-0.603\pm0.049$ & &
  SMC-NGC152 &  31.0 & 12.92 &  2 & $12.946$ & $-0.026\pm0.077$ \\
 LMC-NGC1916 &  22.0 & 10.38 &  2 & $10.868$ & $-0.488\pm0.028$ & &
  SMC-NGC176 &  31.0 & 12.70 &  2 & $13.441$ & $-0.741\pm0.123$ \\
 LMC-NGC1984 &  25.0 &  9.99 &  1 & $12.922$ & $-2.932\pm0.086$ & &
  SMC-NGC330 &  31.0 &  9.60 &  2 & $10.621$ & $-1.021\pm0.065$ \\
 LMC-NGC2004 &  36.0 &  9.60 &  2 & $10.380$ & $-0.780\pm0.090$ & &
  SMC-NGC339 &  31.0 & 12.84 &  2 & $12.873$ & $-0.033\pm0.079$ \\
 LMC-NGC2005 &  12.5 & 11.57 &  2 & $12.052$ & $-0.482\pm0.058$ & &
  SMC-NGC361 &  31.0 & 12.12 &  4 & $12.690$ & $-0.570\pm0.085$ \\
 LMC-NGC2011 &  20.0 & 10.58 &  1 & $13.353$ & $-2.773\pm0.175$ & &
  SMC-NGC411 &  31.0 & 12.21 &  2 & $12.234$ & $-0.024\pm0.088$ \\
 LMC-NGC2019 &  36.0 & 10.86 &  2 & $11.314$ & $-0.454\pm0.043$ & &
  SMC-NGC416 &  31.0 & 11.42 &  2 & $11.609$ & $-0.189\pm0.038$ \\
 LMC-NGC2031 &  36.0 & 10.83 &  2 & $11.560$ & $-0.730\pm0.039$ & &
  SMC-NGC458 &  31.0 & 11.73 &  2 & $11.832$ & $-0.102\pm0.092$ \\
 LMC-NGC2100 &  30.0 &  9.60 &  2 & $10.184$ & $-0.584\pm0.101$ & &
     FORNAX1 &  70.0 & 15.57 &  5 & $15.395$ &  $0$             \\
 LMC-NGC2121 &  31.0 & 12.37 &  2 & $12.581$ & $-0.211\pm0.058$ & &
     FORNAX2 &  32.5 & 13.73 &  6 & $14.017$ & $-0.287\pm0.050$ \\
 LMC-NGC2136 &  30.0 & 10.54 &  2 & $10.733$ & $-0.193\pm0.089$ & &
     FORNAX3 &  32.5 & 12.74 &  6 & $13.177$ & $-0.437\pm0.047$ \\
 LMC-NGC2153 &  50.0 & 13.05 &  1 & $13.472$ & $-0.422\pm0.103$ & &
     FORNAX4 &  32.5 & 13.49 &  6 & $14.069$ & $-0.579\pm0.090$ \\
 LMC-NGC2155 &  31.0 & 12.60 &  2 & $12.542$ &  $0.058\pm0.091$ & &
     FORNAX5 &  32.5 & 13.55 &  6 & $14.009$ & $-0.459\pm0.055$ \\
\enddata

\tablecomments{
Six columns for each of 68 star clusters in the LMC, SMC, and Fornax studied
by \citet[collectively MG03]{mg03a,mg03b,mg03c}:
\hfill\break
{\bf Column (1)}---Cluster name.
\hfill\break
{\bf Column (2)}---Aperture radius, in arcsec, of a ground-based magnitude
measurement from previous literature.
\hfill\break
{\bf Column (3)}---Ground-based $V$ magnitude of the cluster within the
radius in Column (2).
\hfill\break
{\bf Column (4)}---Source of the cited ground-based aperture magnitude.
\hfill\break
{\bf Column (5)}---$V$-band magnitude within $R_{\rm ap}$ obtained by
integrating the published surface-brightness profile of MG03.
\hfill\break
{\bf Column (6)}---Zeropoint correction to be {\it added} to MG03 $V$-band
surface brightnesses to bring their cluster magnitudes into agreement
with the earlier, ground-based data.
}

\tablerefs{
(1)---\citet{bic96};
(2)---\citet{vdb81};
(3)---\citet{eff87};
(4)---\citet{gor83};
(5)---\citet{web85};
(6)---\citet{dev70}.
}

\end{deluxetable*}

With these combined, parsed, and extended $\mu_V(R)$ profiles in hand, and
in effect calibrated by \citet{mg03a,mg03b,mg03c}, we found it necessary
to re-examine this calibration---the zeropoint of their conversion from HST
starcounts to $V$-band fluxes---itself. To check it, we used the 
$\mu_V$ profiles as published in MG03 to calculate the integrated (enclosed)
magnitude, $V(\le R)$, as a function of projected radius on scales
$R\la 100\arcsec$ in each cluster in the total LMC+SMC+Fornax
sample. We then compared these integrated profiles against appropriate
ground-based aperture magnitudes. Photometry within apertures
$R_{\rm ap}\la 100\arcsec$ exists in the literature for all but two of the
clusters under consideration, and the sources that we have used are listed in
Table \ref{tab:apphot}. In principle, we might have simply used true SB
profiles from the literature to compare directly against Mackey \& Gilmore's
surface brightness calibration without any integration; but such profiles
exist for a much smaller fraction of the clusters in this sample, so for
reasons of homogeneity we took the aperture-photometry approach in all cases.

Table \ref{tab:offset} presents the results of this comparison for all 68
clusters from MG03. Each cluster has six columns in this table: first is the
cluster name; next, the aperture size, $R_{\rm ap}$, associated with a
ground-based magnitude measurement; then the $V$ magnitude from the literature
and the specific source of this $V(\le R_{\rm ap})$ combination; next, the
integrated $V$ magnitude within $R_{\rm ap}$ implied by Mackey \& Gilmore's SB
profile as published; and finally, the difference $\Delta\mu_V$ of the
ground-based aperture magnitude {\it minus} the MG03 $V$ magnitude.

For the majority of clusters, the $V$-band magnitudes implied by the MG03
surface-brightness calibration are significantly fainter---sometimes by a full
magnitude or more---than any independent aperture photometry. This appears
to be related to the fact that, in performing their number counts, MG03
necessarily had to mask out some bright stars, whether because of
issues with saturation or scattered light, or because they disturbed the
smoother overall distribution of the underlying, more numerous fainter stars
in the cluster. The flux from these stars was {\it not} added back in when the
stellar counts were converted to mag arcsec$^{-2}$, so that the $V$-band
surface brightnesses published in MG03 are systematically too faint
\citep[see the discussion in][especially; also, Mackey 2003, private
communication]{mg03a}.

\begin{figure}
\epsscale{1.20}
\plotone{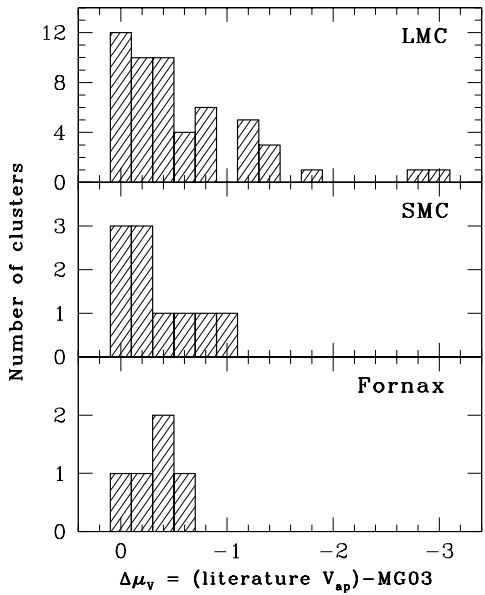}
\caption{\label{fig:zeropts}
Histogram of zeropoint corrections $\Delta\mu_V$ to the $V$-band surface
brightness scale defined for LMC, SMC, and Fornax clusters in
\citet{mg03a,mg03b,mg03c}. Values of $\Delta\mu_V$ are given for individual
clusters in Table \ref{tab:offset}.}
\end{figure}

\begin{figure*}
\epsscale{1.10}
\plotone{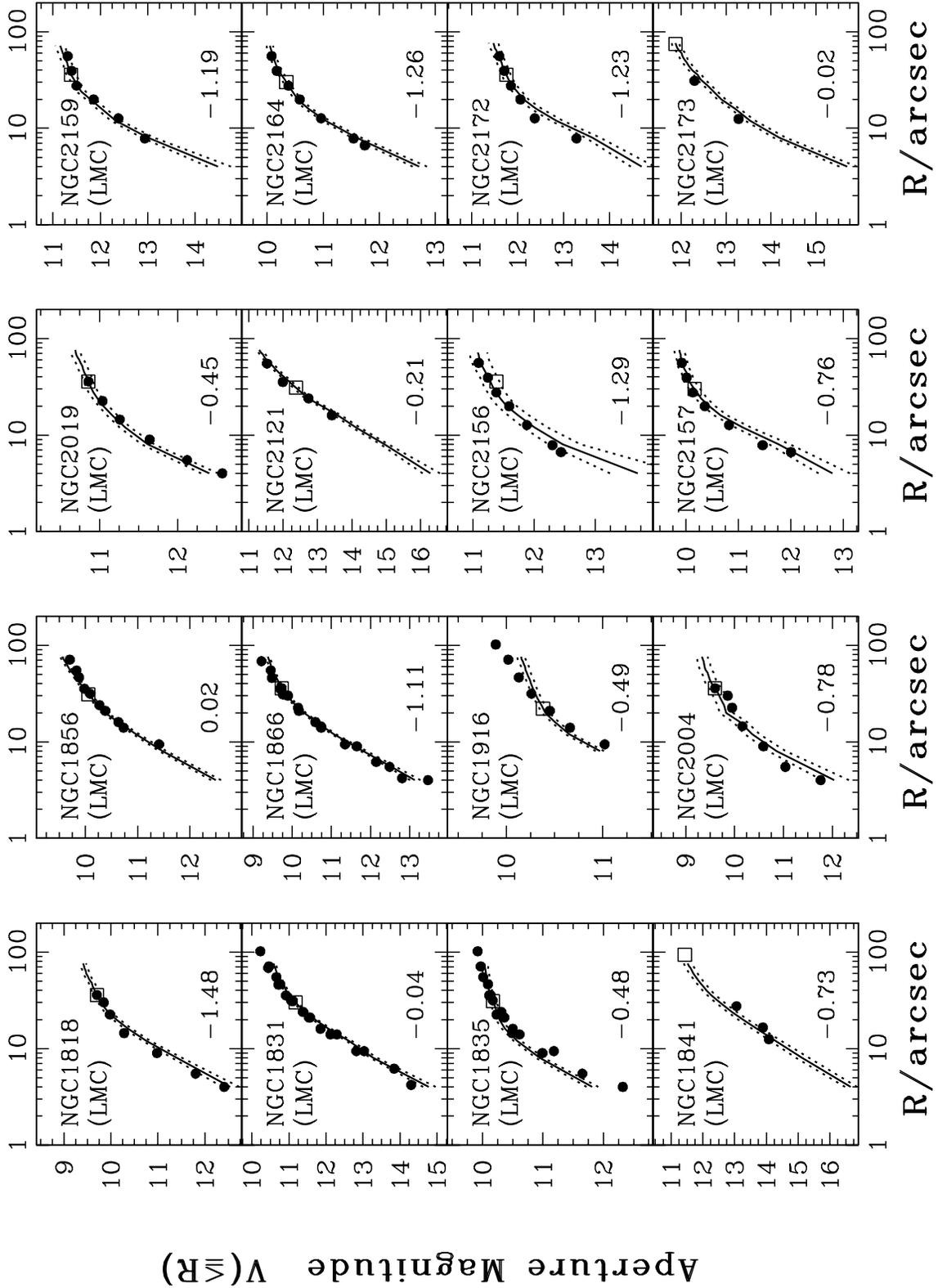}
\caption{\label{fig:apphot}
Aperture magnitude vs.~radius for LMC, SMC, and Fornax clusters from the
sample of \citet{mg03a,mg03b,mg03c}. Only clusters with more than one
ground-based $V_{\rm ap}$ found in the literature are shown. Solid line in
each panel is the integrated magnitude $V(\le R)$ as a function of radius
from the SB profile published by MG03, {\it after} applying the zeropoint
correction given in Table \ref{tab:offset}. Dotted lines demark the
uncertainties derived from the MG03 profiles. Open squares in all panels
denote aperture magnitudes taken from either \citet{bic96} or \citet{vdb81},
filled circles are aperture magnitudes from any other source
(see Table \ref{tab:apphot}), and open circles are measurements cited as
uncertain by the original authors.
}
\end{figure*}

\begin{figure*}
\figurenum{\ref{fig:apphot} [continued]}
\epsscale{1.10}
\plotone{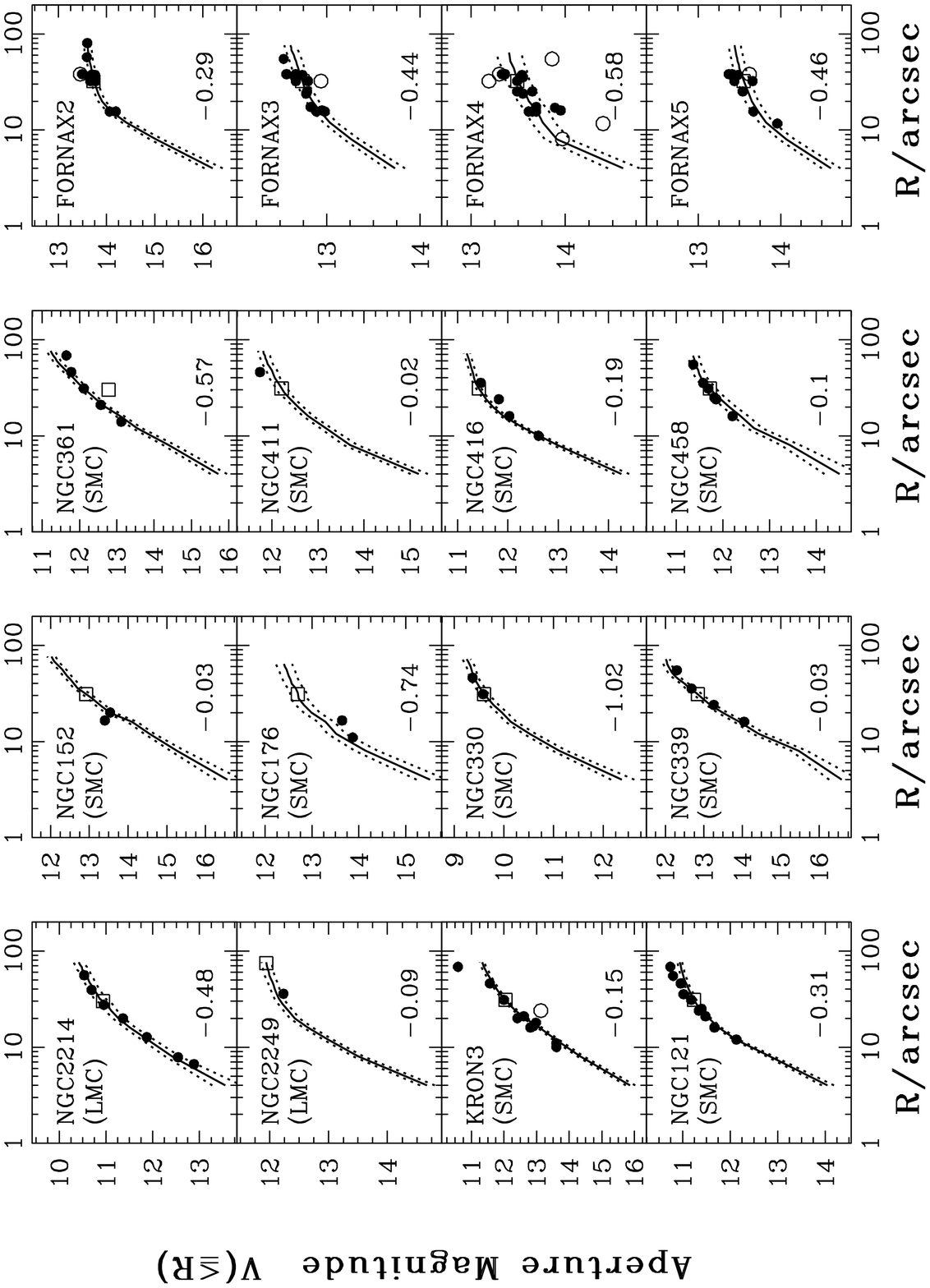}
\caption{}
\end{figure*}

Figure \ref{fig:zeropts} is a histogram of the zeropoint corrections required
to bring the MG03 surface brightnesses into agreement with the independent
aperture photometry we have taken from the literature. Some of these offsets
are rather surprisingly large, particularly in a few of the LMC clusters, and
the possibility exists in principle that the older aperture magnitudes could
be the ones in error---if, say, a bright foreground star were unwittingly
included in the ground-based aperture but properly excluded from the HST
analysis. However, a detailed examination of several specific cases has shown
that the ground-based numbers are {\it not} the ones at fault (Mackey 2003,
private communication). More general support for the same conclusion comes
from the fact that, as we describe just below, there is typically good
agreement between the integrated $V(\le R)$ profile {\it shapes} from the
MG03 and ground-based data in many clusters where multiple aperture magnitudes
can be found in the literature---suggesting that if the ground-based data were
erroneously brightened by foreground stars, these would have had to be
projected essentially onto the very centers of the clusters in an inordinate
number of cases. And the instrumental colors (Table \ref{tab:colors})
of those clusters for which MG03 publish both $B$ and $V$ surface-brightness
profiles do not agree with the aperture-photometry colors from the
literature [Column (4) of Table \ref{tab:poptable} below], even when the
$V$-band zeropoint correction that we infer is negligible (implying that
the missing $V$ flux in MG03 is not generally consistent with a regular
population of foreground stars). Thus, when
fitting models and deriving physical parameters for the Magellanic Cloud and
Fornax clusters (\S\ref{subsec:fits}), we always add the (generally negative)
offsets $\Delta\mu_V$ listed in Table \ref{tab:offset} to the $V$-band
profile numbers published in MG03.

Note that for two of the LMC clusters (SL-663 and SL-855) we
could find no independent $V$ photometry against which to compare the MG03
calibration; while for Globular Cluster 1 in Fornax, various estimates of
the total $V$ magnitude from \citet{web85} and \citet{hod65,hod69}
differ from each other by at least as much as they do from the implied MG03
magnitude. In these three cases, we do not apply any zeropoint
correction to the MG03 surface brightnesses, but we recognize that the
adopted numbers could be in error by tenths of a magnitude in each case.

For 32 of the clusters, ground-based $V$ magnitudes are available in the
literature for more than one aperture---allowing for further checks in these
cases, both of the consistency between the various old magnitude estimates
and of the overall {\it shape} of the new Mackey \& Gilmore profiles. These
checks are presented in Figure \ref{fig:apphot}. Each panel here corresponds
to a different cluster, and all points plotted are aperture magnitudes from
the sources listed in Table \ref{tab:apphot}. Measurements of $V_{\rm ap}$
from \citet{vdb81} and \citet{bic96}---the most comprehensive
listings---are set apart as a large open square in every panel, while the
smaller filled circles refer to reliable data from any other source. The large
open circles, which are particularly evident in the Fornax globular clusters,
denote published magnitudes which the original authors have indicated are
especially uncertain. In all panels of Fig.~\ref{fig:apphot}, the solid line
is the integrated magnitude profile derived from the
$V$-band surface brightness data of MG03, {\it after} a shift brightward by
the offset $\Delta\mu_V$ from Table \ref{tab:offset}, which is listed in each
panel. As was mentioned above, these comparisons show generally rather good
agreement between the shapes of the cluster profiles as measured by MG03, and
what can be inferred from the ground. In the few cases where the two do appear
to differ at some significant level (e.g., NGC 1916 in the LMC), we simply
view the newer HST data as providing an improved estimate of the true relative
density distribution inside these distant clusters.

\begin{deluxetable*}{lccccccrcc}
\tabletypesize{\scriptsize}
\tablecaption{Calibrated $V$-band Surface-Brightness Profiles of
LMC, SMC, and Fornax Clusters  \label{tab:profs}}
\tablewidth{0pt}
\tablecolumns{10}
\tablehead{
\colhead{Cluster}                         & \colhead{$R$}            &
\colhead{$\log\,R$}                       &
\colhead{$\mu_V$\tablenotemark{a}}        &
\colhead{(max)}                           & \colhead{(min)}          &
\colhead{band\tablenotemark{b}}           &
\colhead{$A_V$\tablenotemark{c}}          & \colhead{uncertainty}    &
\colhead{fit flag\tablenotemark{d}}                                  \\
\colhead{}   &   \colhead{[sec]}  &   \colhead{}   &
\multicolumn{3}{c}{[mag arcsec$^{-2}$]}            &
\multicolumn{4}{c}{}                               \\
\colhead{(1)} & \colhead{(2)}  & \colhead{(3)}  & \colhead{(4)}  &
\colhead{(5)} & \colhead{(6)}  & \colhead{(7)}  & \colhead{(8)}  &
\colhead{(9)} & \colhead{(10)}
}
\startdata
LMC-HODGE11 & 1.06  & 0.025 & 18.82 & 19.21 & 18.43 & $V$
            & 0.143 & 0.077 & 1 \\
LMC-HODGE11 & 1.06  & 0.025 & 19.00 & 19.35 & 18.65 & $B$
            & 0.143 & 0.077 & 1 \\
LMC-HODGE11 & 2.37  & 0.374 & 19.53 & 19.78 & 19.28 & $B$
            & 0.143 & 0.077 & 1 \\
  SMC-KRON3 & 1.06  & 0.027 & 19.76 & 20.10 & 19.42 & $V$
            & 0.050 & 0.081 & 1 \\
  SMC-KRON3 & 1.06  & 0.027 & 19.75 & 20.07 & 19.43 & $B$
            & 0.050 & 0.081 & 1 \\
  SMC-KRON3 & 2.37  & 0.375 & 20.08 & 20.26 & 19.90 & $B$
            & 0.050 & 0.081 & 1 \\
    FORNAX1 & 1.41  & 0.150 & 22.84 & 23.29 & 22.39 & $V$
            & 0.220 & 0.077 & 1 \\
    FORNAX1 & 1.41  & 0.150 & 22.61 & 23.08 & 22.14 & $I$
            & 0.220 & 0.077 & 1 \\
    FORNAX1 & 3.16  & 0.500 & 23.41 & 23.60 & 23.22 & $I$
            & 0.220 & 0.077 & 1 \\
\enddata

\tablecomments{A machine-readable version of the complete Table
\ref{tab:profs} is published in the electronic edition of the
{\it Astrophysical Journal Supplement Series};
only a portion is shown here, for guidance regarding its form and content.}

\tablenotetext{a}{Surface brightnesses $\mu_V$ in Column (4) are on our
{\it calibrated} $V$-band scale, with the zeropoint corrections to the
\citet{mg03a,mg03b,mg03c} numbers already applied (see Table \ref{tab:offset}).
Maximum and minimum surface brightnesses in Columns (5) and (6) define the
errorbars on $\mu_V$, which are asymmetric in general. The uncertainty in
the zeropoint correction $\Delta\mu_V$ from Table \ref{tab:offset} is a
source of systematic error; it is {\it not} accounted for in Columns (5) and
(6) but is added in quadrature to the formal, $\chi^2$-based uncertainties
in all fitted and derived cluster parameters relating to surface brightness.}

\tablenotetext{b}{$V$, $B$, or $I$ denotes the original bandpass of a
\citet{mg03a,mg03b,mg03c} surface-brightness datapoint, with $B$ or $I$ data
shifted in zeropoint as in Table \ref{tab:colors}, and then again by an
amount $\Delta\mu_V$ from Table \ref{tab:offset}, to match onto our calibrated
$\mu_V$ scale. $G$ denotes a surface density taken
from ground-based starcounts in the literature (Table \ref{tab:starcount}),
transformed to a surface-brightness and zeropointed to the $\mu_V$ scale
of column (4).}

\tablenotetext{c}{$V$-band extinction $A_V=3.1\,E(B-V)$, with $E(B-V)$
taken from literature values compiled by \citet{mg03c} for the
Fornax globular clusters but inferred from our population-synthesis modeling
in \S\ref{sec:popsyn} (Table \ref{tab:poptable}) for LMC and SMC clusters.
$A_V$ must be {\it subtracted} from columns (4), (5), and (6); it is
spatially constant in each cluster but is listed at every radius as a
convenience.}

\tablenotetext{d}{1 = point used in fitting of structural models in
\S\ref{sec:modeling}; 0 = point {\it not} used in model fitting.}

\end{deluxetable*}

Table \ref{tab:profs} contains the final surface-brightness profiles of the
68 LMC, SMC, and Fornax clusters to which we fit models in
\S\ref{sec:modeling} below. This table is published in its entirety in the
electronic version of ApJS; only a short excerpt is presented here. Each
cluster is given several tens of lines in the table. The first column of each
line is the cluster name, followed by: a radius in arcsec
\citep[the median of an annulus, determined, as mentioned above,
according to][]{king88};
the logarithmic radius; the $V$-band surface brightness measured at that
radius {\it after adjusting the magnitude scale of Mackey \& Gilmore by the
zeropoint offset from Table \ref{tab:offset}}; the faint and bright limits
on $\mu_V$ (reflecting the possibility of asymmetric errorbars, especially
at faint intensity levels); the bandpass from which the datapoint originally
came (with $G$ denoting ground-based number densities scaled as described
above); an estimate of the $V$-band extinction towards the cluster
(assumed spatially constant in each cluster), and its uncertainty; and a
flag indicating whether or not the point was included explicitly when fitting
the models of \S\ref{sec:modeling}.

To determine whether or not any given point should contribute to the weighting
of the model fits---i.e., in setting the ``fit flag'' of Table \ref{tab:profs}
to 1 or 0---we first used all the tabulated $V$--$B/I$--ground-based SB values
for a cluster to produce a smoothed (nonparametric) density profile. Any
individual points that fell more than $2\sigma$ away from this smoothed
approximation were assigned a fit flag of 0.

The $V$-band extinctions in Table \ref{tab:profs} require further explanation,
since direct estimates of the reddening of individual Magellanic Cloud clusters
do not exist for the majority in this sample. However, in order to
accurately determine cluster physical parameters---true central surface
brightnesses, total luminosities and masses, etc.---some knowledge of the
extinction is clearly required. MG03 assign a single, average
extinction to all LMC clusters, and another average to all SMC clusters; but
the reddening is known to vary across these systems. We have instead used
population-synthesis modeling, given an age and a metallicity for each cluster
as compiled from the literature by MG03, to {\it predict} an intrinsic
$(B-V)_0$ color for every object here. Observed $(B-V)$ colors from
the aperture-photometry literature listed in Table \ref{tab:apphot} then imply
reddenings $E(B-V)$, and extinctions $A_V=3.1\,E(B-V)$ follow.
More details of this procedure are given in \S\ref{sec:popsyn}, where we also
use population-synthesis models to produce estimates of cluster mass-to-light
ratios. In the next subsection, we first briefly describe some points related
to our handling of the \citet{tkd95} catalogue data for Milky Way globular
clusters.

\subsection{Milky Way Globular Clusters}
\label{subsec:trager}

\begin{deluxetable*}{lrclrclrclr}
\tabletypesize{\scriptsize}
\tablewidth{0pt}
\tablecaption{Base Errorbars Assigned to Milky Way Globular-Cluster
Surface Brightnesses in Catalogue of Trager et al.~(1995)
\label{tab:MWerrs}}
\tablecolumns{11}
\tablehead{
\colhead{Cluster}   & \colhead{$\sigma_\mu$}  & \colhead{~~} &
\colhead{Cluster}   & \colhead{$\sigma_\mu$}  & \colhead{~~} &
\colhead{Cluster}   & \colhead{$\sigma_\mu$}  & \colhead{~~} &
\colhead{Cluster}   & \colhead{$\sigma_\mu$}
}
\startdata
    AM1 & 0.300 & & NGC5986 & 0.081 & & NGC6397 & 0.199 & & NGC6752 & 0.173 \\
   ARP2 & 0.200 & & NGC6093 & 0.171 & & NGC6401 & 0.119 & & NGC6760 & 0.207 \\
     HP & 0.248 & & NGC6101 & 0.182 & & NGC6402 & 0.137 & & NGC6779 & 0.207 \\
 IC1276 & 0.259 & & NGC6121 & 0.180 & & NGC6426 & 0.238 & & NGC6809 & 0.197 \\
 IC4499 & 0.218 & & NGC6139 & 0.136 & & NGC6440 & 0.110 & & NGC6864 & 0.138 \\
 NGC104 & 0.082 & & NGC6144 & 0.155 & & NGC6441 & 0.076 & & NGC6934 & 0.173 \\
NGC1261 & 0.187 & & NGC6171 & 0.143 & & NGC6453 & 0.150 & & NGC6981 & 0.223 \\
NGC1851 & 0.249 & & NGC6205 & 0.128 & & NGC6496 & 0.307 & & NGC7006 & 0.165 \\
NGC1904 & 0.094 & & NGC6218 & 0.182 & & NGC6517 & 0.098 & & NGC7078 & 0.178 \\
NGC2298 & 0.147 & & NGC6229 & 0.151 & & NGC6522 & 0.140 & & NGC7089 & 0.107 \\
NGC2419 & 0.120 & & NGC6235 & 0.188 & & NGC6528 & 0.283 & & NGC7099 & 0.115 \\
NGC2808 & 0.124 & & NGC6254 & 0.137 & & NGC6535 & 0.348 & & NGC7492 & 0.253 \\
 NGC288 & 0.158 & & NGC6256 & 0.197 & & NGC6539 & 0.235 & &    PAL1 & 0.202 \\
NGC3201 & 0.206 & & NGC6266 & 0.126 & & NGC6541 & 0.201 & &   PAL10 & 0.075 \\
 NGC362 & 0.097 & & NGC6273 & 0.124 & & NGC6544 & 0.205 & &   PAL11 & 0.220 \\
NGC4147 & 0.242 & & NGC6284 & 0.122 & & NGC6553 & 0.162 & &   PAL12 & 0.395 \\
NGC4372 & 0.428 & & NGC6287 & 0.158 & & NGC6558 & 0.117 & &   PAL13 & 0.454 \\
NGC4590 & 0.176 & & NGC6293 & 0.146 & & NGC6569 & 0.240 & &   PAL14 & 0.140 \\
NGC5024 & 0.153 & & NGC6304 & 0.148 & & NGC6584 & 0.148 & &    PAL2 & 0.359 \\
NGC5053 & 0.244 & & NGC6316 & 0.139 & & NGC6624 & 0.130 & &    PAL3 & 0.120 \\
NGC5139 & 0.142 & & NGC6325 & 0.162 & & NGC6626 & 0.137 & &    PAL4 & 0.129 \\
NGC5272 & 0.191 & & NGC6333 & 0.169 & & NGC6637 & 0.093 & &    PAL5 & 0.190 \\
NGC5286 & 0.128 & & NGC6341 & 0.119 & & NGC6638 & 0.111 & &    PAL6 & 0.098 \\
NGC5466 & 0.120 & & NGC6342 & 0.165 & & NGC6642 & 0.140 & &    PAL8 & 0.220 \\
NGC5634 & 0.220 & & NGC6352 & 0.269 & & NGC6652 & 0.190 & & TERZAN1 & 0.142 \\
NGC5694 & 0.120 & & NGC6355 & 0.152 & & NGC6656 & 0.173 & & TERZAN2 & 0.157 \\
NGC5824 & 0.106 & & NGC6356 & 0.074 & & NGC6681 & 0.186 & & TERZAN5 & 0.129 \\
NGC5897 & 0.118 & & NGC6362 & 0.136 & & NGC6712 & 0.218 & & TERZAN6 & 0.099 \\
NGC5904 & 0.104 & & NGC6366 & 0.217 & & NGC6715 & 0.117 & & TERZAN7 & 0.236 \\
NGC5927 & 0.089 & & NGC6380 & 0.173 & & NGC6717 & 0.215 & & TERZAN9 & 0.128 \\
NGC5946 & 0.136 & & NGC6388 & 0.101 & & NGC6723 & 0.212 & &    TON2 & 0.162 \\
\enddata

\tablecomments{$\sigma_{\mu}$ is a constant for each Galactic globular
cluster in the catalogue of \citet{tkd95}, used to estimate errorbars
on individual surface-brightness datapoints. Given a relative weight $w_i$
($0\le w_i\le 1$) from Trager et al.~for each datapoint in a cluster,
we define the uncertainty in $\mu_V$ to be $\sigma_i\equiv \sigma_{\mu}/w_i$.}

\end{deluxetable*}

As was mentioned above, \citet{tkd95} have already taken steps, similar
to those we have outlined above and applied to LMC, SMC, and Fornax
clusters, to combine heterogeneous data into unique, calibrated
$V$-band SB profiles for 124 Galactic globular clusters. However, their
results as published do not include any estimates of uncertainty in the
individual surface-brightness points. The profiles defined above for the MG03
cluster sample {\it do} include errorbars, and we use these during our model
fitting in \S4 to estimate the uncertainties in all basic and derived physical
parameters of the clusters, via a standard analysis of $\chi^2$ variations
over grids of model fits.
We would like to proceed in the same way with the Milky Way GC sample, and
thus we have attempted to estimate errorbars for the $\mu_V$ values given
by \citet{tkd95}.

In lieu of absolute errorbars, \citeauthor{tkd95} give each point in their
brightness profiles a relative weight $w_i \in [0,1]$. As they discuss,
these weights were assigned ``by eye'' to reflect the authors' judgement of
the overall quality
of the source dataset. They are therefore not connected rigorously
to relative errors, and their precise meaning is left somewhat open to
interpretation. Initially, we proceeded under the natural assumption that the
weights were proportional to the inverse square of the surface-brightness
uncertainties, or $\sigma_i \propto 1/\sqrt{w_i}$.
When fitting models by minimizing an error-weighted $\chi^2$
statistic, this led to some cases where the best-fit model parameters were
unduly influenced by just a few discrepant points with low weights from
\citeauthor{tkd95}. Thus, we decided instead to de-weight such points even
further by adopting the heuristic prescription that the relative
surface-brightness errorbars grow as $\sigma_i \propto 1/w_i$.

Given this choice, we estimated the uncertainties as follows.
\citet{tkd95} also tabulate the value at each radius in each cluster, of
interpolating (Chebyshev) polynomials that provide reasonably accurate,
model-independent approximations to the overall SB profile. We assume that the
reduced $\chi^2$ per degree of freedom for these polynomial fits is exactly 1
for every cluster. Typically, the polynomials fit by \citeauthor{tkd95} are of
third order, so for a surface-brightness profile with $N$ datapoints there are
$N-4$ degrees of freedom. Then, with our prescription
$\sigma_i \equiv \sigma_{\mu}/w_i$, where $\sigma_\mu$ is a different
constant for each GC, and writing $C(R_i)$ for the value of the
polynomial fit at each radius $R_i$ in any one cluster, we
have that
$\frac{1}{N-4}
\sum_{i=1}^{N} (w_i^2/\sigma_{\mu}^2)\left[C(R_i)-\mu_V(R_i)\right]^2 = 1$.
We have solved this identity for the ``base'' errorbar $\sigma_\mu$ of each
of the 124 clusters in the collection of \citet{tkd95}, and reported the
results in Table \ref{tab:MWerrs}.
Again, the uncertainty in $\mu_V$ at any single radius $R_i$ in any one
cluster is taken as $\sigma_{\mu}/w_i$, assumed to be symmetric
about the measured surface brightness.

For two globular clusters---Palomar 10 and Terzan 7---Trager et al.~only
present uncalibrated surface brightnesses, $\mu_V$ profiles relative to an
unknown central value. Subsequently, the central brightnesses of these objects
have been determined, and they are tabulated in the catalogue of Harris (1996):
$\mu_{V,0}=22.12$ for Palomar 10, and $\mu_{V,0}=20.69$ for Terzan 7. We have
used these values as zeropoint ``corrections'' to the Trager et al.~data, and
treated them subsequently in the same fashion as our zeropoint shifts to all
of the Mackey \& Gilmore LMC/SMC/Fornax cluster surface brightnesses.

Aside from these two small points, we have proceeded with modeling the Trager
et al.~data as published, with no further embellishments nor any attempted
updates. There are another 26 globular clusters included in the more recent
catalogue of \citet{har96}, but we have not made any attempt to model their
density structures: the raw profiles of these additional clusters have not
been gathered into a uniform collection on par with the Trager et
al.~database, and indeed many of them are rather obscure and not studied well
enough for our purposes in the first place. For completeness, we list them
in Table \ref{tab:MWnotfit} as objects that we have not fit with structural
models in \S\ref{sec:modeling}. We do, however, include most of them in the
population-synthesis modeling of \S\ref{sec:popsyn}, since only estimates of
their metallicities are required there.

\begin{deluxetable*}{ll}
\tabletypesize{\scriptsize}
\tablecaption{Milky Way Globular Clusters not fit by Structural Models
\label{tab:MWnotfit}}
\tablewidth{0pt}
\tablecolumns{2}
\tablehead{
\colhead{Reason}              & \colhead{Clusters}  
}
\startdata

Clusters in \citet{har96} but & 1636--283; 2MS--GC01, GC02; AM4; BH176;
                                DJORG 1, 2;                               \\
not in \citet{tkd95}          & E 3; ERIDANUS; ESO--SC06; IC 1257;
                                LILLER 1; LYNGA 7;                        \\ 
                              & NGC 4833, 6540, 6749, 6838; PAL 15; PYXIS;
                                RUP 106;                                  \\
                              & TERZAN 3, 4, 8, 10, 12; UKS 1             \\

 ~~ & ~~ \\

Clusters in
\citet{tkd95}                 & NGC 4372, 5927, 5946, 6144, 6256, 6284,
                                    6293, 6304, \\
with large fitted $R_h$
(see text)                    & \phantom{NGC }6325, 6342, 6352, 6355,
                                              6380, 6401, 6426, 6453,     \\
                           ~~ & \phantom{NGC }6517, 6522, 6544, 6558,
                                              6624, 6626, 6642, 6717;     \\
                           ~~ & HP; PAL 6, 8, 13; TERZAN 1, 2, 5, 6, 9;
                                TON 2                                     \\

~~ & ~~ \\

Core-collapsed clusters       & NGC 6397, 6681, 6752, 7078, 7099

\enddata
\end{deluxetable*}

We originally attempted to fit all 124 globulars from \citet{tkd95}
with the variety of models that we also fit to the massive clusters in the
Magellanic Clouds and Fornax. In doing so we found that for several clusters
the effective (projected half-light) radius $R_h$ returned by the fits was of
order or larger than the projected radius $R_{\rm last}$ of the outermost
SB datapoint tabulated by Trager et al. In particular, there are 34 GCs
for which $R_h/R_{\rm last}>0.9$,
signifying either that the observations are too sparse to significantly
constrain important aspects of the models, or that the models we fit are
simply poor descriptions of the data. In fact, these 34 objects include
a majority of those designated as core-collapse or ``possible'' core-collapse
candidates by \citet{tkd95}, and subsequently in the catalogues of
\citet{djo93} and \citet{har96}. As Trager et al.~make very clear, their
own estimation of \citet{king66} model parameters for such globulars are
essentially rough guesses and not quantitatively trustworthy. Our fits to 
the 34 GCs with $R_h/R_{\rm last}>0.9$  are likewise rather uncertain, and we
do not present any detailed results for them in \S\ref{sec:modeling} and later
sections. Again, however, we name the clusters in Table \ref{tab:MWnotfit} for
completeness, and we include
them in the population-synthesis analysis that we describe next.\footnotemark
\footnotetext{In \S\ref{subsubsec:goodness} it will be seen that there are
also 4 LMC
clusters in the MG03 sample which have $R_h/R_{\rm last}>1$ and particularly
uncertain fit extrapolations. In these cases, however, this is because the
clusters have rather large cores rather than any hint of a post-collapse
morphology. Given this, and the fact that they have not been fully modeled
before, we do report all of our results for these objects.}
After making this cut, five other core-collapsed globulars remain in the 
Trager et al.~sample. These are known a priori not to be properly
described by King models (or others with constant-density cores) and so we 
remove them as well from consideration for structural modeling, but list them
in Table \ref{tab:MWnotfit} and include them in our population-synthesis
modeling.

\section{Population-Synthesis Models: Reddenings and Mass-to-Light Ratios}
\label{sec:popsyn}

Before proceeding with the structural and dynamical modeling of the clusters
in our Magellanic Cloud, Fornax, and Milky Way sample, we discuss some
aspects of population-synthesis modeling. We have folded this into our
derivation of physical cluster parameters from model fits to the cluster SB
distributions. There are two main issues at hand that necessitate these
considerations:

First, in order to move from a description of the observed surface brightness
profile of a cluster to one of the true luminosity density profile, we need
information on the extinction towards the cluster. For each of the Milky Way
globular clusters that we model, an estimate of the foreground reddening
$E(B-V)$ is available in the catalogue of \citet{har96}; the extinction
follows directly, and the required corrections are straightforward. The same
is true of the globulars in Fornax; \citet{mg03c}, for example, have tabulated
measurements from the literature of the reddenings $E(B-V)$ of each cluster.
But, as was mentioned above, there is no such information in the
literature for many of the 63 LMC and SMC clusters being
analyzed here. Aperture measurements (ground-based) of the clusters'
$(B-V)_{\rm ap}$ colors do exist, however (see the references in Table
\ref{tab:apphot}), as do individual estimates of their ages and
metallicities \citep[compiled from the literature, again, by][]{mg03a,mg03b}.
We therefore use the cluster ages and [Fe/H] values to derive
an expected intrinsic $(B-V)_0$ from population-synthesis models, and then
$E(B-V)=(B-V)_{\rm ap}-(B-V)_0$ and $A_V=3.1\,E(B-V)$. The fact that
direct reddening measurements do exist for the Milky Way and Fornax
globular clusters allows us to perform the same analysis on them, and then
compare our ``theoretical'' $E(B-V)$ values to the known ones---a valuable
check on the method, albeit only in the extreme of old ages.

Second, to go from a description of the de-reddened luminosity density profile
of a cluster to its {\it mass} density distribution, knowledge of an average
mass-to-light ratio is required. Given a sample of clusters of a roughly
common age (and a common stellar IMF), the mass-to-light ratio typically can
also be treated as roughly constant, at least to within a factor of order
unity which depends on [Fe/H] differences. Luminosity differences are then
essentially proportional to mass differences, and detailed knowledge of the
exact mass-to-light ratio is not critical to the accurate discernment of
relative trends in physical cluster properties
(such as mass-radius, mass-velocity
dispersion, or other fundamental-plane correlations). In our case, however, we
aim ultimately to compare these sorts of correlations for the old globular
clusters in the Milky Way and other galaxies (ages $\tau\sim 13$ Gyr for
the most part), against those for the much younger massive clusters in the
Magellanic Clouds ($\tau<10^9$ yr and as young as $\tau\simeq3\times10^6$ yr
for R136=30 Doradus). Trends or correlations in luminosity then reflect a
complex mix of age and mass effects which must be separated to make sense of
the physical situation. Ideally, we would have liked to use measurements of
the stellar velocity dispersions in the clusters to compute their
mass-to-light ratios directly; but, as for the reddenings, such measurements
exist for only a handful of the young LMC/SMC clusters, and they tend to be
highly uncertain. Even in the Milky Way globular cluster system,
reliable velocity-dispersion measurements and dynamical $M/L$ ratios exist
for fewer than half of the 85 clusters that we model here. Therefore, we also
use population-synthesis models to define a $V$-band mass-to-light ratio for
every cluster in our total sample. In \S\ref{sec:dynml} we compare these
predicted values with the dynamical mass-to-light ratios that can be computed
for the minority of (mostly old) clusters with measured velocity dispersions.

The population-synthesis model that we have chosen to use to derive cluster
reddenings and mass-to-light ratios is that of \citet{bru03},
using the Padova 1994 stellar-evolution tracks and assuming a stellar IMF
following that of \citet{cha03} for the Galactic disk population. The
results are presented in \S\ref{subsec:results}. However, since they are so
central to the establishment of all final, physical (mass-based) cluster
properties and to the definition of systematic interdependences between these
properties, we have also carried through the full suite of calculations with
the alternate population-synthesis code P\'EGASE \citep[version 2.0]{frv97}
and under different assumptions on the form of the
stellar IMF. In \S\ref{subsec:comp}, then, we first present point-by-point
comparisons between the $E(B-V)$ and $\Upsilon_V\equiv M/L_V$ values predicted
by various combinations of codes and IMFs. Note, however, that we {\it always}
assume that every cluster is a single-age population formed instantaneously
in one coherent burst of star formation, and that all clusters, of any age or
metallicity in any galaxy, share a common stellar IMF. Also, both
\citet{bru03} and \citet{frv97} have coded
prescriptions for mass loss over time due to stellar-evolution debris
(from winds and supernovae, essentially) that is assumed to be swept out of a
cluster. When using either code, we always employ these prescriptions.

\subsection{Comparison of Codes and Stellar IMFs}
\label{subsec:comp}

\begin{figure}
\epsscale{1.30}
\plotone{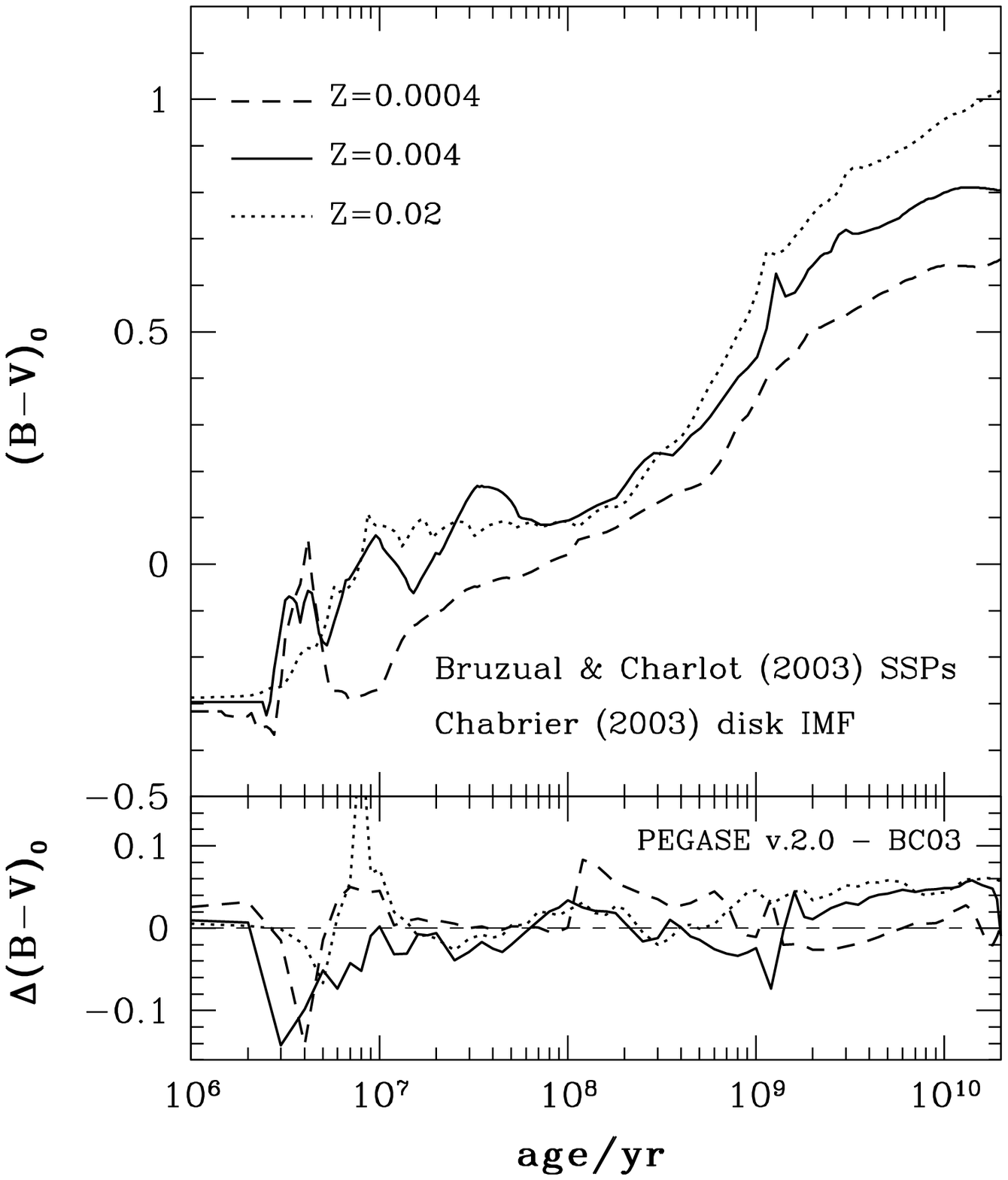}
\caption{\label{fig:popcols}
{\it Top panel}: Intrinsic color as a function of age for single-burst
stellar populations of three different metallicities, as predicted by
the population-synthesis code of \citet{bru03} assuming the
disk-star IMF of \citet{cha03}. {\it Bottom panel}: Difference in $(B-V)_0$
for the same model clusters as predicted by the P\'EGASE (v.2.0)
population-synthesis code of \citet{frv97}.
}
\end{figure}

\begin{figure}
\epsscale{1.30}
\plotone{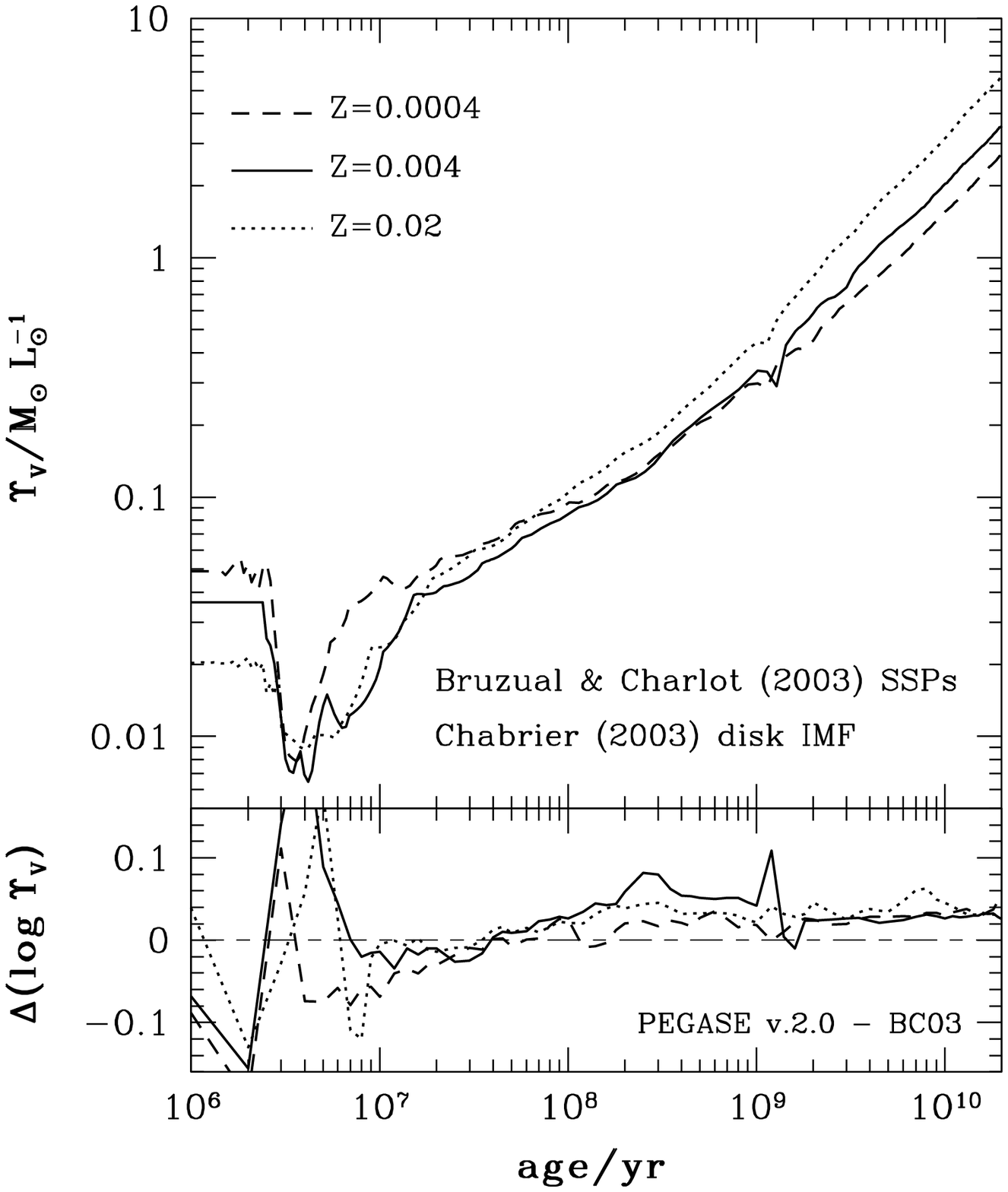}
\caption{\label{fig:popmtol}
{\it Top panel}: $V$-band mass-to-light ratio as a function of age for
single-burst stellar populations of three different metallicities, as
predicted by the population-synthesis code of \citet{bru03}
assuming the disk-star IMF of \citet{cha03}. {\it Bottom panel}: Difference
in $\log\,\Upsilon_V$ for the same model clusters as predicted by the
P\'EGASE (v.2.0) population-synthesis code of \citet{frv97}.
Both codes allow for mass loss from the clusters due to the evacuation of
stellar-wind and supernova debris over time. This amounts to a $\sim30\%$
reduction in total cluster mass after a Hubble time.
}
\end{figure}

The \citet{cha03} disk-star IMF that we adopt for our primary calculations
is a \citet{sal55} power law ($dN/dm\propto m^{-2.35}$) for
stellar masses $m\ge 1\,M_\odot$, and a much flatter, lognormal
distribution below $m\le 1\,M_\odot$. Figures \ref{fig:popcols} and
\ref{fig:popmtol} show, in their larger (upper) panels, the intrinsic colors
$(B-V)_0$ and $V$-band mass-to-light ratios $\Upsilon_V$ predicted by the
\citet{bru03} code for a single-burst (or ``simple'') stellar population with
this IMF, as functions of cluster age for three set heavy-element abundances
roughly spanning the range appropriate for our combined sample of globular
and young massive clusters. The smaller (bottom) panels in these figures show
the {\it differences} in $(B-V)_0$ and $\log\,\Upsilon_V$ as predicted by
the P\'EGASE code of \citet{frv97} given the same IMF
and the same metallicities. Aside from the problematic region around ages
$10^6\la \tau \la 10^7$ yr, there is generally good agreement between the two
codes.

\begin{figure*}
\epsscale{1.10}
\plotone{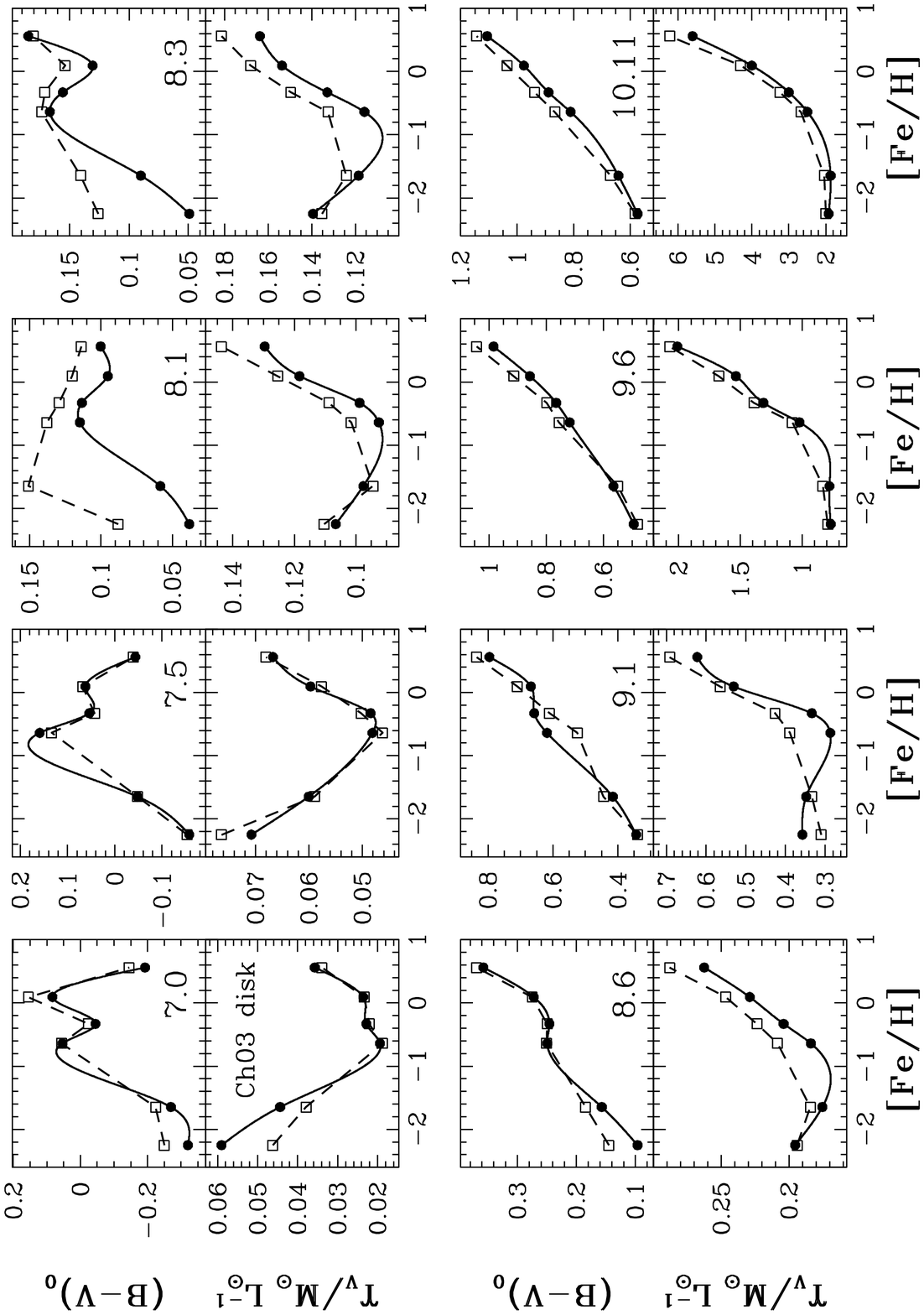}
\caption{\label{fig:popcomp}
Intrinsic $(B-V)$ colors and $V$-band mass-to-light ratios as functions of
metallicity for single-burst stellar populations at a range of fixed ages.
Filled circles and solid lines in every panel refer to predictions from
the population-synthesis code of Bruzual \& Charlot (2003); open squares and
dashed lines refer to predictions from the code of
\citet[P\'EGASE v.2.0]{frv97}. The logarithm of the cluster age is given in
the upper $(B-V)_0$ panel of each pair. All calculations employ the disk-star
IMF of \citet{cha03}.
}
\end{figure*}

Figure \ref{fig:popcomp} expands somewhat on these plots, showing $(B-V)_0$
and $\Upsilon_V$ predicted as functions of cluster [Fe/H] at various ages,
all for a \citet{cha03} disk IMF. The logarithm of cluster age is
labeled in the upper panel of each pair in this figure. Filled circles are
the quantities predicted by \citet{bru03} at the six metallicities
for which their code has stellar-evolution tracks. The solid curve joining
these points is a spline interpolation, which we use to infer $(B-V)_0$
and $\Upsilon_V$ for clusters with metallicities between the few that are
explicitly calculable (at any age) with the Bruzual-Charlot code. The open
squares in every
panel are the synthetic colors and mass-to-light ratios predicted by
P\'EGASE v.2.0, again at the six metallicities for which that code has
explicit stellar-evolution tracks (these metallicities, and indeed the
tracks themselves, are the same as those used by Bruzual \& Charlot). The
dashed lines joining the squares are linear interpolations for determining
quantities at intermediate [Fe/H]. (The P\'EGASE code does this interpolation
internally, given any arbitrary [Fe/H] specified by the user.)

More detailed discussion of the comparison between the P\'EGASE and
Bruzual-Charlot codes---including the difficulties at young ages $\tau<10^7$
years, which fortunately are not relevant to us in general---may be found in
\citet{bru03}. The plots we have presented
have vertical-axis scales chosen deliberately to emphasize the differences
between the codes for one IMF. Evidently, although there are some exceptions,
these differences are generally slight: at the level of $\la 0.05$ mag in
$(B-V)_0$ and $\la 10\%$ in $\Upsilon_V$. Again, we have adopted the Bruzual
\& Charlot code to define the intrinsic colors and mass-to-light ratios
for our clusters, and we can proceed with some confidence that this choice is
not introducing sizeable systematic errors in our subsequent analyses.

The next question concerns the implications of our choice of the \citet{cha03}
disk-star IMF for the population-synthesis calculations. The
Bruzual-Charlot code offers only the choice between this option or a pure
\citet{sal55} power law at all stellar masses (which is well known by now
to be an incorrect description of the true IMF below $\sim 1\,M_\odot$, but is
still commonly used for reference calculations). By contrast, the P\'EGASE
code of Fioc \& Rocca-Volmerange can produce models for any user-defined IMF.
We have therefore used P\'EGASE to calculate $(B-V)_0$ and $\Upsilon_V$
as functions of age for clusters of three different metal abundances
($Z=0.0004$, $Z=0.004$, and $Z=0.02$, as in Figs.~\ref{fig:popcols} and
\ref{fig:popmtol}), given three stellar IMFs besides our preferred
\citet{cha03} disk-star distribution.

\begin{figure}
\epsscale{1.30}
\plotone{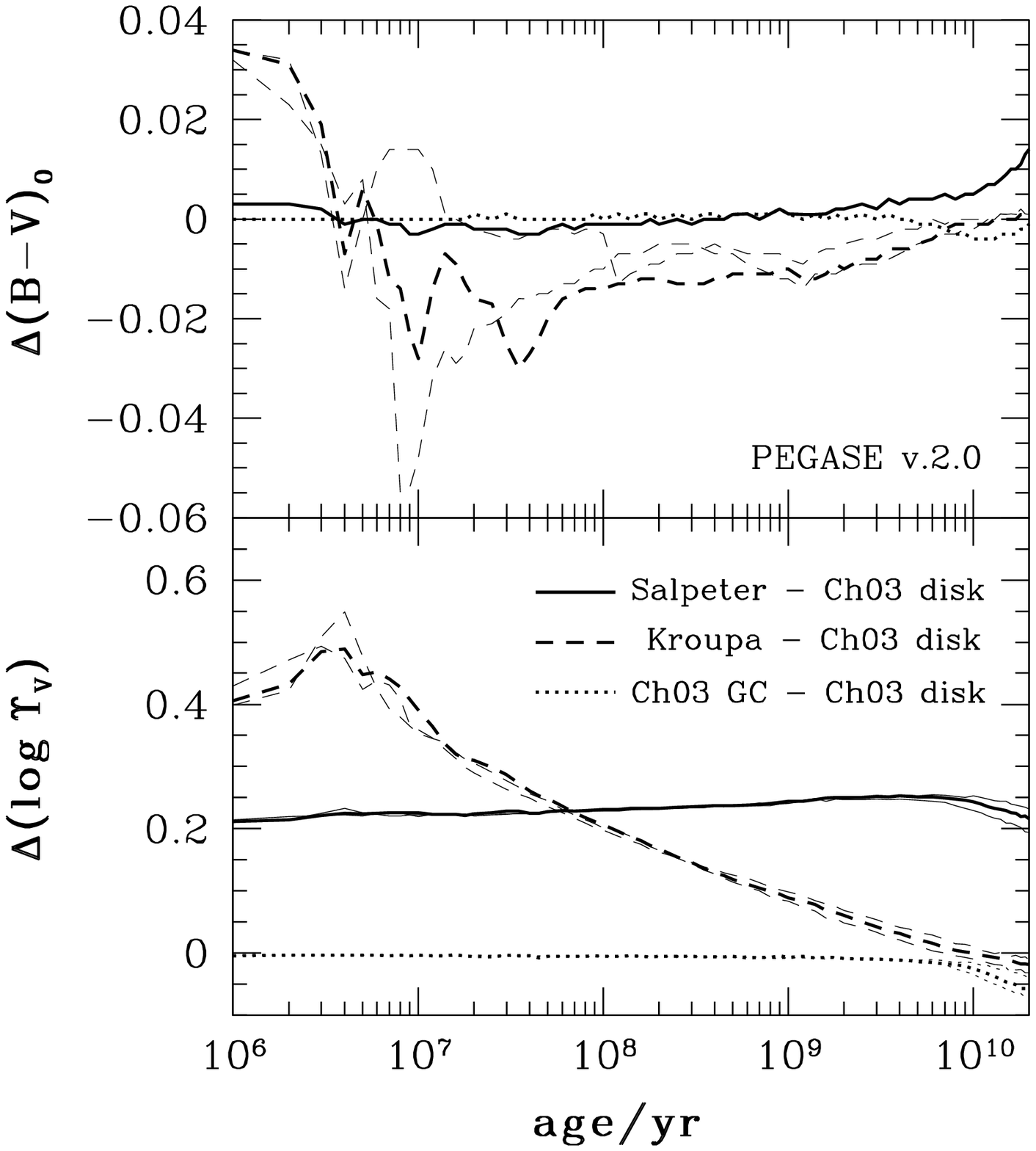}
\caption{\label{fig:imfcomp}
{\it Top panel}: IMF dependence of intrinsic color $(B-V)_0$ of a single-burst
stellar population in the population-synthesis code P\'EGASE (v.2.0)
\citep{frv97}. {\it Bottom panel}: IMF dependence of the logarithm
of $V$-band mass-to-light ratio $\log\,\Upsilon_V$ in the P\'EGASE code.
Results are shown as differences of the computed quantities using the various
IMFs indicated, {\it minus} the ``reference'' values obtained with a
\citet{cha03} disk-star IMF. Bold lines show the differences as functions of
cluster age for a metal abundance $Z=0.004=0.2Z_\odot$; lighter lines, when
shown, correspond to $Z=0.0004$ and $Z=0.02$.
}
\end{figure}

The alternate IMFs that we have examined are those of \citet{sal55} (a single
power law, $dN/dm \propto m^{-2.35}$ at all masses); \citet{ktg93} (a
three-part piecewise power law which, with $dN/dm\propto m^{-2.7}$
above $m>1\,M_\odot$, is significantly steeper than either Salpeter or Chabrier
at high masses, but significantly shallower than Salpeter---and not
substantially different from Chabrier's disk IMF---at lower masses); and
the ``globular cluster'' IMF of \citet{cha03} (which is nearly the same
as his disk-star IMF but for a somewhat narrower and displaced
lognormal peak below $1\,M_\odot$). Figure \ref{fig:imfcomp} shows the
differences in the predicted $(B-V)_0$ and {\it logarithmic} $\Upsilon_V$
predicted by P\'EGASE for these IMFs vs.~the \citet{cha03} disk-star
function.

The bold solid lines in both panels of Figure \ref{fig:imfcomp} compare
the Chabrier disk and \citet{sal55} IMFs; the bold dashed lines, the Chabrier
disk and \citet{ktg93} IMFs; and the bold dotted lines, the Chabrier
disk and ``globular cluster'' IMFs. All of these bold lines correspond to a
cluster metal abundance $Z=0.004=0.2\,Z_\odot$; the thinner lines around them
refer to $Z=0.0004=0.02\,Z_\odot$ and $Z=0.02=Z_\odot$ (for clarity, in the
upper panel the lines for different metallicities are only shown for the
Chabrier disk vs.~Kroupa et al.~IMF comparison). Clearly,
the choice of IMF has little bearing on the intrinsic $(B-V)_0$ color of a
cluster at a given age, with changes of less than a few hundredths of a
magnitude---within the differences between the P\'EGASE and Bruzual-Charlot
codes---typically being implied. Given that the color is the ratio of
the total cluster luminosity in two bandpasses, and the two codes employ
essentially identical treatments of stellar evolution, its robustness against
even major changes in the IMF is expected.

The $V$-band mass-to-light ratio is,
however, naturally more sensitive to details of the IMF. The Salpeter IMF
gives $\Upsilon_V$ consistently higher than the Chabrier disk IMF (by factors
of $\sim60\%-75\%$), because of the much higher proportion of the total cluster
mass that the former distribution places in very faint, low-mass stars. As
we mentioned just above, however, it is known that the Salpeter IMF is not
a good description of the real distribution at low stellar masses
\citep[see, e.g.,][]{cha03}. The difference in $\Upsilon_V$ between
Chabrier's disk and
``globular cluster'' IMFs is essentially negligible until very old ages,
where the GC IMF would predict slightly lower mass-to-light ratios.
This is because the two IMFs are identical at stellar masses $m\ga 1\,M_\odot$,
where both take on the Salpeter power-law shape. It is only after these
relatively massive stars have all evolved significantly that the slight
difference between the two IMFs at lower stellar masses has a measurable
effect. Even then the effect is small, however, and in any case Chabrier's
disk IMF is better constrained empirically than his GC IMF---both arguments
for our use of the disk-star IMF to make predictions for all of our clusters,
GCs as well as the massive young objects in the LMC and SMC.

By contrast, the difference between $\Upsilon_V$ as computed for the
\citet{cha03} disk IMF vs.~the \citet{ktg93} IMF is
rather dramatic and, more worrisome, varies systematically with cluster age.
This is a direct result of the much steeper slope specified by the Kroupa et
al.~IMF for stellar masses above $1\,M_\odot$. For a given total cluster
mass, the Kroupa et al.~model initially puts a much smaller fraction
into massive (and bright) stars, yielding significantly higher mass-to-light
ratios at young ages. Over time, $\Upsilon_V$ for the Kroupa et al.~IMF
approaches $\Upsilon_V$ for the Chabrier IMF, because the two IMFs are
effectively the same at stellar masses $m\la 1\,M_\odot$---which, of course,
contribute all the cluster light at ages of 10 Gyr and more.

Clearly, the choice between these two IMFs has a direct impact on the
mass-to-light ratios and all dependent physical parameters of clusters younger
than $\tau\la 10^{10}$ yr---which is to say, all of the young LMC and SMC
clusters that we model. In principle, given the strong systematics in the
bottom panel of Fig.~\ref{fig:imfcomp}, choosing between these two particular
IMFs might even influence the ultimate inference of correlations between
(mass-based) cluster properties. We much prefer the \citet{cha03}
disk-star IMF, which has stronger and more recent empirical support than the
older Kroupa et al.~function. But it is important to note that
\citet{mg03a,mg03b,mg03c} adopted the Kroupa et al.~IMF for their own
population-synthesis modeling. {\it Our estimates below of the young LMC/SMC
cluster mass-to-light ratios therefore differ significantly from those
presented by Mackey \& Gilmore.} However, the lower $\Upsilon_V$ implied by
our choice of IMF combines with our generally brighter $V$-band
surface-brightness scale (\S\ref{subsec:mackey}) to produce cluster masses
and mass densities that often differ less strongly from those in MG03.

\subsection{Results for our Cluster Sample}
\label{subsec:results}

Table \ref{tab:poptable} (a sample of which can be found at the end of this
preprint, and which can be downloaded in full from the electronic edition of
the {\it Astrophysical Journal Supplement Series})
presents in full the results of our
population-synthesis modeling of the 53 LMC clusters, 10 SMC clusters, and 5
Fornax globular clusters in the combined MG03
database, as well as for 148 Galactic globular clusters with [Fe/H] values
given in the catalogue of \citet{har96} (that is, 58 GCs in addition to the 90
that we fit with structural models in \S\ref{sec:modeling}). In every case,
to predict an intrinsic $(B-V)_0$ color and average (global) $\Upsilon_V$
ratio, we require only an estimate of the cluster age and metallicity. In the
LMC, SMC, and Fornax cases, these quantities have been either derived or
recovered from the literature by MG03, and we have generally taken their
tabulated values, with the single exception that we assign an age
of $\tau=13\pm2$
Gyr ($\log\,\tau=10.11\pm0.07$) to every cluster that they list as older than
13 Gyr. In the Milky Way, we have taken [Fe/H] from the \citet{har96}
catalogue, and assigned a single age of $13\pm2$ Gyr to all clusters.
(Although an age spread of a few Gyr is known to exist in the Galactic GC
system, at such an old average age this makes little difference in the
population-synthesis colors and mass-to-light ratios; see
Figs.~\ref{fig:popcols} and \ref{fig:popmtol}.) The second and third columns
of Table \ref{tab:poptable} list these ages and metallicities. Column 4 gives
the observed $(B-V)$ color of the cluster whenever such a measurement exists
\citep[taken either from the aperture-photometry sources in Table
\ref{tab:apphot} above, or from][for the Milky Way globulars]{har96}). The
rest of the two
lines for each cluster list the intrinsic colors and $M/L_V$ ratios obtained
from each of six combinations of two population-synthesis codes and four
stellar IMFs discussed in \S\ref{subsec:comp}. It is column (6) of Table
\ref{tab:poptable}---the combination of \citet{bru03} model with
\citet{cha03} disk-star IMF---that we draw on for all cluster modeling that
follows. All uncertainties on the population-synthesis quantities in 
Table \ref{tab:poptable} follow directly from uncertainties in the cluster
ages and [Fe/H] values \citep[for which errorbars are cited by MG03 or]
[but not reproduced here]{har96}.

\begin{figure}
\epsscale{1.23}
\plotone{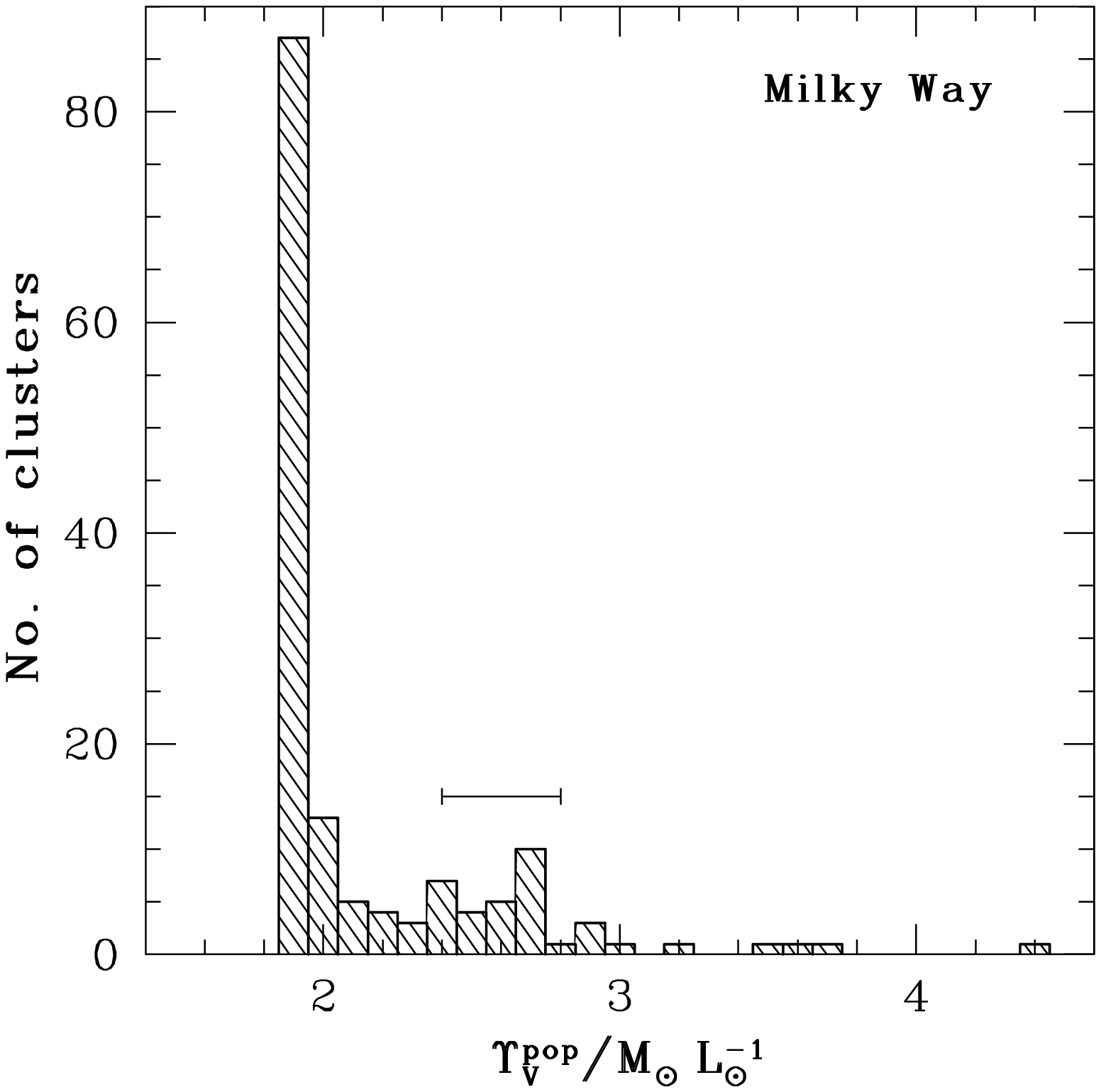}
\caption{\label{fig:galml}
$V$-band mass-to-light ratios of 148 Galactic globular clusters as predicted
by the population-synthesis code of \citet{bru03} using the
disk-star IMF of \citet{cha03}. Values for individual objects are
in Column (6) of Table \ref{tab:poptable}. A common age of $13\pm2$ Gyr has
been assumed for all clusters, so that the spread in $\Upsilon_V^{\rm pop}$
directly reflects the metallicity distribution of Galactic GCs given
[Fe/H] values taken from \citet{har96}. The horizontal errorbar shows
the rms uncertainty in the predicted $\Upsilon_V^{\rm pop}$.
}
\end{figure}

We are not aware of any previous, comprehensive calculation of theoretical
mass-to-light ratios for Galactic globular clusters. In Fig.~\ref{fig:galml}
we show the distribution of population-synthesis $\Upsilon_V$ values for the
Milky Way GC system, taken from Column (6) of Table \ref{tab:poptable}. Note
the strong concentration of clusters at $\Upsilon_V^{\rm pop}\simeq1.9\,
M_\odot\,L_{\odot,V}^{-1}$, which is to be compared to the average
{\it dynamically determined} $\langle\Upsilon_V^{\rm dyn}\rangle=1.45\,
M_\odot\,L_{\odot,V}^{-1}$ for a much smaller sample of clusters
\citep{mcl00}. The tail towards higher $\Upsilon_V^{\rm pop}$ values in
Fig.~\ref{fig:galml} is a direct reflection of the metal-rich (bulge-cluster)
tail in the distribution of GC metallicities in the Galaxy. A more detailed,
cluster-by-cluster comparison of population-synthesis and dynamically-measured
mass-to-light ratios is given in \S\ref{sec:dynml} below, following the bulk of
our model fitting in \S\ref{sec:modeling}.

\begin{figure}
\epsscale{1.23}
\plotone{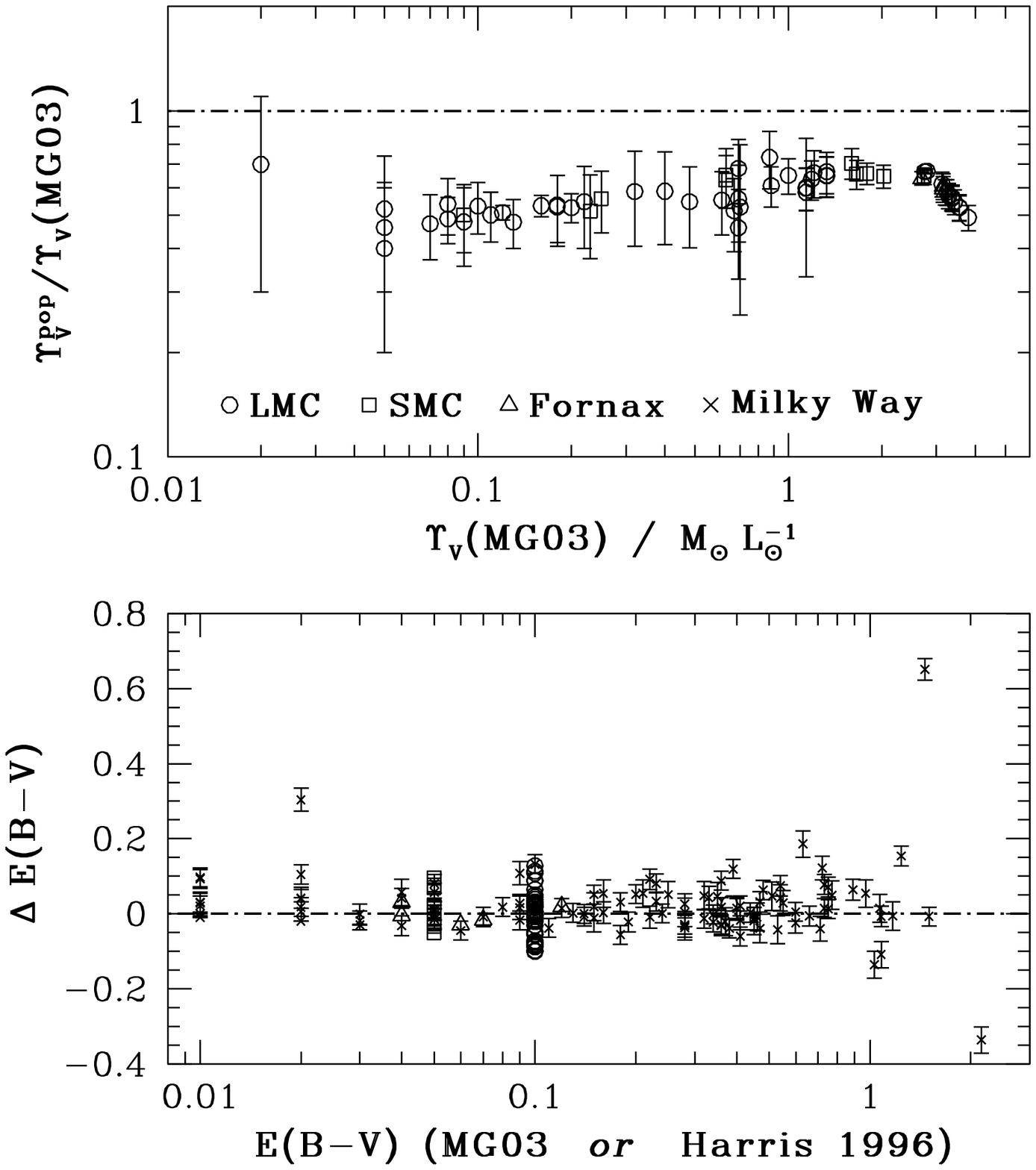}
\caption{\label{fig:redmtol}
{\it Top panel}: Ratio of our computed population-synthesis $V$-band
mass-to-light ratios for LMC, SMC, and Fornax clusters (from Column 6 of
Table \ref{tab:poptable}), to those calculated by \citet{mg03a,mg03b,mg03c}.
Our values are always lower because MG03 adopt a significantly steeper stellar
IMF and do not account for gradual cluster mass loss from stellar-wind and
supernova debris. {\it Bottom panel}: Difference of our computed
population-synthesis reddening {\it minus} published reddenings for
individual clusters in the LMC, SMC, Fornax, and the Milky Way. Reddenings
of Fornax and Milky Way GCs are taken from the compilations of
\citet{mg03c} and \citet{har96} and are largely based on CMD
studies of individual clusters. Reddenings of LMC clusters are compared to an
average $\langle E(B-V)\rangle=0.1$ assumed by \citet{mg03a},
except for R136 = 30 Doradus. SMC reddenings are compared to an average
$\langle E(B-V)\rangle=0.05$ assumed by \citet{mg03b}.
}
\end{figure}

In the top panel of Fig.~\ref{fig:redmtol}, we compare our adopted
population-synthesis mass-to-light ratios for the LMC+SMC+Fornax cluster
sample, with those computed by \citet{mg03a,mg03b,mg03c}. Ours are
always lower than theirs, in part because of the much steeper stellar IMF
that MG03 assumed---as discussed in \S\ref{subsec:comp}---and in part because
they appear not to have allowed for the loss of cluster mass through
massive-star ejecta in their application of the P\'EGASE model (a $\sim30\%$
cumulative effect over a Hubble time, in either the Bruzual-Charlot or Fioc
\& Rocca-Volmerange codes). Further still, MG03 give the ages of some of their
oldest LMC clusters---the genuine globulars there---as greater than 13 Gyr,
but we have fixed such ages at $\tau=13\pm2$ Gyr, resulting again in lower
$\Upsilon_V^{\rm pop}$ ratios. We note in passing that the leftmost point
in this plot---the cluster with the lowest mass-to-light ratio in MG03, and
the largest errorbar in our own assessment---is R136=30 Doradus, which at
$\tau\simeq3\times10^6$ yr is the youngest cluster in our entire sample and the
one most susceptible to uncertainties in the current population-synthesis
models.

Finally, the bottom panel of Fig.~\ref{fig:redmtol} plots the difference
between our population-synthesis derived $E(B-V)=[(B-V)-(B-V)_0]$---that is,
the difference of Columns (4) and (6) in Table \ref{tab:poptable}---and the
published reddenings for individual clusters in the LMC, SMC, Fornax,
and the Milky Way. This plot is dominated by the Galactic GC sample, where the
published $E(B-V)$ are those from \citet{har96}. These are generally in very
good agreement with our model calculations; the strongest outliers are Palomar
6 and Terzan 5, which have large $E(B-V)=1.46$ and 2.15, respectively. This
reflects well both on the calibration of the
population-synthesis models against old, single-burst stellar populations and
on the quality of the reddening estimates in the \citet{har96} catalogue.
The comparisons for the old Fornax globulars
\citep[with true measurements culled from the literature by][]{mg03c} are
likewise very favorable. In the LMC and SMC samples, the ``published''
$E(B-V)$ used for comparison in Fig.~\ref{fig:redmtol} are generally the
rough average values adopted by \citet{mg03a,mg03b}: a constant $E(B-V)=0.10$
mag in the LMC, and constant $E(B-V)=0.05$ mag in the SMC. It is encouraging
that our more detailed modeling returns values clustered well around these
reasonable averages. The one LMC cluster to which MG03 assign a non-average
reddening is R136=30 Doradus, for which they quote $E(B-V)=0.38$ mag from the
CMD analysis of \citet{hun95}. Our model value for 30 Dor is 0.407 mag.

In what follows, then, we make use of the population-synthesis $E(B-V)$ values
that we have derived for each LMC and SMC cluster to correct their observed
surface brightnesses and magnitudes for extinction. There are two LMC
clusters (SL-663 and SL-855) for which we were unable to find $(B-V)$
colors in the literature to estimate their reddenings, so we simply assign
to them the average (0.096) of the other LMC model $E(B-V)$ values. This
also happened for the SMC cluster NGC 361, to which we therefore assign the
average (0.069) of the model results for the other SMC clusters. For the
Fornax and Milky Way GC samples, we use the {\it measured} $E(B-V)$
tabulated by \citet{mg03c} and \citet{har96} (see these papers for
references to the original determinations of the reddenings, most of which
come from direct study of the cluster CMDs). For every cluster in all four
galaxies we apply our population-synthesis $\Upsilon_V$ ratios whenever we
need to convert between luminosity and mass.

\section{Dynamical Models and Fits}
\label{sec:modeling}

\subsection{Dynamical Models}
\label{subsec:mods}

We fit three types of model to each cluster:

First is the usual \citet{king66} single-mass, isotropic, modified isothermal
sphere, which is defined by the stellar distribution function
\begin{equation}
f(E)\propto\left\{
\begin{array}{ll}
\exp[-E/\sigma_0^2] -1 &,\ E<0 \\
0 &,\ E\ge 0\ ,
\end{array}
\right.
\label{eq:king}
\end{equation}
where $E$ is the stellar energy. Under certain restrictive conditions, this
formula roughly approximates a steady-state solution of the Fokker-Planck
equation \citep[e.g.,][]{king65}. It has, of course, already been fit to all
of the Galactic GCs in the surface-brightness profile database of
\citet{tkd95}, and more. These fits are the basis of the standard catalogues
\citep{djo93,har96} of globular cluster structural and
dynamical properties. \citet{king66} models
have also been fit to a number of old globular clusters in other galaxies,
including the LMC, and a few younger Magellanic Cloud clusters; but a
systematic and uniform comparison of these standard models against all of the
high-quality profile data that we now have for the LMC, SMC, and Fornax
clusters has not yet been made.

Second is a power law density profile with a core (also variously known
as Moffat profiles, modified Hubble laws, or ``Elson-Fall-Freeman'' [EFF]
profiles). With $I(R)$ the luminosity surface density of a cluster, such that
$\mu=constant-2.5\,\log\,[I(R)/L_\odot\,{\rm pc}^{-2}]$, these models are
defined by
\begin{equation}
I(R)={{(\gamma-3) L_{\rm tot}}\over{2\pi r_0^2}}\,
\left[1+\left(R/r_0\right)^2\right]^{-(\gamma-1)/2}\ ,
\label{eq:power}
\end{equation}
corresponding to a three-dimensional luminosity density profile
$j(r)\propto [1+(r/r_0)^2]^{-\gamma/2}$. Since the density is non-zero even as
$r\rightarrow\infty$, $\gamma>3$ is required if the integrated luminosity is
to be finite. Equation (\ref{eq:power}) is strictly
an ad hoc fitting function with no underlying basis in theory, but it is
frequently fit to massive young clusters in the Magellanic Clouds and other
galaxies, following the seminal work of \citet{eff87} showing that
\citet{king66} models cannot always account for the spatially extended halos
of such objects. MG03 accordingly fit this model to all of the clusters in
their sample---including the old globulars in the Clouds and the Fornax
dwarf, and even a handful of Milky Way GCs \citep{mg03c}. However,
equation (\ref{eq:power}) has not been fit to any modern surface-brightness
data for large numbers of Galactic globulars such as we have at hand.

The physical parameters extracted for young massive clusters from fits of
equation (\ref{eq:power}) by \citet{eff87}, MG03, and others
\citep[e.g.,][]{lar04} are
generally confined to the central intensity $I_0$, the scale radius $r_0$
(and perhaps an associated core radius $R_c$ and/or effective radius $R_h$),
the power-law exponent $\gamma$, and the total luminosity $L_{\rm tot}$. This
small set of quantities falls well short of the wide array that has been
calculated for \citet{king66} models, including relaxation times, central
escape velocities, global binding energies, and more. A full comparison
between the systemic properties of globular and other massive star clusters
would seem to
require comparable levels of detail in the modeling of all objects, no matter
what specific model is actually applied; and a clear understanding of how
model-dependent are the values of the derived physical parameters is critical.
These issues can only be tackled by completely fitting both of the model
types just described to all of the clusters in all four galaxies of our sample.

Certainly a good reason why power-law models have not been fit
systematically to Galactic globular clusters is that they lack
the tidal cut-off radius so important to \citet{king66} models and obviously
present in real GCs. We therefore expect a priori that the power-law models
may provide poorer fits than King models to many globular cluster density
profiles. Conversely, we already know for a fact \citep[e.g.,][]{eff87} that
power laws provide better fits than \citet{king66} models to the outer
envelopes of some
young LMC clusters. We therefore wish also to fit all of our clusters with a
third model which is intermediate between these two, one which can afford more
extended halos than \citet{king66} models but still goes to zero density at a
finite radius. Another reason to consider such an additional model is the
original suggestion of \citet{eff87}
\citep[see also the recent review of][]{sch04}, that the
power-law fits to some young LMC clusters represent unbound halos of stars
around relatively recently formed systems that could be stripped away over Gyr
timescales, leaving behind more standard King-type configurations. In
assessing this idea, it is worthwhile to ask whether
self-consistent, non-King models with extended but {\it bound} stellar halos
are capable at all of describing these data.

The third model that we have fit is again based on a specified stellar
distribution function, motivated by the work of \citet{wil75} on modeling
elliptical galaxies:
\begin{equation}
f(E)\propto\left\{
\begin{array}{ll}
\exp[-E/\sigma_0^2] -1 + E/\sigma_0^2 &,\ E<0 \\
0 &,\ E\ge 0\ ,
\end{array}
\right.
\label{eq:wilson}
\end{equation}
which is a different type of single-mass and isotropic modified isothermal
sphere.
\citet{wil75} included a multiplicative term in the distribution function
depending on the angular momentum $J_z$, in order to create axisymmetric model
galaxies. We have dropped this term from $f(E)$ to make spherical and
isotropic cluster models, but we still refer to equation (\ref{eq:wilson}) as
Wilson's model. The connection with the \citet{king66} model in equation
(\ref{eq:king}) is clear: the extra $+ E/\sigma_0^2$ in the first
line of equation (\ref{eq:wilson}) is simply taking away the linear term in
the Taylor series expansion of the fundamental $\exp(-E/\sigma_0^2)$
near the zero-energy (tidal) boundary of the cluster. Although patently an ad
hoc thing to do, the net effect of this more gradual lowering of the isothermal
sphere is to produce clusters which are spatially more extended than
\citet{king66} models, but still finite.

It will be noted that there is a slight asymmetry between equations
(\ref{eq:king}) and (\ref{eq:wilson}) and equation (\ref{eq:power}): the
former explicitly involve a velocity scale parameter $\sigma_0$, while the
latter incorporates an explicit radial scale $r_0$. In the formulation of his
model, \citet{king66} defined a radial scale associated with
$\sigma_0$:
\begin{equation}
r_0^2\equiv {{9\sigma_0^2}\over{4\pi G \rho_0}} \ ,
\label{eq:rscale}
\end{equation}
where $\rho_0$ is the central mass density of the model. We adopt the same
definition for our single-mass, isotropic Wilson models; and we use it also
in our construction of power-law models to define a velocity scale
$\sigma_0$ in terms of $r_0$ from equation (\ref{eq:power}). It is important
to recognize that $r_0$ and $\sigma_0$ are {\it not} equivalent, in general,
to typically observed quantities such as a core (half-power) radius or central
velocity dispersion---although the connections between the theoretical
scales and these or other observables are straightforward to calculate
for any member of these model families.

In their dimensionless form, King and Wilson models are characterized by the
profiles of $\rho/\rho_0$ and $\sigma/\sigma_0$ as functions of $r/r_0$. Both
profiles are fully specified by the value of the dimensionless central
potential, $W_0 \equiv -\phi(0)/\sigma_0^2 > 0$.
In principle $W_0$ can take
on any real value between 0 and $\infty$, with the latter limit corresponding
to a regular isothermal sphere of infinite extent. $W_0$ bears a one-to-one
relationship with the more intuitive {\it concentration parameter}: $c\equiv
\log\,(r_t/r_0)$, where $r_t$ is the tidal radius of the model cluster
$[\rho(r_t)=0]$. For more details of the relations
between these model parameters, and of the construction of King models in
general, see \citet{king66} or \citet{bt87}. We have essentially
followed the prescription of \citet{bt87}---with the obvious
substitution of the distribution function $f(E)$ in equation (\ref{eq:wilson})
for that in equation (\ref{eq:king})---to compute Wilson models with arbitrary
$W_0$ or $c$. The result is normalized three-dimensional density and
velocity-dispersion profiles, which are then projected onto the plane of the
sky \citep[using standard integrals that can also be found, e.g., in][]{bt87}
for fitting to data.

\begin{figure}
\epsscale{1.25}
\plotone{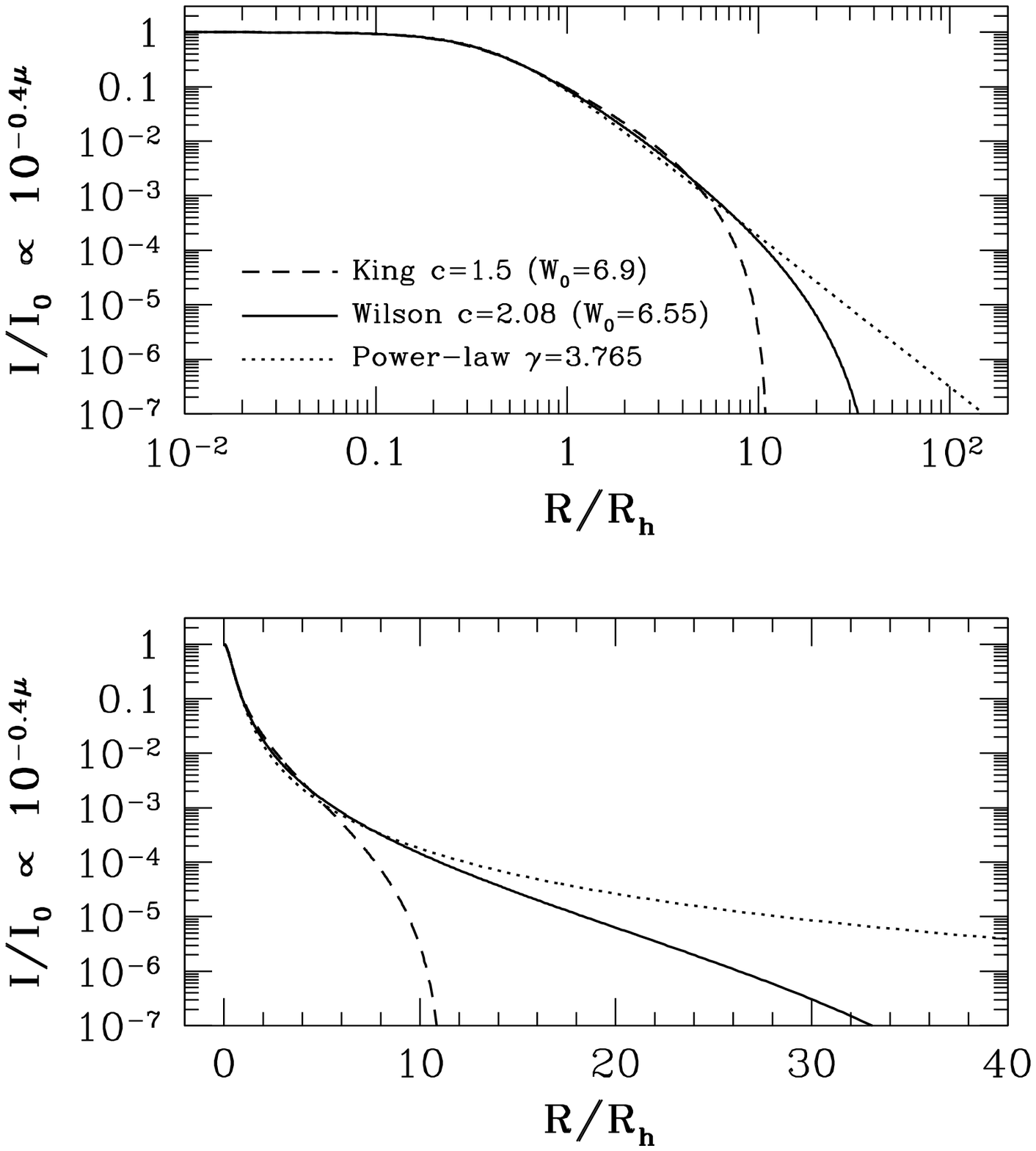}
\caption{\label{fig:models}
Comparison of the projected density/surface-brightness profiles of a
single-mass \citet{king66} model
cluster (dashed curves) defined by equation (\ref{eq:king}); a spherical and
isotropic \citet{wil75} model (solid curves) defined by equation
(\ref{eq:wilson}); and a cored power-law model (dotted curves) defined by
equation (\ref{eq:power}). The three examples shown all have the same
dimensionless total luminosity $L_{\rm tot}/I_0 R_h^2$. {\it Top panel}
is a log-log plot of the models highlighting their core sturctures;
{\it bottom panel} is a log-linear representation emphasizing their halos.
Structural differences between the models are most significant beyond a
few projected half-light (effective) radii.
}
\end{figure}

Fixing $W_0$ or $c$ at some value essentially defines the overall shape of the
internal density profile (and the velocity-dispersion profile) of a King or
Wilson model. The analogous shape parameter
for the cored power-law models of equation (\ref{eq:power}) is the index
$\gamma>3$, the exponent of the power law hypothesized to describe the
asymptotic behavior of a cluster's density distribution.
Clearly in this case $\gamma$ does not correspond to any measure of a tidal
radius---$r_t$ is always infinite in these models---but it still has a
one-to-one connection with the (finite) central potential of a cluster.
Starting from equation (\ref{eq:power}) for the observable surface-density
profile in these models, which is fitted directly to real cluster data,
it is a simple matter to compute the deprojected volume density $j(r)$
analytically and then solve the spherical Jeans equation \citep{bt87},
assuming unit mass-to-light ratio and velocity isotropy, to obtain a
normalized velocity-dispersion profile for any $\gamma$. The complete
structural and dynamical details of power-law clusters are then known in full,
and all of the derived physical parameters that we present in
\S\ref{subsec:fits} can be evaluated equally well within any of the three
models that we fit.

It is instructive first to consider how these different types of models compare
with one another. Thus, in Figure \ref{fig:models} we show normalized
surface-density profiles of one example each of a King, Wilson, and
power-law model cluster. We first calculated a $c=1.5$ \citet{king66}
model---representative of an average Galactic globular cluster---and
found its dimensionless total luminosity, $L_{\rm tot}/I_0 R_h^2$, in terms of
an arbitrary central surface density $I_0$ and projected half-light (effective)
radius $R_h$. The surface-density profile $I(R)$, normalized by $I_0$ and
scaled in projected radius by $R_h$ is shown as the dashed curves in both
panels of Fig.~\ref{fig:models}. The top panel here is a log-log
plot of the density vs.~radius; the bottom is a log-linear plot, a
representation which emphasizes the outer {\it halo} structure over the inner
core regions.

\begin{figure}
\epsscale{1.25}
\plotone{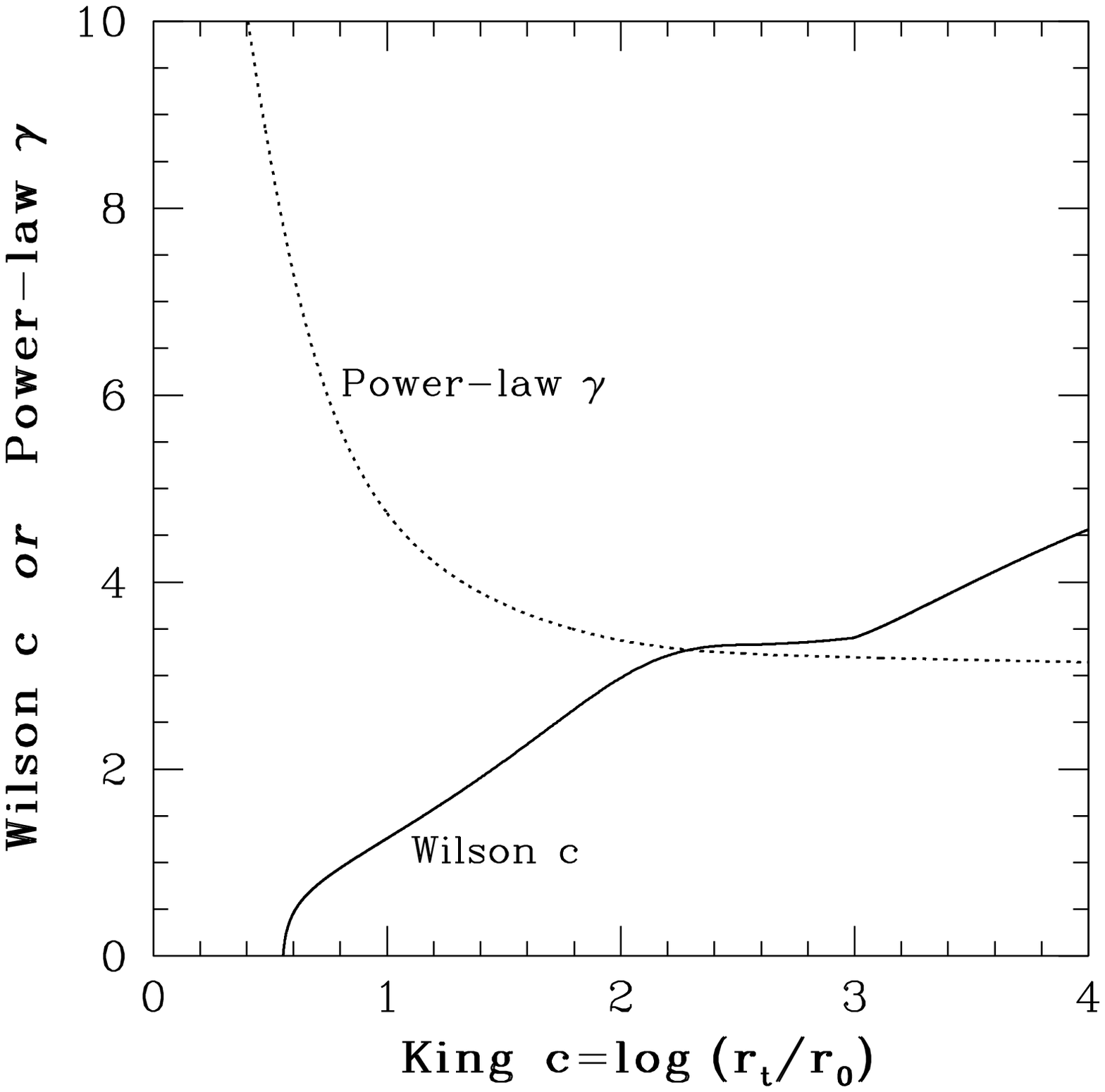}
\caption{\label{fig:conc}
Relation between King-model concentration $c\equiv\log\,(r_t/r_0)$,
Wilson-model $c$, and power-law exponent $\gamma$ when a cluster is
required to have the same total luminosity and effective radius and central
surface-brightness in all three models. (Each $c$ or $\gamma$ corresponds to
a unique value of $L_{\rm tot}/I_0 R_h^2$.) King and Wilson models with
$c<0$ exist
in principle ($c\rightarrow-\infty$ as the dimensionless central potential
$W_0\rightarrow0$) but are not shown here, as real clusters of such low
concentration are rare. The same is true of power-law models with $\gamma>10$.
}
\end{figure}

We then sought the \citet{wil75} model with the same dimensionless
$L_{\rm tot}/I_0 R_h^2$ as the $c=1.5$ \citet{king66} cluster. This turns out
to be given by a Wilson $c=2.08$, and the projection of this model onto the
plane of the sky is shown as the solid curves in Fig.~\ref{fig:models}.
Similarly, the $\gamma=3.765$ power-law model shown as the dotted curves
has the same $L_{\rm tot}/I_0 R_h^2$ again. Thus, Fig.~\ref{fig:models}
illustrates the relative spatial extent predicted by these three models for
a cluster with fixed total luminosity, central surface brightness, and
projected half-light radius (all quantities that tend to be observationally
well-determined in real clusters). Clearly apparent is the intermediacy of the
Wilson model between the more sharply truncated King model and the infinite
power law. Also evident is that the differences between these models are
generally largest beyond a few effective radii, in the outer halos of
clusters.

Figure \ref{fig:conc} extends this comparison to general \citet{king66} cluster
concentrations. [A more comprehensive discussion of the defining features of
King models vs.~Wilson models in particular can be found in \citet{hun77}.]
Here, we have calculated the dimensionless
$L_{\rm tot}/I_0 R_h^2$ for each of a large number of King models with
$0.3\le c\le 4$, and then found the unique Wilson $c$ and power-law $\gamma$
that give the same dimensionless luminosity. Although we have done this
formally without any restrictions on the parameter $c$, note that the
majority of King-model fits to real GCs return $c\la 2$, and indeed the model
itself is unstable to the gravothermal catastrophe at concentrations higher
than this.

It is noteworthy from this figure that, first,
{\it low}-concentration King or Wilson models, which are characterized
by a sharp decline in density beyond a dominant, nearly constant-density core,
find their analogue in
{\it high}-$\gamma$ power laws. Second, while the shape parameters $c$ or
$\gamma$ are certainly useful indicators of a cluster's global structure,
their numerical values are highly model-dependent. This makes it of interest to
consider a more generally applicable ``concentration index'' able to represent
the spatial extent or potential depth of any cluster in a more
model-independent way (so as to allow, e.g., for a combined analysis of
clusters which may not all be described well by the same type of model). We
return to this point below in \S\ref{subsec:fits}, where we now present the
fits of all models to our cluster sample.

\subsection{Surface-Brightness Fits and Cluster Physical Parameters}
\label{subsec:fits}

Our fitting procedure involves computing in full large numbers of King,
Wilson, and power-law structural/dynamical models, spanning a wide range of
fixed values of the appropriate shape parameter $W_0$ or $\gamma$. Separately
for each family in turn, we then fit every model on the appropriate $W_0$- or
$\gamma$-grid to a cluster's observed surface-brightness profile,
$\mu_V=\mu_{V,0}-2.5\,\log\,[I(R/r_0)/I_0]$, finding the radial scale $r_0$
and central SB $\mu_{V,0}$ which minimize
$\chi^2$ for every given value of $W_0$ or $\gamma$. The
$(W_0,r_0,\mu_{V,0})$ or $(\gamma,r_0,\mu_{V,0})$ combination that yields
the global minimum $\chi_{\rm min}^2$ over the grid used defines the best-fit
model of that type. Estimates of the one-sigma
uncertainties on these basic fit parameters, and those on all associated
derived quantities, follow from their extreme values over the subgrid of fits
with $\chi^2\le \chi_{\rm min}^2+1$. For the most part, we calculate and
minimize $\chi^2$ as the weighted sum of squared differences between model and
observed intensities $I$ (in units of $V$-band $L_\odot\,{\rm pc}^{-2}$),
rather than  logarithmic surface-brightness units; although in a few cases
more stable fits resulted from defining $\chi^2$ as the weighted sum of
squared surface-brightness deviations. The vast
majority of our fits are error-weighted, with the uncertainties on individual
datapoints either taken from the original sources of HST and ground-based
starcounts in the LMC, SMC, and Fornax clusters, or estimated by us from
the Trager et al.~(1995) catalogue data (\S\ref{subsec:trager} and Table
\ref{tab:MWerrs}).

\subsubsection{Fits to $\omega$ Centauri}
\label{subsubsec:wcen}

\setcounter{table}{8}
\begin{deluxetable*}{lcrclll}[b]
\tabletypesize{\scriptsize}
\tablewidth{0pt}
\tablecaption{Model Fits to $\omega$ Centauri = NGC 5139  \label{tab:wcen}}
\tablecolumns{7}
\tablehead{
\colhead{Model}                                       &
\colhead{Reduced $\chi_{\rm min}^2$\tablenotemark{a}} &
\colhead{$W_0$/$\gamma$}                    & \colhead{$c=\log(r_t/r_0)$}   &
\colhead{$\mu_{V,0}$\tablenotemark{b}}      & \colhead{$r_0$ [sec]}         &
\colhead{$r_0$ [pc]\tablenotemark{c}}                   \\
\colhead{(1)}  & \colhead{(2)}  & \colhead{(3)}  & \colhead{(4)}  &
\colhead{(5)}  & \colhead{(6)}  & \colhead{(7)}
}
\startdata

King      &  1.73 & $W_0=6.20^{+0.20}_{-0.10}$ & $1.31^{+0.05}_{-0.03}$
          & $16.44^{+0.06}_{-0.11}$ &  $141.20^{+6.50}_{-12.69}$
          & $3.63^{+0.17}_{-0.33}$ \\

Wilson    &  0.47 & $W_0=4.70^{+0.10}_{-0.10}$ & $1.34^{+0.03}_{-0.03}$
          & $16.52^{+0.05}_{-0.05}$ & $196.81^{+7.08}_{-7.01}$
          & $5.06^{+0.18}_{-0.18}$ \\

Power-law &  1.47 & $\gamma=5.55^{+0.10}_{-0.10}$ & $\infty$               
          & $16.65^{+0.05}_{-0.04}$ & $327.58^{+11.48}_{-11.71}$
          & $8.42^{+0.29}_{-0.30}$
\enddata

\tablenotetext{a}{There are 54 datapoints in the fitted profile, and we fit by
  finding the $\mu_{V,0}$ and $r_0$ which minimizes $\chi^2$ over a large
  grid of fixed $W_0$ or $\gamma$ values. Thus, the reduced
  $\chi_{\rm min}^2$ per degree of freedom is just $\chi_{\rm min}^2/52$.}

\tablenotetext{b}{Corrected for extinction given in \citet{har96}.}

\tablenotetext{c}{Assumes a heliocentric distance of $D=5.3$ kpc
  \citep{har96}.}

\end{deluxetable*}

\begin{figure*}
\epsscale{1.10}
\plotone{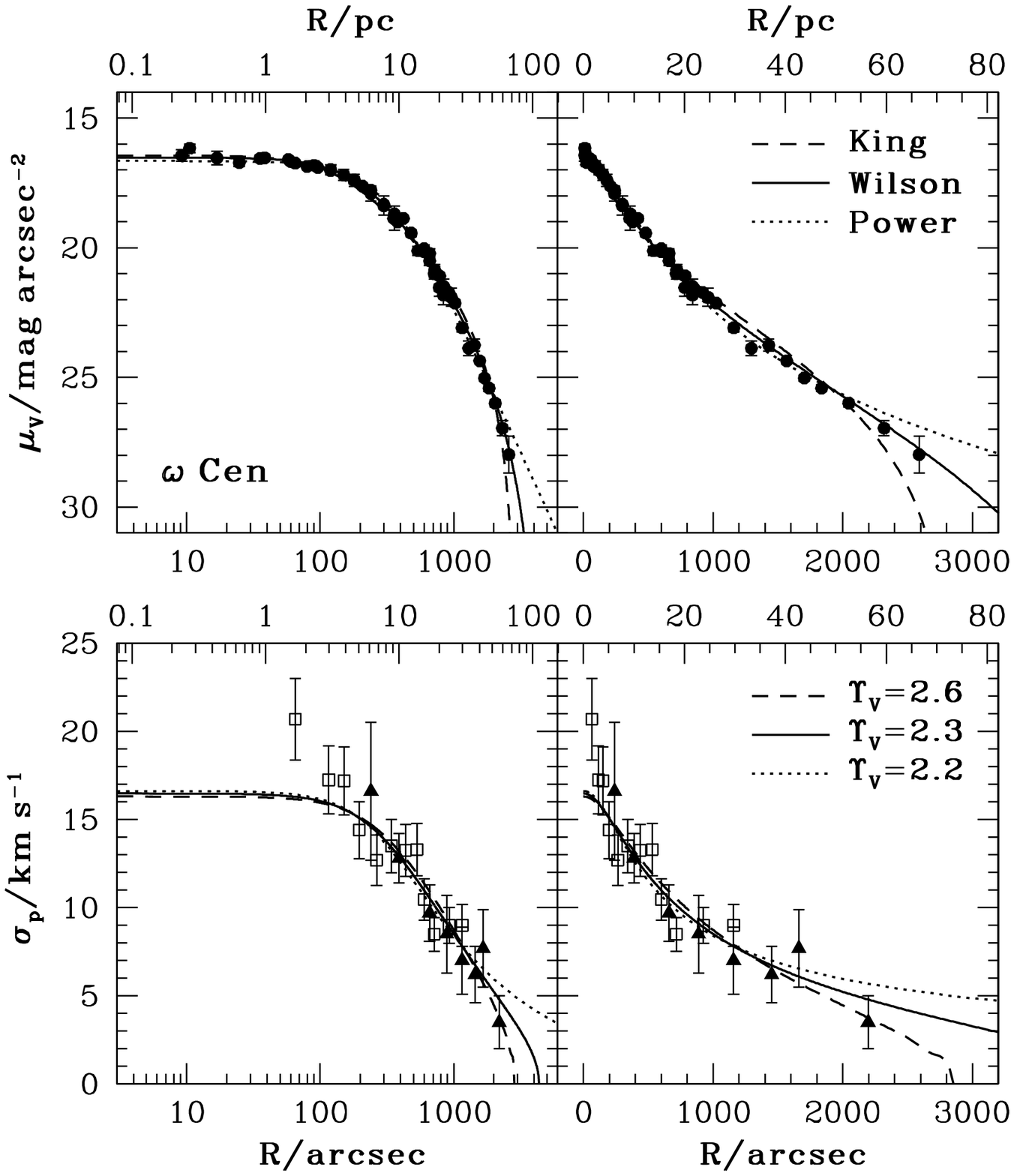}
\caption{\label{fig:wcen}
Detailed fitting of \citet{king66}, \citet{wil75}, and power-law models to the
Galactic globular cluster $\omega$ Centauri. {\it Top panels} show the
surface-brightness profile as a function of logarithmic projected radius on
the left, linear projected radius on the right. {\it Bottom panels} show the
observed internal line-of-sight velocity-dispersion profile from
\citet[open squares]{mey95} and \citet[filled triangles]{sei83}, and the
model profiles after normalization by the $V$-band mass-to-light ratios
indicated.
}
\end{figure*}

As a detailed example, Figure \ref{fig:wcen} shows our best
King, Wilson, and power-law fits
to the well-studied Galactic globular cluster $\omega$ Centauri = NGC 5139
\citep[see also][]{mcl03}. The main parameters and minimum $\chi^2$ values
for the three fits are listed in Table \ref{tab:wcen}. These are also given,
along with many other derived structural and dynamical parameters, in our
results below for the full cluster sample (Tables \ref{tab:basic} through
\ref{tab:kappa}).

The upper panels of Figure \ref{fig:wcen} display the cluster's
model and observed surface-brightness profiles, while the lower panels show
the predicted and observed internal velocity-dispersion profiles, which we
consider in detail for this one object only. The $V$-band surface-brightness
datapoints are those from \citet{tkd95} with our estimated errorbars
attached. The radial scale is given in units of arcseconds
along the lower horizontal axes and parsecs
\citep[for an assumed distance of 5.3 kpc taken from][]{har96} along the
upper axes. The left-hand upper panel
is a log-log representation of density vs.~radius, which focuses on the core
structure of the cluster; the right-hand panel is in log-linear format to
highlight its halo structure.

The main point to be taken from the top of Fig.~\ref{fig:wcen} is that even
among Milky Way globular clusters, where \citet{king66} models are generally
taken as the best physical description, alternate models can fit even the most
comprehensive data at least as well and sometimes, as in the case of the
Wilson model here, better. To be sure, the single-mass and isotropic King
models that we fit have long been known to be inadequate for many old GCs, a
fact that is usually attributed {\it by hypothesis} to the influences of
energy equipartition and mass segregation and/or velocity anisotropy in
well-relaxed multimass stellar populations.
Multimass and anisotropic models still based on the King
distribution function (eq.~[\ref{eq:king}]) have therefore been developed
\citep{dac76,gg79} and fit to a good number of Galactic globulars---including
$\omega$ Centauri \citep[e.g.,][]{mey95}------with much better success. The
case of $\omega$ Cen is instructive, however, in that it is an {\it unrelaxed}
cluster showing no evidence of advanced mass segregation \citep{and97},
nor of velocity anisotropy \citep{mer97,lee00}. Compensating for the
shortcomings of the simple
\citet{king66} model in Fig.~\ref{fig:wcen} is therefore better done with
entirely different single-mass and isotropic models---even if they are ad
hoc to some degree---than with multimass and anisotropic variations on the King
distribution function. This is an additional justification for our fitting of
Wilson and power-law models to a full suite of Galactic globulars.
$\omega$ Cen is the most massive GC in the Milky Way, and an
uncharacteristically diffuse one. It might therefore have been thought,
justifiably, to be a rather
special case; but as we shall see, our single-mass, isotropic Wilson
models in particular do provide fits of quality comparable to or better than
King models for a majority of GC surface-brightness profiles. We also note
that the cluster tidal radius implied by the Wilson-model fit in $\omega$ Cen
is only 50\% larger than that of the King model (see Table
\ref{tab:struc} below), meaning that the generally greater extent of Wilson's
model relative to King's need not imply severe inconsistencies
with the expected sizes of GCs for a given Galactic tidal field.

The bottom panels of Fig.~\ref{fig:wcen} show the observed velocity dispersion
as a function of radius in $\omega$ Cen \citep{mey95,sei83},
again with a logarithmic radial scale on the left and a linear radial scale on
the right. The model curves now are the profiles predicted (after projection
along the line of sight) by solving the spherical Jeans equation with the
de-projections of the best-fit density profiles in the upper panels.
The model predictions are inherently dimensionless in form, yielding
$\sigma_{\rm p}/\sigma_0$ as a function of projected radius $R$ where
$\sigma_0$ is the velocity scale {\it defined} by
$\sigma_0^2 \equiv (4\pi G \Upsilon_V j_0 r_0^2)/9$
(cf.~eq.~[\ref{eq:rscale}]). Knowing the radial scale $r_0$ and the
central three-dimensional luminosity density $j_0$ from fitting to
the cluster surface brightness, the theoretical velocity-dispersion curves are
scaled to match the data essentially by fixing a (spatially constant)
mass-to-light ratio $\Upsilon_V$. The value of $\Upsilon_V$ that gives the
best agreement with the velocity dispersions in $\omega$ Cen is listed for
each model type in the lower right-hand panel of Fig.~\ref{fig:wcen}. The
main conclusion is that our different surface-brightness fits are, in
general, associated with self-consistent internal dynamics that are closely
similar both in relative terms (the shapes of the model velocity-dispersion
profiles) and in an absolute sense (the implied value of $\Upsilon_V$).

\subsubsection{Fits to All Clusters}
\label{subsubsec:allcl}

Figure \ref{fig:sbfits}, which can be found at the end of the paper,
displays the best-fit King, Wilson, and power-law
models for the $V$-band surface-brightness profiles of our 53 LMC clusters,
10 SMC clusters, the 5 globulars in Fornax, and 84
Galactic GCs besides $\omega$ Centauri
\citep[recall that we do not present fits to the entire database of]
[for reasons discussed in \S\ref{subsec:trager}]{tkd95}.
Each cluster is presented in two panels formatted as
in the top of Fig.~\ref{fig:wcen}, one with both axes on logarithmic scales
and one in log-linear form. Along the top of every panel, projected radius is
given in pc, obtained by assuming a distance of 50.1 kpc to all clusters in
the LMC, 60.0 kpc to all clusters in the SMC, and 137 kpc to Fornax;
heliocentric distances to individual Milky Way GCs are taken from the
catalogue of \citet{har96}. In all cases, dashed curves are the King-model
fits; solid curves are Wilson models; and dotted lines are power laws with
cores.

In the LMC, SMC, and Fornax cluster panels, the open {\it circles} correspond
to surface-brightness data taken from \citet{mg03a,mg03b,mg03c}, {\it after}
applying the zeropoint corrections discussed
in \S\ref{subsec:mackey} (see Table \ref{tab:offset}) and correcting for
the $V$-band extinction $A_V=3.1\left[(B-V)-(B-V)_0\right]$ inferred from
our population-synthesis modeling in \S\ref{subsec:results}
(see Table \ref{tab:poptable}). There
are generally two such points at each radius, one coming from the primary
$V$-band counts of MG03 and the other from their secondary $B$ or $I$ profiles
shifted by the color terms in Table \ref{tab:colors}. The open
{\it squares} in the plots for LMC and SMC clusters denote the ground-based
starcount data that we have collected and matched onto the
re-zeropointed MG03 profiles. Asterisks refer to datapoints that we have not
included in calculating and minimizing $\chi^2$ to identify the best model
fits (i.e., those with ``fit flags'' of 0 in Table \ref{tab:profs}).

In the Milky Way GC panels, the open circles are the datapoints from
\citet{tkd95}, taken exactly as published except for Palomar 10 and Terzan
7, which we have calibrated as described in \S\ref{subsec:trager}. The
errorbars on these points are our own estimates based on Table
\ref{tab:MWerrs}. The asterisks in these cases
are surface-brightness measurements also published by Trager at al.~but
given relative weightings $w_i<0.15$ by them; we did not include these
points when fitting our models.

Table \ref{tab:basic} (also at the end of the paper)
gives a few important numbers for each cluster and lists
the main parameters of every fit. The first column here is the cluster
name with a short prefix identifying its parent galaxy. Column (2) is the
zeropoint shift that we have applied to published surface-brightness
magnitudes (copied directly from Table \ref{tab:offset} for the MG03 cluster
sample; 0 for all but Palomar 10 and Terzan 7 in the Milky Way GC sample).
Column (3) is the estimated $V$-band extinction, derived as discussed in
\S\ref{subsec:results}. Next are the assumed distance to the cluster, the
number of datapoints ultimately included in fitting any model to the cluster,
and a code for the weighting scheme used to minimize $\chi^2$.

Subsequent columns in Table \ref{tab:basic} cover three lines for each
cluster, one line for each type of model fit. Column (7) identifies the model;
Column (8) gives the minimum {\it unreduced} $\chi^2$ obtained for
that class of model; Column (9) gives the appropriate ``shape'' parameter
$W_0$ or $\gamma$ at which $\chi^2$ is minimized; Column (10) gives the
concentration $c\equiv\log(r_t/r_0)$ of the best fit (this is related uniquely
to $W_0$ for King and Wilson models but is always $\infty$ for power-law
models); Column (11) gives the best-fit central surface brightness {\it after
correction} for both extinction and any zeropoint change; Column (12) gives
the best-fit model scale radius, $r_0$, in arcseconds; and Column (13) gives
the value of $r_0$ in parsecs. The uncertainties in $W_0$, $c$, $\gamma$,
and $r_0$ reflect their variations among model fits with $\chi^2\le
\chi_{\rm min}^2+1$. The errorbars on $\mu_{V,0}$ derive from these formal
fitting uncertainties combined in quadrature with the uncertainties in the
zeropoint offsets $\Delta\mu_V$ and extinctions $A_V$ (the $\chi^2$ of any fit
is calculated using the published uncertainties in the individual SB datapoints
before any systematic corrections are applied).

Table \ref{tab:struc} (at the end of the paper)
contains a number of other structural cluster
properties derived from the basic fit parameters:
\begin{itemize}

\item $\log\,r_t=c+\log\,r_0$ is the model tidal radius.

\item $\log\,R_c$ refers to the projected {\it core} radius of the model
fitting a cluster. It is defined by $I(R_c)=I_0/2$, or
$\mu_V(R_c)\simeq\mu_{V,0}+0.753$, and is {\it not} the same
in general as the radial scale $r_0$ in Table \ref{tab:basic}. For King and
Wilson models, $r_0$ is simply a convenience defined by dimensional analysis
of Poisson's equation, although it does bear a uniqe relationship to the
observable $R_c$ through well-defined functions of $W_0$ or $c$. For power-law
models, the connection between $r_0$ and $R_c$ stems from the defining
equation (\ref{eq:power}) above.

\item $\log R_h$ refers to the half-light, or effective, radius of a model:
that radius containing half the total luminosity in projection. It is related
to $r_0$ by one-to-one functions of $W_0$ or $\gamma$.

\item $\log\,(R_h/R_c)$ is a measure of cluster concentration that is
relatively more model-independent than $W_0$ or $c$ or $\gamma$, in the sense
that it is generically well defined and its physical meaning is always the
same. We consider it a more suitable quantity to use when intercomparing the
overall properties of clusters which may not all be fit by the same kind of
model (cf.~our earlier discussion around Fig.~\ref{fig:conc}).

\item $\log\,I_0=0.4\,(26.422-\mu_{V,0})$ is the logarithm of the best-fit
central luminosity surface density in the $V$ band, in units of $L_\odot\,
{\rm pc}^{-2}$. The surface-brightness zeropoint of 26.422 corresponds to a
solar absolute magnitude $M_{V,\odot}=+4.85$ \citep[e.g.,][]{lan99}.

\item $\log\,j_0$ is the logarithmic central luminosity {\it volume} density
in the $V$ band, in units of $L_\odot\,{\rm pc}^{-3}$.
It is given by $j_0={\cal J}I_0/r_0$, where ${\cal J}$ is a smooth,
model-dependent function of $W_0$ or $\gamma$, which we have calculated in
detail for King, Wilson, and power-law models.

\item $\log\,L_{\rm tot}$ is the logarithm of the total integrated model
luminosity in the $V$ band. It is related to the product $I_0r_0^2$ by
model-dependent functions of $W_0$ or $\gamma$.

\item $V_{\rm tot}=4.85-2.5\,\log\,(L_{\rm tot}/L_\odot)+
5\,\log\,(D/10\,{\rm pc})$ is the total, {\it extinction-corrected} apparent
magnitude of a model cluster.

\item $\log I_h \equiv \log\,(L_{\rm tot}/2\pi R_h^2)$ is the $V$-band
luminosity surface density averaged over the half-light/effective radius, in
units of $L_\odot\,{\rm pc}^{-2}$.

\item $\langle \mu_V\rangle_h \equiv
26.42-2.5\,\log\,(I_h/L_\odot\,{\rm pc}^{-2})$ is the average surface
brightness inside the half-light radius, in $V$-band mag arcsec$^{-2}$.

\end{itemize}
Again, the uncertainties on all of these derived parameters have been
estimated (separately for each given model family) by calculating
them in every model which yields $\chi^2$ within 1 of the global minimum for
a cluster, and then taking the differences between the extreme and best-fit
values of the parameters.

Table \ref{tab:dyn} (see the end of the paper)
next lists a number of cluster properties derived from
the structural parameters already given {\it plus} a mass-to-light ratio.
The first column of this table contains the cluster name, as usual. Column
(2) shows the mass-to-light ratio, in solar units, that we have adopted for
each object from the analysis in \S\ref{sec:popsyn}---that is, on the basis of
population-synthesis modeling given individual ages and metallicities for the
clusters, using the model code of \citet{bru03} and assuming the
disk-star IMF of \citet{cha03}. The values of $\Upsilon_V^{\rm pop}$ in
Table \ref{tab:dyn} have been copied directly from Column (6) of Table
\ref{tab:poptable}. The remaining entries in Table \ref{tab:dyn} are, for each
type of model fit to each cluster:
\begin{itemize}

\item $\log\,M_{\rm tot}=\log\,\Upsilon_V^{\rm pop}+\log\,L_{\rm tot}$, the
integrated model mass in solar units.

\item $\log\,E_b$, the integrated binding energy in ergs, defined through
$E_b\equiv -(1/2)\int_{0}^{r_t} 4\pi r^2 \rho \phi\,dr$.
Here the minus sign makes $E_b$
positive for gravitationally bound objects, and $\phi(r)$ is the potential
generated (through Poisson's equation) by the model mass-density distribution
$\rho(r)$. $E_b$ can be written in terms of the fitted central luminosity
density $j_0$, scale radius $r_0$, a model-dependent function of $W_0$ or
$\gamma$, and $\Upsilon_V^{\rm pop}$. A more detailed outline of this
procedure for King models may be found in \citet{mcl00}, which we have
followed closely to evaluate $E_b$ for our Wilson and power-law fits as well.

\item $\log\,\Sigma_0=\log\,\Upsilon_V^{\rm pop}+\log\,I_0$, the central
{\it mass} surface density of the model fit in $M_\odot\,{\rm pc}^{-2}$.

\item $\log\rho_0=\log\,\Upsilon_V^{\rm pop}+\log\,j_0$, the central mass
volume density in $M_\odot\,{\rm pc}^{-3}$.

\item $\log\Sigma_h=\log\,\Upsilon_V^{\rm pop}+\log\,I_h$, the model mass
density averaged over the half-light radius $R_h$ (which is equal to the
half-mass radius under our assumption of single-mass stellar populations,
i.e., spatially constant mass-to-light ratios).

\item $\sigma_{{\rm p},0}$, the predicted line-of-sight velocity dispersion
at the cluster center, in km~s$^{-1}$. As was already suggested above, the
solution of Poisson's and Jeans' equations for any model yields a dimensionless
$\sigma_{{\rm p},0}/\sigma_0$, and with $\sigma_0$ given by the fitted $r_0$
and $\rho_0$ through equation (\ref{eq:rscale}), the predicted observable
dispersion follows immediately.

\item $v_{{\rm esc},0}$, the predicted central ``escape'' velocity in
km~s$^{-1}$. A star
moving out from the center of a cluster with speed $v_{{\rm esc},0}$ will just
come to rest at infinity. In general, then, $v_{{\rm esc},0}^2/\sigma_0^2=
2\left[W_0+GM_{\rm tot}/r_t\sigma_0^2\right]$. Note that the second term
on the right-hand side of this definition vanishes
for power-law models, in which $r_t\rightarrow\infty$. In these models
a (finite) dimensionless $W_0$ is associated with every value of $\gamma>3$
by solving Poisson's equation with $\phi(\infty)=0$.

\item $\log\,t_{\rm rh}$, the two-body relaxation time at the model
projected half-mass radius. This is estimated as
$t_{\rm rh}/{\rm yr}=\left[2.06\times10^6/\ln(0.4M_{\rm tot}/m_\star)\right]
m_\star^{-1} M_{\rm tot}^{1/2} R_h^{3/2}$
\citep[eq.~8-72]{bt87}, if $m_\star$ (the average stellar mass in a cluster)
and $M_{\rm tot}$ are both in solar units and $R_h$ is in pc. We have
evaluated this timescale assuming an average $m_\star=0.5\,M_\odot$ in all
clusters.

\item
$\log f_0\equiv \log\,\left[\rho_0/(2\pi \sigma_c^2)^{3/2}\right]$,
a measure of the model's central phase-space density in units of
$M_\odot\,{\rm pc}^{-3}\,({\rm km}\,{\rm s}^{-1})^{-3}$. In this expression,
$\sigma_c$ refers to the central one-dimensional velocity dispersion
{\it without} projection along the line of sight. The ratio $\sigma_c/\sigma_0$
is obtained in general from the solution of the Poisson and  Jeans equations
for given $W_0$ or $\gamma$, and the fitted $\sigma_0$ again is known from
equation (\ref{eq:rscale}). With the central relaxation time $t_{\rm rc}$
of a cluster defined as in equation (8-71) of \citet{bt87},
taking an average stellar mass of $m_\star=0.5\,M_\odot$ and a typical
Coulomb logarithm $\ln\Lambda\approx12$ leads to the
approximate relation $\log\,(t_{\rm rc}/{\rm yr})\simeq 8.28-\log\,f_0$.

\end{itemize}
The uncertainties in these derived dynamical quantities are estimated from
their variations around the minimum of $\chi^2$ on the model grids we fit,
as above, combined in quadrature with the population-synthesis model
uncertainties in $\Upsilon_V^{\rm pop}$, which in turn reflect the estimated
uncertainties in published cluster ages and metallicities.

\subsection{Fit Comparisons}
\label{subsec:fitcomp}

\subsubsection{Our Fits vs.~Published Catalogues}
\label{subsubsec:pubcat}

An important check on the array of cluster properties presented in
\S\ref{subsec:fits} is provided by comparing our basic fit parameters against
those in published catalogues for the Milky Way and LMC/SMC/Fornax clusters.
We confine such comparisons to one between our power-law fit parameters and
those reported by \citet[MG03]{mg03a,mg03b,mg03c} for LMC, SMC, and
Fornax clusters; and one between our King-model fits to the Galactic GCs and
the parameters in the catalogue of \citet{har96} \citep[itself ultimately
based in large part on the profile database of][]{tkd95}.
The generally good results give us
confidence that there are no problematic biases or any other procedural
issues with our modeling or fitting techniques; thus, we have not searched the
literature to compare with all other model fits that might have been performed
on any of these clusters.

The left-hand panels of Fig.~\ref{fig:complit} show our power-law exponents
$\gamma$, scale radii $r_0$, and central surface brightnesses $\mu_{V,0}$
against those published by MG03 for their cluster sample. The overall
agreement between our numbers and theirs is apparent---although note that
the ``Mackey \& Gilmore'' surface brightnesses we compare to in the bottom
left panel have already been {\it corrected} for our extinctions and zeropoint
offsets in Table \ref{tab:basic}. There is somewhat more scatter in the plots
of $\gamma$ and $r_0$ values than in the $\mu_{V,0}$ graph, for the simple
reason that the latter parameters are more sensitive to the ground-based
starcounts which we have added to about two-thirds of the MG03 cluster sample.

It is worth noting a frequent tendency for our power-law fits to return
somewhat steeper $\gamma$, and correspondingly larger $r_0$,
than those of MG03. This can also be traced to our inclusion of ground-based
data, as the specific example of the LMC cluster NGC 2121 illustrates well.
MG03 quote a value of $\gamma=3.25$ for this object, whereas we have obtained
$\gamma=6.00$; it is clearly visible as an ``outlier'' in the upper left-hand
panel of Fig.~\ref{fig:complit}. Referring back to the plotted fits in
Fig.~\ref{fig:sbfits}, however, we see that the SB profile from \citet{mg03a}
extends only to $R\simeq70\arcsec=17$ pc in this case---just
barely outside the constant-density core region of NGC 2121---while our
additional groundbased data reach to $R\simeq100\arcsec\simeq24$ pc and are
critical to accurately constraining any model fit. Fortunately this example is
the most extreme in our sample, but the point is made that when a model
extrapolates significantly beyond the limit of the fitted data, care must be
taken in using the results. This is of most concern for power-law models,
which are innately the most spatially extensive of the ones we fit.

The right-hand panels of Fig.~\ref{fig:complit} show our fitted King-model
concentrations, scale radii, and central surface brightnesses for 85 Galactic
globular clusters, against those parameters taken from the Harris
catalogue. For the most part, the agreement is again quite good; but there are
a number of clusters for which we claim significantly different $c$ values
from those indicated by \citet{har96}, and these differences propagate into
the $r_0$ and $\mu_{V,0}$ plots (given the same data, a lower fitted $c$
corresponds in general to a measurably larger $r_0$ and somewhat fainter
$\mu_{V,0}$).

\setcounter{figure}{12}
\begin{figure}
\epsscale{1.25}
\plotone{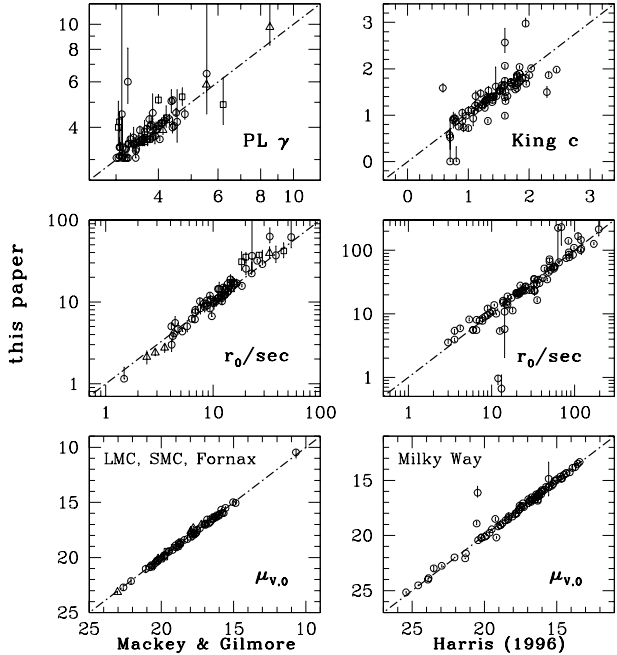}
\caption{\label{fig:complit}
Comparison of our power-law fit parameters for LMC, SMC, and Fornax clusters
({\it left-hand panels}) against those from \citet{mg03a,mg03b,mg03c},
and of our King-model fit parameters for Milky Way globular clusters
({\it right-hand panels}) against those catalogued by \citet{har96}. Broken
lines in all panels indicate equality. The comparison of our central
surface brightnesses with the ``Mackey \& Gilmore'' values in the lower
left-hand panel uses their data {\it after correction} for both
the zeropoint offsets $\Delta\mu_V$ in Column (2) of Table \ref{tab:basic}
(also Table \ref{tab:offset}) and the $V$-band extinctions in Column (3) of
Table \ref{tab:basic}. The Milky Way GC central surface brightnesses plotted
in the bottom right panel are corrected for the $A_V$ in Table
\ref{tab:basic}, which are the same as those in \citet{har96}.
}
\end{figure}

For the two GCs NGC 6101 and NGC 6496, we find King concentrations $c\approx0$
while Harris gives $c\simeq0.7$--0.8. Looking at these objects in
Fig.~\ref{fig:sbfits}, they are unquestionably low-concentration clusters,
with steep declines in surface brightness beyond relatively large cores. The
exact details of any fit are heavily influenced by the outermost one or two
datapoints in each case (and by irregular structure at the center of NGC 6496),
and our errorbars on $c$ (and $r_0$ and $\mu_{V,0}$) reflect this.

\begin{figure*}
\epsscale{1.00}
\plotone{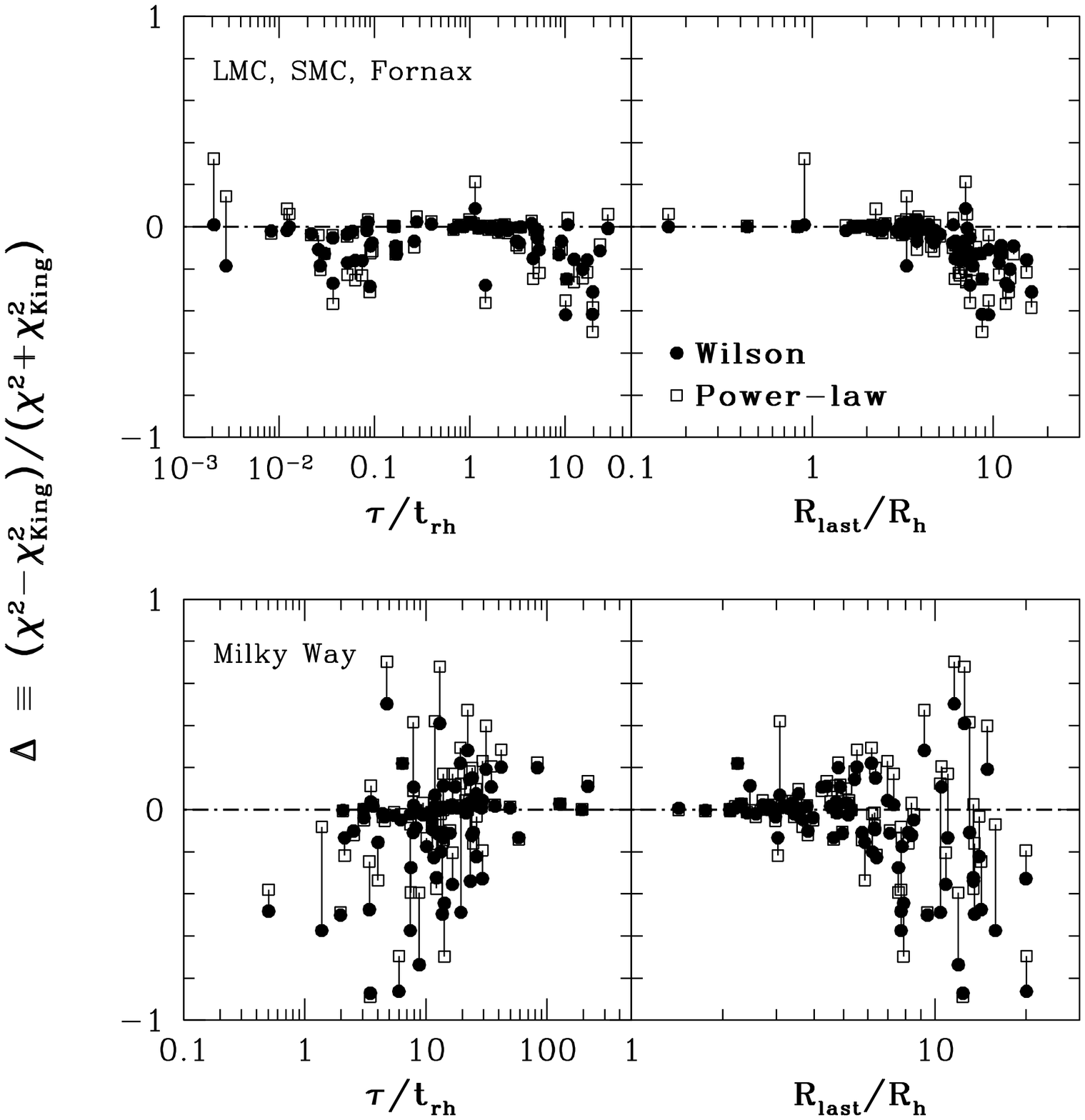}
\caption{\label{fig:chi2}
Goodness of fit of \citet{wil75} spheres and power-law models, relative to
standard \citet{king66} models, for 68 clusters in the LMC, SMC, and
Fornax ({\it top
panels}) and for 85 globular clusters in the Milky Way ({\it bottom panels}).
Solid points mark the relative $\chi^2$ index, $\Delta$, for the Wilson-model
fits; open squares denote $\Delta$ for the power-law fits. Solid lines connect
the Wilson and power-law $\Delta$ values for each cluster. $\Delta$ is
shown as a function of cluster dynamical age in the left-hand panels, and
as a function of the radial extent of the observed surface-brightness profiles
in the right-hand panels. The inherently more extended Wilson model tends to
fit clusters of any age better ($\Delta<0$) when their outer halos are better
defined empirically.
}
\end{figure*}

Somewhat similarly, we find $c=1.5\pm0.12$ for NGC 6528, whereas \citet{har96}
states $c=2.3$. There are many points from \citet{tkd95} for this cluster
to which we assigned zero weight when doing our fits. The model
parameters are apparently quite dependent on which of the data are taken
into account for this relatively poorly defined brightness profile.

The three points falling highest above the line of equality in the top
right-hand panel of Fig.~\ref{fig:complit} correspond to Palomar 10 (Harris
$c=0.58$; our $c=1.59$), Palomar 1 (Harris $c=1.6$; our $c=2.57$), and
Palomar 12 (Harris $c=1.94$; our $c=2.98$). Inspection of these in
Fig.~\ref{fig:sbfits} shows no obvious problem with any of our models for
Pal 10, although clearly the range of fitted datapoints from
\citet{tkd95} is less than ideal; in fact, Harris' tabulated parameters are
based on observations by \citet{khm97} which
supersede those of Trager et al. The data for Pal 1 do not
show any clear evidence for an isothermal core in the first place,
allowing this cluster to be fit by a high-concentration King model with a
small predicted scale radius and bright central surface brightness that are
in fact unobserved. Pal 12 shows what might be described as a double-core
structure, and it is therefore also fit relatively best by high-$c$, low-$r_0$
and bright-$\mu_{V,0}$ King and Wilson models---although the fits are
certainly not ``good'' in an absolute $\chi^2$ sense (see
Table \ref{tab:basic}). Any catalogued fit parameters (including ours) for
these three Palomar clusters are probably best viewed as only provisional.

\subsubsection{Goodness-of-Fit for Different Models}
\label{subsubsec:goodness}

We next compare the $\chi^2$ values of the different model fits to every
cluster in our sample. We make this comparison relative to $\chi^2$ of the
best-fit King model in each case by computing
$\Delta \equiv (\chi^2-\chi_{\rm King}^2)/(\chi^2+\chi_{\rm King}^2)$
for the best Wilson and power-law fits. This index ranges from a minimum of
$\Delta=-1$ for an alternate model with $\chi^2\ll \chi_{\rm King}^2$, to
a maximum of $\Delta=+1$ when $\chi_{\rm King}^2\ll \chi^2$.

The top panels of Fig.~\ref{fig:chi2} show the distribution of $\Delta$ values
for the Wilson (filled circles) and power-law (open squares) fits to the
MG03 cluster sample in the LMC, SMC, and Fornax. The points for the
Wilson and power-law fits of any one cluster are connected by a solid line.
On the left we plot $\Delta$ as a function of cluster dynamical age, i.e., the
chronological age $\tau$ in units of the King-model half-mass relaxation
time $t_{\rm rh}$ from Table \ref{tab:dyn}. On the right, we show $\Delta$ as
a function of the ratio $R_{\rm last}/R_h$, where $R_{\rm last}$ is the
clustercentric radius of the outermost surface-brighntess datapoint observed
in a cluster and $R_h$ is the (King-model) half-mass radius.

It is immediately apparent that, in every LMC/SMC/Fornax cluster studied here,
\citet{wil75} models fit at least as well as \citet{king66} models, and very
often substantially better. Power-law models generally fit roughly as well as
Wilson models, sometimes slightly better and sometimes somewhat worse. On one
level, the top panels of Fig.~\ref{fig:chi2} are therefore a re-statement of
the appreciated fact that many clusters in the Magellanic Clouds are more
extended than classic King models \citep[e.g.,][]{eff87,mg03a,mg03b}. New
here is the demonstration that, although they are usually acceptable fits,
untruncated power-law forms specifically are {\it not required} to describe
the density distributions of these objects. We expect that the same is likely
true of young massive clusters in other disk galaxies
\citep[and references therein]{lar04,sch04}.

Also new is our quantification of the improvement in fit yielded by the more
extended models as a function of cluster age. \citet{eff87} originally
suggested that power-law models fit {\it young} LMC clusters better than King
models because of the presence of unbound stellar halos which are relics
of the cluster formation process and simply have not had time to be stripped
away by tides. The chronologically young clusters in the current sample
\citep[which includes that of][]{eff87} generally have $\tau/t_{\rm rh}\la 1$
in the upper left panel of Fig.~\ref{fig:chi2}. Many of them have $\Delta<0$
and are undoubtedly fit better by Wilson or power-law models than by King
models; but about as many have $\Delta\approx0$ and are equally well fit
by any model type. Moreover, there is a comparable number of
chronologically and dynamically {\it old} clusters (including the Magellanic
Cloud globulars and those in Fornax) which, at
$\tau/t_{\rm rh}\sim10$, also have $\Delta<0$ and are characterized by
spatially extended halos not easily reproduced by King models. [Working
from completely independent data, \citet{rod94} have already
claimed this for three of the five globular clusters in Fornax.] Age
appears not to be the main factor in determining whether or not any of these
clusters can be well described by regular King models.

Instead, the upper right-hand panel of Fig.~\ref{fig:chi2} shows simply that
King models provide progressively less satisfactory fits as surface-brightness
and starcount data extend farther and farther into cluster halos. All of
the clusters here---young or old, in the disk or halo populations of any of
these three small galaxies---are better fit by Wilson or power-law models if
the models are forced to fit beyond $\simeq4$--5 half-light radii in the
clusters. (It will be recalled from Fig.~\ref{fig:conc} above that this is
the point where the King, Wilson, and power-law model structures
begin to differ appreciably.) We suggest that the halos of massive star
clusters are {\it generically} more extensive than the stellar distribution
function of \citet{king66} allows, and that the development of a physically
motivated model accounting for this (one less ad hoc than a Wilson or
power-law prescription) could lend substantial new insight into questions of
cluster formation and evolution.

The fundamental physical question remaining from \citet{eff87}
is whether the stars at the largest observed radii in the young clusters
particularly are gravitationally bound in a model-independent sense: do
these objects overflow the Roche lobes defined by the potentials of their
parent galaxies?
Simply fitting power laws to the clusters does not in fact address this
issue, as equally good or better fits of spatially limited Wilson models
can be found; but at the same time, it remains to be shown that the fitted
$r_t$ from the Wilson models actually correspond to the true tidal limits
imposed by the galaxies. Moreover, the question now has to be extended to
many old globulars. To properly answer it for any cluster requires
{\it at a minimum} not only a highly precise empirical estimate of $r_t$
itself, but also a detailed understanding of the total mass distribution and
gravitational field of the parent galaxy; good information on the present-day
galactocentric position of the cluster; and knowledge of its orbital energy
and pericenter. The interplay between these ingredients makes for a subtle
problem, fraught with uncertainty even in the Milky Way
\citep[see, e.g.,][]{inn83}, which is beyond the scope of this paper.
However, we do touch briefly on a comparison between the fitted ``tidal''
radii in the King and Wilson models for our cluster sample, in
Fig.~\ref{fig:tidal} below.

The bottom panels of Fig.~\ref{fig:chi2} are plots of the relative $\chi^2$
index $\Delta$ against both dynamical age and spatial extent of the SB
data for the 85 Galactic GCs that we have modeled.\footnotemark
\footnotetext{Note the absence of Milky Way globulars
with $R_{\rm last}/R_h<1$ in Fig.~\ref{fig:chi2}, contrasting with the
presence of such objects in the LMC/SMC/Fornax cluster sample. As we
discussed in \S\ref{subsec:trager}, such large fitted $R_h$ in the Galactic
sample tend to be associated either with relatively poor data or with clusters
identified as core-collapsed by \citet{tkd95}. These GCs are named in
Table \ref{tab:MWnotfit} but have been left out of Fig.~\ref{fig:chi2} (and
Tables \ref{tab:basic}--\ref{tab:dyn}) by construction.}
It is clear that many globulars with well-observed halos
($R_{\rm last}/R_h\ga 5$) are again relatively better fit by Wilson models
than by single-mass, isotropic King models, regardless of their dynamical
age $\tau/t_{\rm rh}$. In most of these cases power laws are worse fits
than Wilson models, which likely reflects the inability of the former to
describe tidal limits to a cluster. But now there are also about ten
globulars with data extending beyond $5 R_h$, which are better fit by King
models than either Wilson or power-law spheres. A good example is NGC 104 = 47
Tucanae ($\Delta=+0.5$ for the Wilson model fit), although in
Fig.~\ref{fig:sbfits} this profile actually appears to prefer some kind of
description {\it intermediate} to \citet{king66} and \citet{wil75}.

The bottom left panel in this plot gives the visual impression
that---unlike in the
LMC, SMC, and Fornax cluster sample---there may be some correlation between
our $\Delta$ statistic and the dynamical age $\tau/t_{\rm rh}$. A test
using the Spearman rank correlation coefficient for the Wilson-model fits
specifically shows that these quantities are indeed correlated, with
a formal confidence level of $>99\%$. However, the correlation
between $\Delta$ and $R_{\rm last}/R_h$ for the Milky Way GCs is still
stronger and more significant than any correlation with $\tau/t_{\rm rh}$.
In any case, all of this is going on over a narrow range of extreme age
relative to the rather larger spread in the LMC, SMC, and Fornax sample,
where the situation is much clearer.

Some of these differences in the bottom panels of Fig.~\ref{fig:chi2}
relative to the upper panels may simply be a reflection of the more
heterogeneous nature of the original data compiled by \citet{tkd95}.
It could also be indicative of the ultimate limitations of our assumptions of
velocity isotropy and (probably more important) a single-mass stellar
population in the clusters: NGC 104, for example, is known to exhibit mass
segregation \citep[e.g.,][]{and97} and can be fit well with multimass King
models \citep{mey88,mey89}. On the other hand, we find here that the cluster
NGC 5272 = M 3, which was the original motivation for the development of
multimass King models \citep{dac76,gg79}, can
be perfectly well fit by a single-mass and isotropic Wilson model; it is
the filled circle at $R_{\rm last}/R_h=12.4$ and $\Delta=-0.87$ in the lower
right panel of Fig.~\ref{fig:chi2}.

The totality of the results presented here still suggest to us that a
fundamental alteration to the \citet{king66} distribution function is required
to account for the halo structure and dynamics of massive star clusters in
general. That the ad hoc, single-mass, and isotropic \citet{wil75} model
is not the perfect solution should come as no surprise; but, as our discussion
of $\omega$ Centauri concluded (\S\ref{subsubsec:wcen}), neither can multimass
and/or anisotropic variations on the standard \citet{king66} distribution
function correctly explain the structure of all globular clusters.

This issue aside, we now turn to consider whether the observable, physical
properties---core and half-light radii, total luminosities, and the
like---that we have derived for our clusters are reasonably model-independent.
In particular, we would like some assurance that they are robust enough to
allow useful characterizations of parameter interdependences and trends
between clusters.

\subsubsection{Physical Cluster Properties in Different Models}
\label{subsubsec:physprop}

\begin{figure*}
\epsscale{1.00}
\plotone{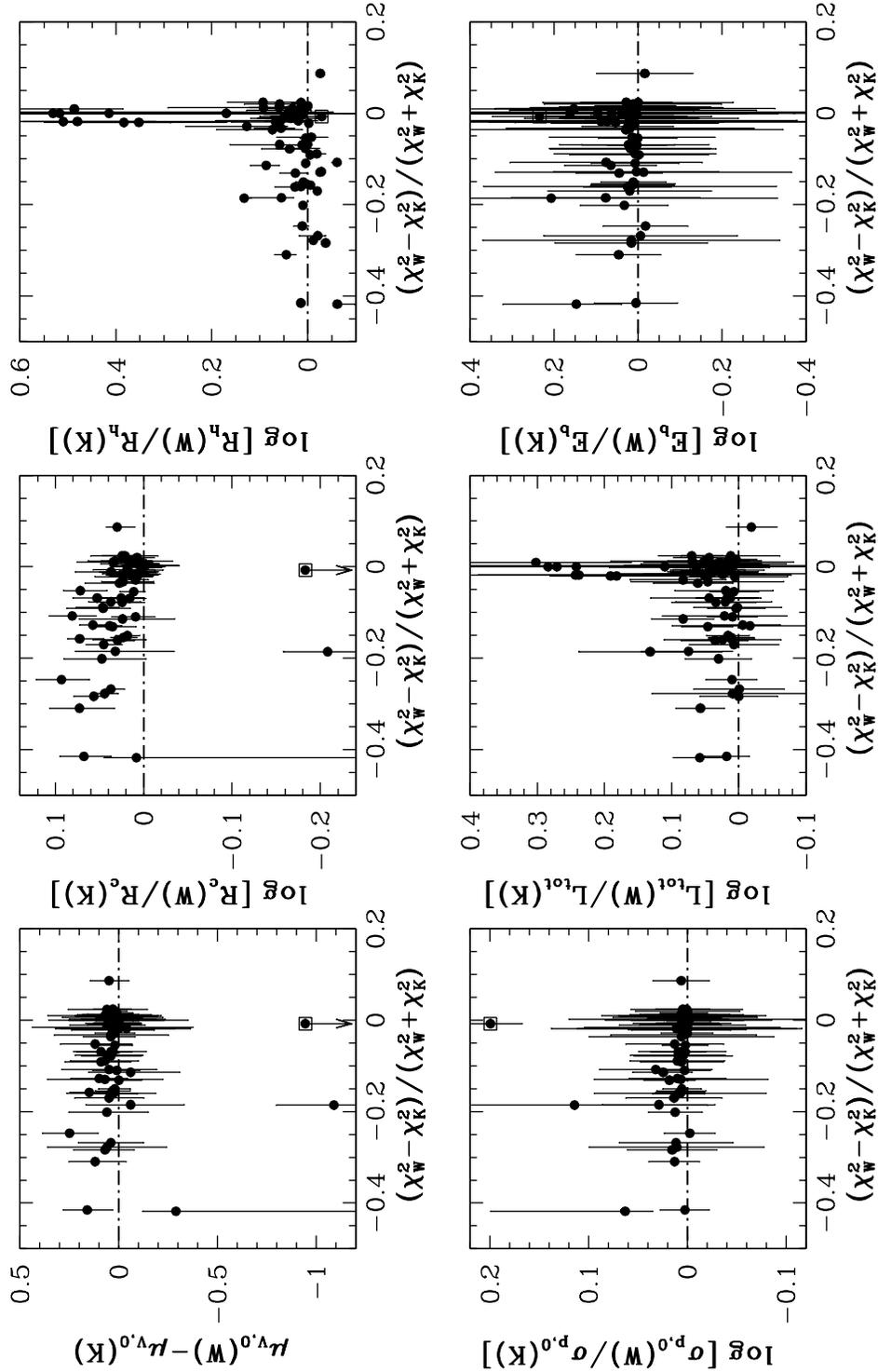}
\caption{\label{fig:mgwk}
Comparison of physical cluster properties derived from Wilson-model fits
to LMC, SMC, and Fornax clusters, vs.~those derived from King-model fits.
The point enclosed in a square in every panel is the LMC globular cluster
NGC 2005. See text for details.
}
\end{figure*}

Figure \ref{fig:mgwk} compares the Wilson- and King-model values for a
number of parameters of the LMC, SMC, and Fornax (MG03)
clusters in our sample (see Tables \ref{tab:basic}--\ref{tab:dyn}). The
overriding conclusion to be drawn is that these two types of model fits tend,
with some understandable exceptions, to agree within $\sim5\%$ ($\sim0.02$
dex) on the values of basic cluster properties.
We have plotted the differences in fitted and derived properties between
the two models as functions of the relative $\chi^2$ index $\Delta\equiv
(\chi^2_{\rm Wilson}-\chi^2_{\rm King})/
(\chi^2_{\rm Wilson}+\chi^2_{\rm King})$.

The upper left-hand panel of Fig.~\ref{fig:mgwk} shows that in most
cases the central surface brightness $\mu_{V,0}$ is very well determined,
stable at the $\sim0.03$-mag level on average no matter which model is fit.
(Note, however, that when the Wilson model fits better, it tends to return a
slightly fainter central brighntess than a King-model fit.) One of the
most obvious exceptions is the LMC cluster NGC 2005, which is the point
enclosed in an open square; it in fact lies off the vertical scale at
$\mu_{V,0}({\rm W})-\mu_{V,0}({\rm K})=-5.01$. Although the $\chi^2$ of the
two fits are essentially the same, the Wilson model for this cluster is of
much higher concentration---and therefore brighter central surface
brightness---than the King model, due to the influence of a single datapoint
at $R\simeq 1\arcsec\simeq 0.24$ pc (see Fig.~\ref{fig:sbfits}). [NGC 2005
is an old globular that has been cited by other authors as a core-collapsed
object; e.g., \citet{mat87}. See \citet{mg03a} for discussion of other
possible core-collapse candidates in this sample of LMC GCs.] The other main
outlier in this plot, with
$\mu_{V,0}({\rm W})-\mu_{V,0}({\rm K})=-1.09$, is R 136 = 30 Doradus in the
LMC. This very young object's extremely compact configuration is not
particularly well
fit by any of our models \citep[for further discussion, see, e.g.,][]{mg03a}.

\begin{figure*}
\epsscale{1.00}
\plotone{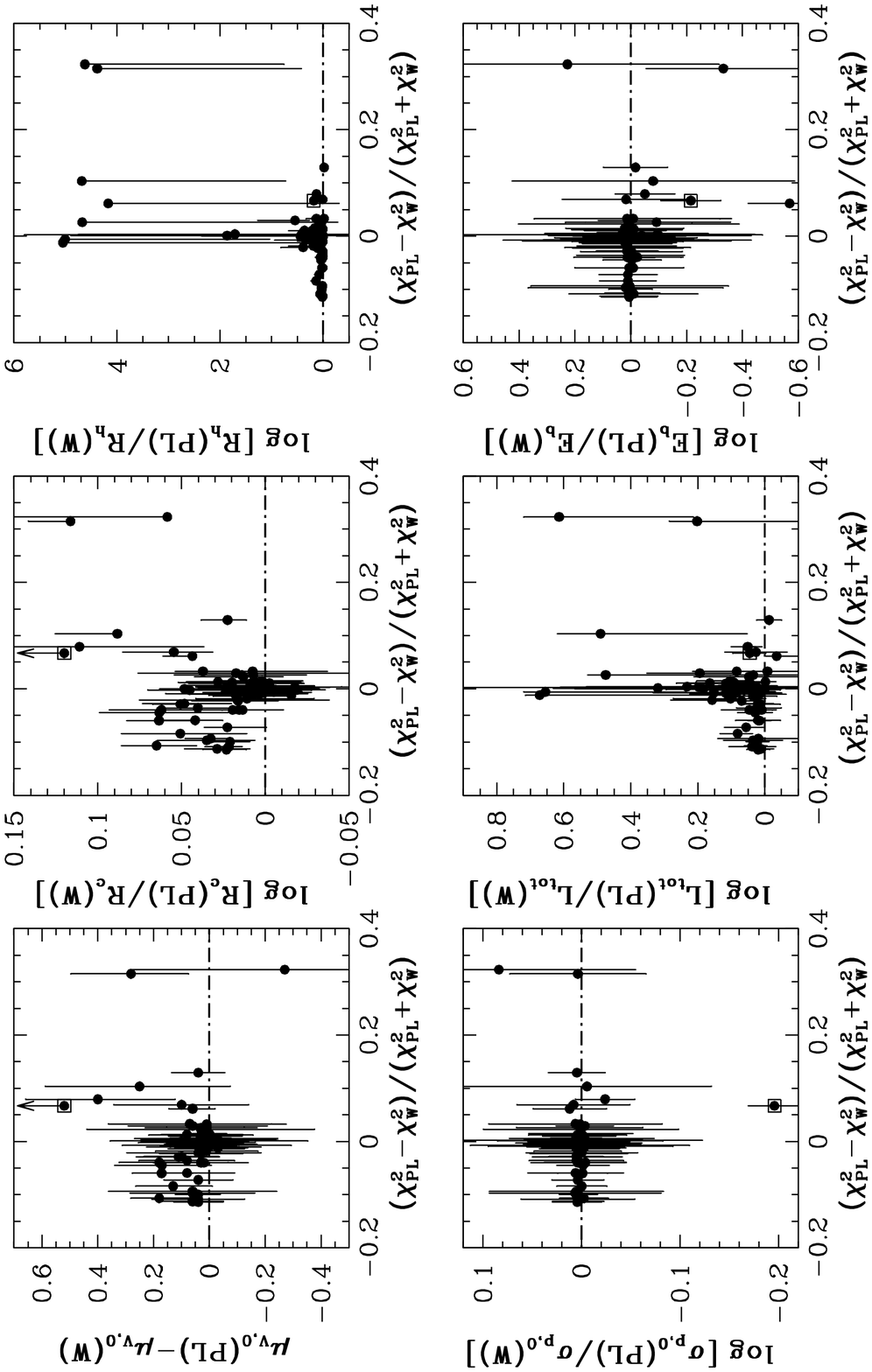}
\caption{\label{fig:mgpw}
Comparison of physical cluster properties derived from power-law model fits
to LMC, SMC, and Fornax clusters, vs.~those derived from Wilson-model fits.
The point enclosed in a square in every panel is the LMC globular cluster
NGC 2005. See text for details.
}
\end{figure*}

\begin{figure*}
\epsscale{1.00}
\plotone{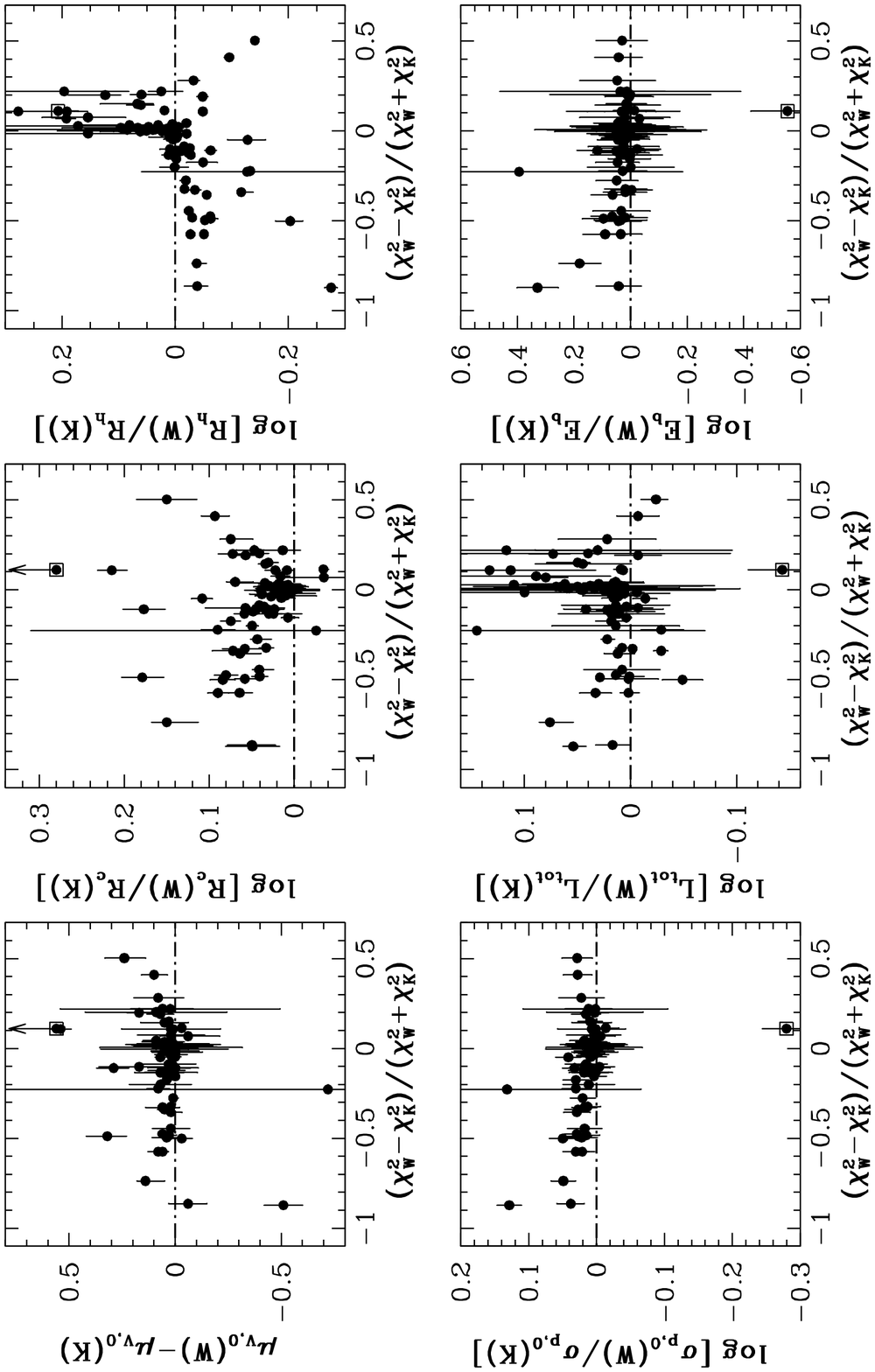}
\caption{\label{fig:galwk}
Comparison of physical cluster properties derived from Wilson-model fits
to Galactic globular clusters, vs.~those derived from King-model fits.
The point enclosed in a square in every panel is Palomar 1. See text for
details.
}
\end{figure*}

The upper middle panel shows a tendency for the projected core radius to be
larger in the Wilson models for these clusters, by $\sim 5\%$ on average
but up to $\sim15\%$ when Wilson spheres fit very much better than King
models. This might be viewed as a demonstration that fitting the ``wrong''
type of model to a cluster introduces possible error of this order in the
fitted $R_c$. Note that NGC 2005 and R 136 are again outliers here, with their
much {\it smaller} core radii in the Wilson fits corresponding to their much
brighter $\mu_{V,0}$.

It can be seen in the upper right-hand panel of Fig.~\ref{fig:mgwk} that, when
Wilson models fit better than King in the LMC/SMC/Fornax sample, the model
half-light radius $R_h$ is essentially the same---again to within about 5\% on
average---in either model. The reason is that observations out to
$R\ga5 R_h$ are generally required (Fig.~\ref{fig:chi2}) in order to show a
clear preference for one model or the other. In such cases $R_h$ is very well
constrained by the data themselves and must be reproduced by essentially
{\it any} model fit. When Wilson- and King-model fits have comparable
$\chi^2$, on the other hand, it occasionally happens that their $R_h$
values differ by factors of 2.5--3. When this occurs, it is most often because
$R_{\rm last}/R_h$ is of order 1 or smaller, meaning that few if any
data are available at large cluster radii to constrain the very different
extrapolations of the two types of model. It then becomes unclear which $R_h$
is correct---although Fig.~\ref{fig:mgwk} shows that our estimated
errorbars in these cases are also larger than average, properly signaling the
problem.

The bottom panels of Fig.~\ref{fig:mgwk} are all very similar.  The predicted
central velocity dispersions for these clusters are (except for NGC 2005 and
R 136) the same to within $\sim1\%$, on average, whether Wilson or King models
are fit (and regardless of which is the better fit); total cluster
luminosities are about as well determined as the cluster half-light radii
(with the same potential for some large Wilson--King discrepancies and large
errorbars when the two models have comparable $\chi^2$ values); and global
cluster binding energies differ by $<5\%$ on average between the two models.

Figure \ref{fig:mgpw} is analogous to Fig.~\ref{fig:mgwk} but compares the
fits of power laws to those of Wilson models for the MG03 cluster set. NGC
2005 is
again enclosed by an open square in every panel, and the rightmost datapoint
is R 136 = 30 Dor. Similar comments apply to these comparisons as to the
previous ones, with the important exception that the half-light radii
implied by power-law fits can sometimes be {\it orders of magnitude
larger} than the $R_h$ obtained from Wilson- or King-model fits. This reflects
a fundamental difficulty with power-law models: extrapolation of a fit which
happens to fall too near $\gamma=3$ over some available (too small) range of
data is barely convergent (see eq.~[\ref{eq:power}]) and clearly unphysical.
Nevertheless, our estimated errorbars on $R_h$ (and on $L_{\rm tot}$) even
for unrealistic power-law fits such as these do reasonably reflect the
situation. And in many of the cases seen here, the power-law fit is,
after all, measurably worse ($\Delta>0$) than the best Wilson-model fit.

Figure \ref{fig:galwk} next compares the King and Wilson fit parameters for
Galactic globular clusters. The point enclosed in an open square in every
panel is Palomar 1, which has $\mu_{V,0}({\rm W})-\mu_{V,0}({\rm K})=5.81$
and $\log[R_c({\rm W})/R_c({\rm K})]=1.74$ as a result of the much higher
concentration of the King model fit in this case (see Fig.~\ref{fig:sbfits}
and recall the discussion around Fig.~\ref{fig:complit} above). The other
main outliers in the upper left-hand panel of Fig.~\ref{fig:galwk} are
NGC 5272 = M 3, at $\Delta=-0.87$ and $\mu_{V,0}({\rm W})-\mu_{V,0}({\rm K})=
-0.51$, and Palomar 2, at $\Delta=-0.23$ and
$\mu_{V,0}({\rm W})-\mu_{V,0}({\rm K})=-0.72$ with a large errorbar. Aside
from these objects, Wilson fits to Galactic GCs tend to imply central surface
brightnesses slightly fainter than King-model fits, but by only 0.02 mag on
average and less than $\simeq0.1$ mag in most cases. Wilson models are also
associated with slightly
larger projected half-intensity radii $R_c$ and smaller half-light radii
$R_h$, and thus relatively lower concentrations as measured by the ratio
$R_h/R_c$. This is simply a result of the more extended cluster structure
essentially assumed {\it ab initio} in the model. It is apparent again that the
disagreement between King- and Wilson-model estimates of $R_h$, and
subsequently $L_{\rm tot}$, is potentially largest when the two models fit a
given cluster about equally well.
As above, equal-quality fits with disparate $R_h$
typically occur when the available cluster data do not extend
far enough in clustercentric radius to definitively constrain the
large-$R$ extrapolation of either model; but our estimated uncertainties on
the derived quantities generally reflect this. The predicted
central velocity dispersion and global cluster binding energy are as
well-behaved as in our LMC/SMC/Fornax cluster fits.

\begin{figure}
\epsscale{1.25}
\plotone{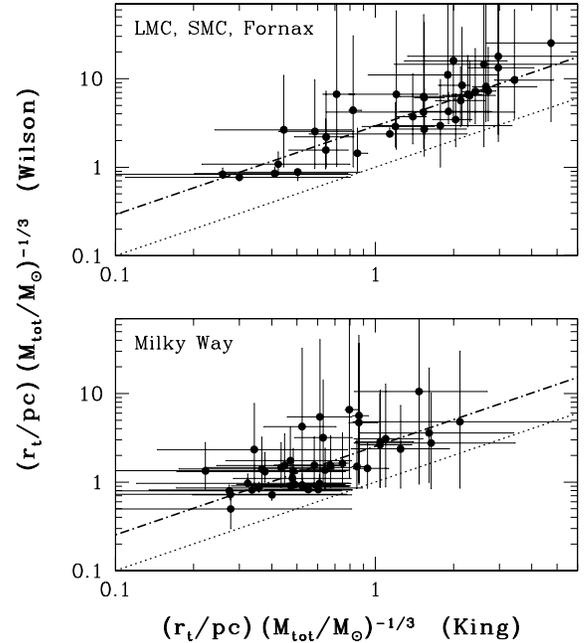}
\caption{\label{fig:tidal}
Mass-normalized tidal radii inferred from Wilson-model fits, vs.~those from
King-model fits, for 37 clusters in the LMC, SMC, and Fornax
({\it top panel}) and 43 globular clusters in the Milky Way ({\it bottom
panel}). Only clusters with $R_{\rm last}/R_h\ge 4$ and
$\Delta=(\chi^2_{\rm Wilson}-\chi^2_{\rm King})/
(\chi^2_{\rm Wilson}+\chi^2_{\rm King}) \le 0.1$
are plotted. Dotted line in each panel indicates equality,
$r_t M_{\rm tot}^{-1/3}({\rm Wilson})=r_t M_{\rm tot}^{-1/3}({\rm King})$.
Bolder, dash-dot lines are at the median ratios
$r_t M_{\rm tot}^{-1/3}({\rm Wilson})/r_t M_{\rm tot}^{-1/3}({\rm King})=2.9$
for the LMC/SMC/Fornax sample, and
$r_t M_{\rm tot}^{-1/3}({\rm Wilson})/r_t M_{\rm tot}^{-1/3}({\rm King})=2.5$
for the Milky Way globulars.
}
\end{figure}

Finally, in Fig.~\ref{fig:tidal} we compare the extrapolated tidal radii
from King- and Wilson-model fits to 37 of the LMC, SMC, and Fornax clusters
(top panel) and 43 Milky Way globulars (bottom panel). We have only included
clusters in these graphs if they have $R_{\rm last}/R_h\ge 4$---so that the
extrapolations of the fits to $r_t$ are constrained as well as possible---and
$\Delta=(\chi^2_{\rm Wilson}-\chi^2_{\rm King}) /
(\chi^2_{\rm Wilson}+\chi^2_{\rm King}) \le 0.1$---in which case \citet{wil75}
spheres fit the surface-brightness data at least as well as \citet{king66}
models.

What we have actually plotted in Fig.~\ref{fig:tidal} is the mass-normalized
tidal radius evaluated for each of these clusters within each of the two
model fits: $r_t M_{\rm tot}^{-1/3}$, the inverse cube root of the average
cluster density. This quantity is related directly to the tidal field in
which a cluster is embedded; most simply, in the case of a cluster at radius
$r_{\rm gc}$ in a spherical galaxy,
$M_{\rm tot}/r_t^3 \propto M_{\rm gal}(r_{\rm gc})/r_{\rm gc}^3$.
Ideally, we would like to compare estimates of $r_t M_{\rm tot}^{-1/3}$
from fits of structural models, to the value expected for any
cluster in a given galactic tidal field (in order, for example, to assess
whether an extended cluster halo is ``unbound;''
cf.~\S\ref{subsubsec:goodness}). However, to do this requires detailed
knowledge of the parent galaxy mass profile $M(r_{\rm gc})$ (including dark
matter); of the instantaneous three-dimensional position of the cluster,
$r_{\rm gc}$; and of the shape and energy of the cluster orbit, which
together set the coefficient connecting the mean cluster density to the
average galaxy density at $r_{\rm gc}$ \citep[e.g.,][]{king62,inn83}.
Estimating all of these quantities individually for each object in our sample
is clearly out of the question, and even a statistical treatment of the
cluster ensemble in each galaxy \citep[such as in][]{inn83} is beyond the
scope of our analysis.

We simply point out that---as expected---the mass-normalized tidal radii
implied by the Wilson-model fits to our clusters are systematically larger
than those implied by King-model fits. The dotted line in each panel of
Fig.~\ref{fig:tidal} indicates equality between the Wilson and King values
for $r_t M_{\rm tot}^{-1/3}$, while the bolder, dash-dot lines show the
median ratios of the two estimates: $\simeq 2.9$ in the LMC/SMC/Fornax
sample, and $\simeq2.5$ for the Milky Way globular clusters. We emphasize
again that each of the points
plotted represents a cluster which is fit {\it at least as well or better} by
a Wilson sphere vs.~a King model. But the question remains open as to whether
the limiting radii (or mean densities) of the former models are
quantitatively consistent with a naive association of {\it fitted} $r_t$
values with the {\it true} tidal radii of the real clusters. The large
errorbars on both King and Wilson values of $r_t M_{\rm tot}^{-1/3}$ for all
the clusters in Fig.~\ref{fig:tidal} only stress further the difficulty of
precision work along these lines. Even for these best-observed clusters,
$r_t$ is almost always inferred from extrapolation rather than measured
directly; it is the most uncertain cluster parameter that we estimate.

\section{Observed Velocity Dispersions and Dynamical Mass-to-Light Ratios}
\label{sec:dynml}

As we have described, the derivation of total cluster masses (and all
the dependent quantities in Table \ref{tab:dyn} above) from our
surface-brightness fits has been facilitated by adopting a mass-to-light
ratio for each cluster based on the population-synthesis modeling of
\S\ref{sec:popsyn}. This is a necessary step towards examining physical
trends and dependences among the properties of star clusters spanning a wide
range of ages. It remains to be shown, however, that the population-synthesis
models we use predict mass-to-light ratios that are consistent with what can
be inferred directly for the minority of clusters which have measured
velocity dispersions. Such a demonstration is the purpose of this Section.

To make this check, we have compiled velocity-dispersion data for as many
of our modeled LMC, SMC, Fornax, and Milky Way clusters as we could
easily find in the published literature. We have not made an attempt at a
comprehensive collection, but simply one including enough clusters to address
meaningfully the question at hand
\citep[in fact, for Milky Way globulars we have relied exclusively on the
work of][which is itself a compilation of earlier studies]{pry93}.
A drawback is that very few young
Magellanic Cloud clusters have well-determined velocity dispersions:
among those in our sample, we have found data only for NGC 1850, 1866, 2157,
2164, and 2214 in the LMC, and NGC 330 in the SMC. Although bright, these
objects are usually of
relatively low mass compared to the average old GC, and their velocity
dispersions are intrinsically low and difficult to measure. Thus, our
comparison of population-synthesis and dynamical mass-to-light ratios really
speaks most clearly to the old-age limit of the models; but the good results
in that limit are encouraging.

Our analysis is detailed in Table \ref{tab:popvdyn}. After the cluster name,
Column (2) of this table repeats the $V$-band mass-to-light ratio predicted
by population-synthesis models [from Column (6) of Table \ref{tab:poptable}
or Column (2) of Table \ref{tab:dyn}]. Column (3) is the observed velocity
dispersion in the cluster, as reported in the literature. Some of these
dispersions are based on radial-velocity measurements of individual stars
spread throughout the cluster; others, on integrated-light
spectroscopy within a finite slit width. Either way, every
$\sigma_{\rm p, obs}$ is in effect a weighted average of the cluster's
projected velocity-dispersion profile over some area on the sky. From the
details of each observation in the original papers, we have estimated the
effective radius of a circular aperture with roughly the appropriate area.
[Note that this is always $R_{\rm ap}=0$ for the Milky Way globular clusters,
since the observed dispersions in this case have already been extrapolated
to their central values by \citep{pry93}.]
This is reported in Column (4) of Table \ref{tab:popvdyn}. Column (5) gives
the reference to the source of the data.

Given a King, Wilson, or power-law model with fitted $W_0$ (or $\gamma$)
and $r_0$ for any cluster, solving Poisson's and Jeans' equations and
projecting
along the line of sight yields a dimensionless velocity-dispersion profile,
$\widetilde{\sigma}_{\rm p}=\sigma_{\rm p}(R)/\sigma_{\rm p}(R=0)$ as a
function of projected clustercentric radius $R$. The weighted average
${\cal S}^2(R_{\rm ap})\equiv
\left[\int_{0}^{R_{\rm ap}} R\, I(R)\, \widetilde{\sigma}_{\rm p}^2\, dR
\right]
\left[\int_{0}^{R_{\rm ap}} R\, I(R)\, dR \right]^{-1}$
then gives the predicted mean-square velocity dispersion within any circular
aperture of radius $R_{\rm ap}$. We have calculated ${\cal S}$
for each of the model fits to each of the clusters in Table \ref{tab:popvdyn}
given the aperture radii estimated in the table, and obtained
the line-of-sight velocity dispersions at the cluster centers as
$\sigma_{\rm p}(R=0)=\sigma_{\rm p, obs}/{\cal S}(R_{\rm ap})$. In general,
the value of $\sigma_{\rm p}(R=0)$ depends on the model used to compute
${\cal S}$, but the differences in our case are usually small and in Column
(6) of Table \ref{tab:popvdyn} we report only the mean of our
three determinations.

The observed $\sigma_{\rm p}(R=0)$ values are to be compared with the
{\it predicted} $\sigma_{{\rm p},0}$, based on our population-synthesis
$M/L_V$ ratios for each cluster, in Table \ref{tab:dyn}. As described
above for the calculation of these predictions, our fitted models with known
$W_0$ or $\gamma$ also provide the dimensionless ratio
$\sigma_{\rm p}(R=0)/\sigma_0$, for $\sigma_0$ the theoretical scale velocity
appearing in equation (\ref{eq:rscale}). We therefore compute
$\sigma_0$ and use our fitted $r_0$ (Table \ref{tab:basic}) in equation
(\ref{eq:rscale}) to compute the central {\it mass} density $\rho(r=0)$ of
every cluster in Table \ref{tab:popvdyn}. A ``dynamical'' estimate of the
$V$-band mass-to-light ratio follows immediately as
$\Upsilon_V^{\rm dyn}\equiv \rho(r=0)/j_0$, with the luminosity density
$j_0$ taken from Table \ref{tab:struc}. Our estimates of
$\Upsilon_V^{\rm dyn}$ for every model fit, and the comparisons
$\Delta(\log\,\Upsilon_V)=\log\,(\Upsilon_V^{\rm dyn}/\Upsilon_V^{\rm pop})$,
constitute the rest of Table \ref{tab:popvdyn}.

\begin{figure}
\epsscale{1.25}
\plotone{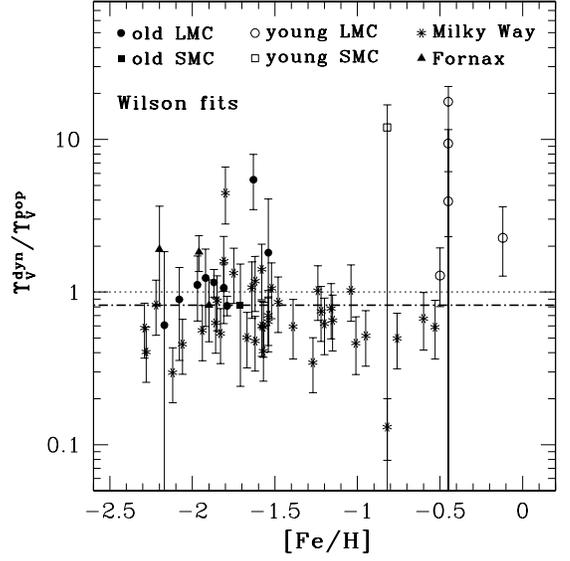}
\caption{\label{fig:upsrat}
Ratio of dynamical $V$-band mass-to-light ratio to population-synthesis model
prediction, as a function of cluster metallicity, for all clusters with
measured central velocity dispersions in Table \ref{tab:popvdyn}. The
dynamical $\Upsilon_V$ used are those calculated from $\sigma_{\rm p, obs}$
using Wilson-model structural fits to each cluster. Population-synthesis
mass-to-light ratios are those predicted by the model of \citet{bru03}
using the disk-star IMF of \citet{cha03}. The bold, dash-dot line
indicates the median $\Upsilon_V^{\rm dyn}/\Upsilon_V^{\rm pop}=0.82$,
which has a standard error of $\simeq\pm0.07$.
}
\end{figure}

Inspection of Table \ref{tab:popvdyn} shows, first, that the dynamical
estimates
of $\Upsilon_V$ are generally very model-independent. The single greatest
exception is NGC 2005 in the LMC, the cluster for which our Wilson-model fit
has a much higher central surface brightness and smaller scale radius, and
thus a smaller inferred mass-to-light ratio, than the King- or power-law model
fits (see Fig.~\ref{fig:mgwk} and Fig.~\ref{fig:sbfits}). Second, the
comparison between $\Upsilon_V^{\rm dyn}$ and $\Upsilon_V^{\rm pop}$ is
favorable. Figure \ref{fig:upsrat} shows this graphically, with the ratio
of the two mass-to-light values plotted as a function of cluster metallicity
(from Table \ref{tab:poptable} above). For definiteness, the results for
$\Upsilon_V^{\rm dyn}$ from our Wilson modeling have been used in this
plot, but it makes no significant difference if the King- or power-law model
numbers are used instead. In all cases, the median $\Upsilon_V^{\rm dyn}/
\Upsilon_V^{\rm pop}\simeq0.82\pm0.07$.

The few young massive clusters in the LMC and SMC for which we have obtained
dynamical mass-to-light estimates are shown as open symbols in
Fig.~\ref{fig:upsrat}. Although these tend to fall nominally above the
line $\Upsilon_V^{\rm dyn}=\Upsilon_V^{\rm pop}$, their measured
$\sigma_{\rm p, obs}$ in Table \ref{tab:popvdyn} are relatively uncertain,
as our associated errorbars attest. In fact, in three of the six cases
only {\it upper limits} to $\sigma_{\rm p, obs}$ are given by the original
authors. All other points in Fig.~\ref{fig:upsrat}  refer to old
($\tau>10^{10}$ yr) globular-type clusters and, as mentioned above, provide a
direct check only on that extreme of the population-synthesis models.
Overall, however, we feel confident that our use of $\Upsilon_V^{\rm pop}$ in
general to infer mass-based cluster properties from simple surface-brightness
modeling is well justified.

\section{$\kappa$-Space Parameters and Galactocentric Distances}
\label{sec:kappa}

We anticipate a main use of our results in Tables \ref{tab:basic},
\ref{tab:struc}, and \ref{tab:dyn} above to be in the definition and
interpretation of correlations between the primary physical properties of
star clusters. Ultimately, such correlations can constrain theories of
cluster formation and evolution. They have been identified and discussed in
many forms in the literature for old, globular clusters in the Milky Way and
a few other galaxies. So far as we are aware, our work here is the first to
allow for systematic investigation of the effects of fitting GCs with models
other than that of \citet{king66}. It is also the first to put a significant
number of young massive clusters on a completely equal footing with the old
globulars.

Correlations among GCs are typically couched in terms of a structural
``fundamental plane'' analogous to that originally defined by
\citet{djo87} and \citet{dre87} for dynamically
hot galaxies. There are at least three {\it equivalent}
formulations of the globular cluster fundamental plane in the literature:

First, \citet{djo95} presents strong bivariate correlations involving
$\sigma_{{\rm p},0}$, $r_0$, $\mu_{V,0}$, $R_h$, and $\langle\mu_V\rangle_h$,
showing Galactic GCs to be an essentially two-parameter family.

Second, \citet{mcl00} works with $\Upsilon_V$, $L_{\rm tot}$, King-model $c$,
and global binding energy $E_b$ to arrive, in different form and with different
physical emphasis, at the same basic conclusion. McLaughlin shows explicitly
the equivalence between his and Djorgovski's formulations of the GC fundamental
plane. In either, the Galactocentric positions $R_{\rm gc}$ of the globulars
are an important external influence; it is well known, for example, that GC
half-mass radii and binding energies correlate significantly with their
location in the Galaxy
\citep[$R_h\propto R_{\rm gc}^{0.4}$ and $E_b\propto R_{\rm gc}^{-0.4}$:][]
{vdb91,mcl00}.

To bring young massive clusters, such as those we have modeled in the LMC
and SMC, into analyses along these lines, it is preferable to work in terms
of $M_{\rm tot}$, $\Sigma_0$, and $\Sigma_h$---rather than their luminosity
or surface-brightness equivalents---so as to avoid purely age-related
effects. All but one of the required fundamental-plane variables for our full
cluster sample are then given in Tables \ref{tab:basic} through \ref{tab:dyn}
above. The last---cluster positions within their parent galaxies---is listed
in Table \ref{tab:kappa}, discussed below.

A third equivalent formulation of the fundamental plane is that of
\citet{bbf92} and \citet{bur97}, who manipulate
the basic observables of velocity dispersion, surface density, and half-mass
radius to define an orthonormal set of derived parameters,
\begin{equation}
\begin{array}{rcl}
\kappa_1 & \equiv & (\log\,\sigma_{{\rm p},0}^2 + \log\,R_h)/\sqrt{2} \\
\kappa_2 & \equiv & (\log\,\sigma_{{\rm p},0}^2 + 2\,\log\Sigma_h
                      - \log\,R_h)/\sqrt{6} \\
\kappa_3 & \equiv & (\log\,\sigma_{{\rm p},0}^2 - \log\,\Sigma_h
                      - \log\,R_h)/\sqrt{3} \\
\end{array}
\label{eq:kspace}
\end{equation}
and find tight distributions in $\kappa_3$ vs.~$\kappa_1$ for
early-type galaxies and (separately) for globular clusters.
\citet{bbf92} and \citet{bur97} actually define this
``$\kappa$ space'' using the luminosity intensity $I_h$ averaged over the
half-light radius $R_h$; but in order to remove the influence of age from
comparisons of cluster structures, we instead use the average mass
density $\Sigma_h = \Upsilon_V I_h = M_{\rm tot}/2\pi R_h^2$. Then
$\kappa_1 \leftrightarrow \log\,(\sigma_{{\rm p},0}^2 R_h)$
is related to the total mass of a system, and
$\kappa_3 \leftrightarrow \log\,(\sigma_{{\rm p},0}^2 R_h/M_{\rm tot})$
contains the exact details of this relationship---that is, information
on the internal density profile. In fact, the mass-based $\kappa_3$ of
equation (\ref{eq:kspace}) can be viewed as a replacement for King- or
Wilson-model concentrations $c$ or power-law indices $\gamma$, or any other
model-specific shape parameter. As such, any trends
involving $\kappa_3$ are directly of relevance to questions concerning
cluster (non)homology. The definition of $\kappa_2$ is chosen simply to make
the three $\kappa$ parameters mutually orthogonal; it results in the
correspondence $\kappa_2 \leftrightarrow \log\,(\Sigma_h^3)$
\citep[see][for further discussion]{bbf92}.

Table \ref{tab:kappa} gives the values of $\kappa_1$, $\kappa_2$, and
$\kappa_3$ for all the young and globular clusters that we fit with
structural models in \S\ref{subsec:fits}. In calculating these parameters,
we have used the $\sigma_{{\rm p},0}$ and $\Sigma_h$ values as predicted
in Table \ref{tab:dyn} by our adoption of population-synthesis mass-to-light
ratios. Equations (\ref{eq:kspace}) are evaluated for $\sigma_{{\rm p},0}$ in
units of km~s$^{-1}$, $\Sigma_h$ in $M_\odot\,{\rm pc}^{-2}$, and $R_h$ in
kpc (following Bender et al.~and Burstein et al., who originally had galaxies
in mind).

Table \ref{tab:kappa} also contains the observed distance, in kpc, of each
cluster from the center of its parent galaxy. These are projected
galactocentric radii for the LMC, SMC, and Fornax clusters, and
three-dimensional radii for the Milky Way globular clusters. For the
Galactic GCs, we have simply copied $R_{\rm gc}$ directly from
the catalogue of \citet{har96}. For the LMC, SMC, and Fornax clusters, we
have computed $R_{\rm gc}$ ourselves, using the right ascensions and
declinations of the clusters as given by \citet{mg03a,mg03b,mg03c}
and taking the galaxy centers to be
\begin{equation}
\begin{array}{lclcl}
    \alpha_{2000}= 5^{\rm h} 25^{\rm m}  6^{\rm s}
  & ~~ & \delta_{2000}=-69^\circ 47^\prime
  & ~  & {\rm (LMC)}    \\
    \alpha_{2000}= 0^{\rm h} 52^{\rm m} 45^{\rm s}
  & ~~ & \delta_{2000}=-72^\circ 49^\prime 43^{\prime\prime}
  & ~  & {\rm (SMC)}    \\
    \alpha_{2000}= 2^{\rm h} 39^{\rm m} 59\fs3
  & ~~ & \delta_{2000}=-34^\circ 26^\prime 57^{\prime\prime}
  & ~  & {\rm (Fornax)\ .} \\
\end{array}
\label{eq:centers}
\end{equation}
The H{\small I} and optical centers of the LMC are offset from each other,
and here we have more or less arbitrarily adopted the optical center of the
bar from \citet{vdm01}. The SMC center is taken from \citet{wes97}, and
the Fornax center from SIMBAD. In converting angular distances to
kpc, we assume a constant distance of 50.1 kpc to all clusters in the LMC;
60.0 kpc to the SMC; and 137 kpc to Fornax.

\section{Summary}
\label{sec:summ}

We have fit three distinct dynamical models to $V$-band surface-brightness
profiles for each of 68 massive star clusters (50 of which have young ages,
between several Myr and a few Gyr) in the LMC, SMC, and Fornax dwarf
spheroidal, and to 85 old globular clusters in the Milky Way. We have also
applied publicly available population-synthesis models to infer the expected
intrinsic $(B-V)_0$ colors (and thus reddenings) and $V$-band mass-to-light
ratios for all of these clusters plus another 63 Galactic globulars.
Combining these with the surface-brightness model fits, we have calculated a
wide range of structural and dynamical parameters characterizing the clusters.
Our main results are contained in Tables \ref{tab:poptable}, \ref{tab:basic},
\ref{tab:struc}, \ref{tab:dyn}, and \ref{tab:kappa} above. We have also taken
velocity-dispersion measurements from the literature for a subset of the full
cluster sample and calculated dynamical mass-to-light ratios for comparison
with the predicted population-synthesis values. The results of this are in
Table \ref{tab:popvdyn}, which shows quite good agreement in general.

The $V$-band surface-brightness data we have employed for LMC, SMC, and
Fornax star clusters derive primarily from HST-based starcounts made in the
inner regions by \citet[collectively MG03]{mg03a,mg03b,mg03c}. We have
supplemented these wherever possible with ground-based starcounts and $BV$
aperture photometry from the literature, both to extend the MG03 cluster
profiles to larger projected radii and to re-calibrate the $V$-band magnitude
scale of MG03. This re-calibration turns out to be rather significant,
amounting to several tenths of a magnitude (in the sense that the surface
brightnesses published in MG03 are generally too faint) in many cases. Full
details of our analyses on the 68 MG03 clusters are in \S\ref{subsec:mackey};
the re-calibrated surface-brightness profiles, with all ground-based data
properly incorporated, are given in Table \ref{tab:profs}. All
surface-brightness profiles for Milky Way globular clusters were taken
from the database constructed by \citet{tkd95}, with only minor modifications
described in \S\ref{subsec:trager}.

Our population-synthesis modeling, which includes a comparison of results from
two separate codes \citep{bru03,frv97} using a variety of assumed stellar
IMFs, is described in \S\ref{sec:popsyn}. Table \ref{tab:poptable} lists
intrinsic cluster colors and theoretical mass-to-light ratios obtained
with six different code+IMF combinations, although for our subsequent modeling
(requiring estimates of cluster $V$-band extinctions and conversions between
luminosity and mass) we adopted numbers from the code of \citet{bru03} using
the disk-star IMF of \citet{cha03}. As described in \S\ref{subsec:results},
we have thus obtained systematically lower predicted mass-to-light
ratios, at any age, than MG03 inferred for their LMC/SMC/Fornax clusters.
Our values nevertheless compare well with the dynamical mass-to-light ratios
calculated directly from velocity-dispersion measurements for 57 clusters
(mostly old globulars) in \S\ref{sec:dynml}. Additionally, our model-based
extinctions agree very well with direct measurements for the old globular
clusters in the Galaxy and the Fornax dwarf.

The three models that we fit to each cluster are described in some detail
in \S\ref{sec:modeling}, which also contains the fits themselves and the
bulk of our derived structural and dynamical parameters. The models are:
(1) the single-mass, isotropic, modified isothermal sphere of \citet{king66};
(2) an asymptotic power law with a constant-density core; and (3) an alternate
modification of the isothermal sphere (still single-mass and isotropic) based
on the stellar distribution function developed by \citet{wil75}.
For otherwise similar clusters (e.g., given a fixed total luminosity, central
surface brightness, and effective radius), \citet{wil75} spheres are spatially
more extended than \citet{king66} models, although both have finite tidal
radii; the untruncated power-law models we fit are formally
infinite in extent and, in some cases, barely convergent in their integrated
properties. Even so, the structural differences between these models are
most important at relatively large clustercentric distances, in the outer
halos beyond a few effective radii. As a result, we have verified
(\S\ref{subsec:fitcomp}) that for {\it most} clusters, {\it most} of the
physical properties we have calculated are reasonably well constrained no
matter which particular model is taken to fit the surface-brightness data.

We have looked with particular interest at the question of which of the
three fits---the spatially limited King model; a more extended but still
finite Wilson sphere; or an infinite power-law model---provides the best
description of these clusters in a $\chi^2$ sense (see
\S\ref{subsubsec:goodness}). {\it All} of the 68 LMC, SMC, and Fornax
clusters from MG03 are fit {\it at least} as well, and in several cases
significantly better, by the larger envelopes of \citet{wil75} models rather
than the more compact \citet{king66} models. This is true for the 18 old
globular clusters as well as of the 50 younger objects in this sample. It
also holds for the majority ($\approx 70/85$) of the old Galactic globulars
to which we have fit all three models; a particularly clear example is
provided by $\omega$ Centauri (\S\ref{subsubsec:wcen}). In all cases,
asymptotic
power laws are not vast improvements over the Wilson-model fits; in fact,
they are somewhat worse for many clusters.

In the LMC, SMC, and Fornax sample especially (where the
surface-brightness data are most homogeneous), there is no correlation between
cluster age and the relative quality of fit for King vs.~Wilson or power-law
models. Instead, we have shown that the primary factor in determining whether
an extended-halo model describes a cluster better than a \citet{king66} model
is simply the spatial extent of the available surface-brightness data being
fit. Specifically, it is only when a
cluster's observed density profile reaches to more than $\sim4$--5 effective
radii that it becomes possible to decide conclusively whether or not it has a
non-King envelope structure; and when such data do exist, a more distended
but finite Wilson model is most often the better option of the three we have
examined---whatever the cluster age.

Thus, we conclude that the extended halos which are known to surround many
young massive star clusters in the Magellanic Clouds and other galaxies
{\it do not require} description by untruncated power laws which would
necessarily have all the clusters overfill the Roche lobes defined by their
parent galaxies \citep[see][]{eff87,lar04,sch04}---although whether or not
this does happen in individual cases is a complicated question that we have
not undertaken to address here. More generally, despite the ad hoc nature of
the Wilson and power-law models that we have fit, it is a clear fact that
self-gravitating clusters {\it commonly} have envelope structures which do
not match the extrapolations of simple \citet{king66} models fitting the
cluster cores. This phenomenon is not confined exclusively to young clusters
and is not obviously only transient; it may point instead to generic,
internal cluster physics not captured by King's stellar distribution function.

This interesting issue aside, the work we have presented provides the
basic information
required to define and compare physical parameter correlations and the
fundamental plane(s) of young and old massive star clusters in the Milky
Way, LMC, SMC, and Fornax. It also offers a starting point for
careful examination of potential model-dependent artifacts in these
correlations, and ultimately it should allow for direct contact to be made
with the well-established fundamental plane of elliptical galaxies and
bulges. We plan to address these issues in future work.

\acknowledgments

We thank Dougal Mackey for providing his LMC, SMC, and Fornax cluster data to
us in electronic form, and for helpful discussions. Support for this work was
provided by NASA through a grant associated with HST project 10323, awarded by
STScI, which is operated by AURA, Inc., under NASA contract NAS 5-26555.


\vfill

\section*{\bf Online Material and Figure 12}

The next seven pages show extracts from Table \ref{tab:poptable} and
Tables \ref{tab:basic}--\ref{tab:kappa}, which are discussed in
\S\ref{subsec:results}, \S\ref{subsubsec:allcl}, 
\S\ref{sec:dynml}, and \S\ref{sec:kappa}.
Each of these tables can be downloaded in its entirety, in machine-readable
format, from the online edition of the {\it Astrophysical Journal Supplement
Series}, where this paper is published as

\

McLaughlin, D. E., \& van der Marel, R. P. 2005, ApJS, 161, 304.

\

Following the tables is Figure \ref{fig:sbfits} (\S\ref{subsubsec:allcl}),
which shows the full set of fits to each of the LMC, SMC, Fornax, and
Milky Way star clusters that we have modeled in detail. The full-resolution
versions of these plots are available on the ApJ website, and also from
\ {\bf http://www.astro.le.ac.uk/${\mathbf \sim}$dm131/clusters.html}\ . 

\clearpage

\setcounter{table}{7}
\begin{landscape}
\begin{deluxetable*}{lcccccccccc}
\tabletypesize{\scriptsize}
\tablewidth{0pt}
\tablecaption{Population-Synthesis Colors and Mass-to-Light Ratios
\label{tab:poptable}}
\tablecolumns{11}
\tablehead{
\colhead{}                                              &
\multicolumn{3}{c}{Published Data\tablenotemark{a}}       &
\colhead{}                                                & 
\multicolumn{6}{c}{SSP Model Code/IMF\tablenotemark{b}}   \\
\cline{2-4} \cline{6-11}                 \\
\colhead{Cluster}                        &
\colhead{age}  & \colhead{[Fe/H]}    & \colhead{($B$$-$$V$)}  &
\colhead{Property}                       &
\colhead{BC/ChD}   & \colhead{BC/Salp}   &
\colhead{PEG/ChD}  & \colhead{PEG/Salp}  &
\colhead{PEG/ChGC} & \colhead{PEG/KTG}   \\
\colhead{(1)} & \colhead{(2)}  & \colhead{(3)}  & \colhead{(4)} & 
\colhead{(5)} & \colhead{(6)}  & \colhead{(7)}  & \colhead{(8)} & 
\colhead{(9)} & \colhead{(10)} & \colhead{(11)}
}
\startdata
LMC-HODGE11  & $10.11$ & $-2.06$ &  0.64 &  ($B$$-$$V$)$_0$ 
           & $ 0.594\pm0.015$ & $ 0.616\pm0.014$ & $ 0.609\pm0.017$ 
           & $ 0.626\pm0.016$ & $ 0.606\pm0.017$ & $ 0.610\pm0.017$    \\
               &         &         &       & $ \Upsilon_V^{\rm pop}$ 
           & $ 1.893\pm0.160$ & $ 3.133\pm0.226$ & $ 2.013\pm0.166$ 
           & $ 3.279\pm0.237$ & $ 1.778\pm0.126$ & $ 1.923\pm0.146$    \\
SMC-KRON3   & $ 9.78$ & $-1.16$ &  0.68 &  ($B$$-$$V$)$_0$ 
           & $ 0.664\pm0.017$ & $ 0.675\pm0.017$ & $ 0.700\pm0.021$ 
           & $ 0.706\pm0.021$ & $ 0.699\pm0.020$ & $ 0.698\pm0.021$    \\
               &         &         &       & $ \Upsilon_V^{\rm pop}$ 
           & $ 1.129\pm0.113$ & $ 2.018\pm0.194$ & $ 1.281\pm0.140$ 
           & $ 2.269\pm0.240$ & $ 1.236\pm0.130$ & $ 1.312\pm0.127$    \\
FORNAX1 & $10.11$ & $-2.20$ &  0.65 &  ($B$$-$$V$)$_0$ 
           & $ 0.579\pm0.015$ & $ 0.601\pm0.015$ & $ 0.589\pm0.017$ 
           & $ 0.608\pm0.016$ & $ 0.586\pm0.017$ & $ 0.590\pm0.017$    \\
               &         &         &       & $ \Upsilon_V^{\rm pop}$ 
           & $ 1.915\pm0.163$ & $ 3.155\pm0.228$ & $ 2.004\pm0.167$ 
           & $ 3.249\pm0.239$ & $ 1.763\pm0.127$ & $ 1.911\pm0.147$    \\
MW-1636-283 & 10.11 & $-1.50$ & \nodata & ($B$$-$$V$)$_0$ 
           & $ 0.659\pm0.016$ & $ 0.678\pm0.016$ & $ 0.698\pm0.025$ 
           & $ 0.710\pm0.024$ & $ 0.694\pm0.025$ & $ 0.698\pm0.025$    \\
            &       &         &         & $ \Upsilon_V^{\rm pop}$ 
           & $ 1.888\pm0.155$ & $ 3.169\pm0.224$ & $ 2.134\pm0.187$ 
           & $ 3.542\pm0.286$ & $ 1.914\pm0.149$ & $ 2.058\pm0.170$    \\
\enddata

\tablecomments{A machine-readable version of Table \ref{tab:poptable} is
published in its entirety in the
electronic edition of the {\it Astrophysical Journal Supplement Series};
only a portion is shown here, for guidance regarding its form and content.}

\tablenotetext{a}{Column (2) is the logarithm of cluster age in years.
Ages, metallicities, and their uncertainties for
LMC, SMC, and Fornax clusters are taken from the compilation and
analysis by \citet{mg03a,mg03b,mg03c}. All ages given as greater than
13 Gyr by Mackey \& Gilmore have been re-set here to $13\pm2$ Gyr
($\log\,\tau=10.11\pm0.07$). Aperture $(B-V)$ colors for LMC, SMC, and
Fornax clusters are from the literature summarized in Table \ref{tab:apphot}.
Milky Way globular cluster ages are set to a uniform $13\pm2$ Gyr
($\log\,\tau=10.11\pm0.07$); their metallicities are taken from the catalogue
of \citet{har96} and assigned uncertainties of $\pm$0.2 dex. $(B-V)$ colors
are also from \citet{har96}, where available.}

\tablenotetext{b}{Key to the combinations of population-synthesis codes and
assumed stellar IMFs:
BC/ChD---code of \citet{bru03} using the disk IMF of \citet{cha03};
BC/Salp---code of \citet{bru03} using the \citet{sal55} IMF;
PEG/ChD---P\'EGASE v.~2.0 code of \citet{frv97} using the disk IMF
of \citet{cha03};
PEG/Salp---P\'EGASE code using the \citet{sal55} IMF;
PEG/ChGC---P\'EGASE code using the Globular Cluster IMF of \citet{cha03};
PEG/KTG---P\'EGASE code using the IMF of \citet{ktg93}.}

\end{deluxetable*}
\setcounter{table}{9}

\clearpage
\end{landscape}

\clearpage

\begin{landscape}
\begin{deluxetable*}{lrrrrccrrrrrr}
\tabletypesize{\scriptsize}
\tablewidth{0pt}
\tablecaption{Basic Model Fit Parameters  \label{tab:basic}}
\tablecolumns{13}
\tablehead{
\colhead{Cluster}                    & \colhead{$\Delta\mu_V$}    &
\colhead{$A_V$}                      & \colhead{$D$ [kpc]}        &
\colhead{$N_{\rm pts}$}              & \colhead{weight}           &
\colhead{model}                      & \colhead{$\chi^2$}         &
\colhead{$W_0$ / $\gamma$}           &
\colhead{$c$}                        &
\colhead{$\mu_{V,0}$}                & \colhead{$r_0$ [sec]}  &
\colhead{$r_0$ [pc]}                   \\
\colhead{(1)}  & \colhead{(2)}  & \colhead{(3)}  & \colhead{(4)}  &
\colhead{(5)}  & \colhead{(6)}  & \colhead{(7)}  & \colhead{(8)}  &
\colhead{(9)}  & \colhead{(10)} & \colhead{(11)} & \colhead{(12)} &
\colhead{(13)}
}
\startdata
LMC-HODGE11   & $-0.23^{+0.05}_{-0.05}$ & $0.14^{+0.08}_{-0.08}$ &   50.1
          &   50 &  l &  K
          &  17.62 & $6.0^{+0.6}_{-0.6}$ & $1.25^{+0.16}_{-0.14}$   
                   & $19.30^{+0.10}_{-0.09}$ & $14.42^{+1.17}_{-0.94}$   
                   & $3.50^{+0.28}_{-0.23}$   \\
         &         &         &        &         &         &  W
          &  17.06 & $6.1^{+0.8}_{-0.7}$ & $1.84^{+0.47}_{-0.28}$   
                   & $19.32^{+0.09}_{-0.10}$ & $15.21^{+1.30}_{-1.24}$   
                   & $3.69^{+0.32}_{-0.30}$   \\
         &         &         &        &         &         & PL
          &  16.73 & $3.65^{+0.35}_{-0.25}$ & $\infty$   
                   & $19.32^{+0.10}_{-0.10}$ & $16.52^{+2.40}_{-1.83}$   
                   & $4.01^{+0.58}_{-0.44}$   \\
SMC-KRON3   & $-0.15^{+0.05}_{-0.05}$ & $0.05^{+0.08}_{-0.08}$ &   60.0
          &   72 &  l &  K
          &  56.99 & $5.5^{+0.2}_{-0.2}$ & $1.14^{+0.05}_{-0.04}$   
                   & $20.04^{+0.10}_{-0.10}$ & $23.24^{+1.26}_{-1.31}$   
                   & $6.76^{+0.37}_{-0.38}$   \\
         &         &         &        &         &         &  W
          &  67.83 & $4.4^{+0.3}_{-0.2}$ & $1.27^{+0.08}_{-0.05}$   
                   & $20.09^{+0.09}_{-0.10}$ & $28.15^{+1.32}_{-1.94}$   
                   & $8.19^{+0.38}_{-0.56}$   \\
         &         &         &        &         &         & PL
          &  87.98 & $5.10^{+0.25}_{-0.15}$ & $\infty$   
                   & $20.13^{+0.09}_{-0.09}$ & $37.46^{+2.76}_{-1.75}$   
                   & $10.90^{+0.80}_{-0.51}$   \\
FORNAX1 & 0                     & $0.22^{+0.08}_{-0.08}$ &  137.0
          &   51 &  l &  K
          &  31.25 & $2.6^{+0.5}_{-0.6}$ & $0.61^{+0.08}_{-0.10}$   
                   & $23.06^{+0.10}_{-0.10}$ & $21.50^{+4.07}_{-2.61}$   
                   & $14.28^{+2.70}_{-1.73}$   \\
\enddata
\end{deluxetable*}

\clearpage
\end{landscape}

\begin{landscape}
\begin{deluxetable*}{lcrrrrrrrrrr}
\tabletypesize{\scriptsize}
\tablewidth{0pt}
\tablecaption{Derived Structural Parameters \label{tab:struc}}
\tablecolumns{12}
\tablehead{
\colhead{Cluster}                    & \colhead{mod}                        &
\colhead{$\log\,r_t$}                & \colhead{$\log\,R_c$}                &
\colhead{$\log\,R_h$}                & \colhead{$\log\,(R_h/R_c)$}          &
\colhead{$\log\,I_0$}                & \colhead{$\log\,j_0$}                &
\colhead{$\log\,L_{\rm tot}$}        & \colhead{$V_{\rm tot}$}              &
\colhead{$\log\,I_h$}                & \colhead{$\langle\mu_V\rangle_h$}    \\
\colhead{}       & \colhead{}        & \colhead{[pc]}   &  \colhead{[pc]}   &
\colhead{[pc]}   & \colhead{}        & \colhead{[$L_\odot\,{\rm pc}^{-2}$]} &
\colhead{[$L_\odot\,{\rm pc}^{-3}$]} & \colhead{[$L_\odot$]}                &
\colhead{}       & \colhead{[$L_\odot\,{\rm pc}^{-2}$]} & \colhead{}        \\
\colhead{(1)}  & \colhead{(2)}  & \colhead{(3)}  & \colhead{(4)}  &
\colhead{(5)}  & \colhead{(6)}  & \colhead{(7)}  & \colhead{(8)}  &
\colhead{(9)}  & \colhead{(10)} & \colhead{(11)} & \colhead{(12)}
}
\startdata
LMC-HODGE11   &  K & $1.80^{+0.13}_{-0.11}$ & $0.510^{+0.023}_{-0.021}$   
               & $0.843^{+0.063}_{-0.042}$ & $0.333^{+0.083}_{-0.065}$   
               & $2.85^{+0.04}_{-0.02}$ & $2.03^{+0.05}_{-0.06}$   
               & $4.89^{+0.06}_{-0.05}$ & $11.12^{+0.12}_{-0.15}$   
               & $2.41^{+0.06}_{-0.09}$ & $20.40^{+0.22}_{-0.15}$   \\
         &  W & $2.41^{+0.43}_{-0.25}$ & $0.518^{+0.020}_{-0.023}$   
               & $0.900^{+0.135}_{-0.063}$ & $0.381^{+0.158}_{-0.083}$   
               & $2.84^{+0.04}_{-0.02}$ & $2.01^{+0.05}_{-0.05}$   
               & $4.94^{+0.10}_{-0.06}$ & $11.00^{+0.15}_{-0.24}$   
               & $2.34^{+0.09}_{-0.19}$ & $20.57^{+0.46}_{-0.22}$   \\
         & PL & $\infty$ & $0.522^{+0.025}_{-0.023}$   
               & $1.039^{+0.259}_{-0.138}$ & $0.517^{+0.282}_{-0.163}$   
               & $2.84^{+0.04}_{-0.02}$ & $2.01^{+0.05}_{-0.06}$   
               & $5.03^{+0.13}_{-0.10}$ & $10.77^{+0.24}_{-0.32}$   
               & $2.16^{+0.19}_{-0.40}$ & $21.03^{+0.99}_{-0.48}$   \\
SMC-KRON3   &  K & $1.97^{+0.02}_{-0.02}$ & $0.787^{+0.019}_{-0.021}$   
               & $1.065^{+0.000}_{-0.001}$ & $0.278^{+0.021}_{-0.019}$   
               & $2.55^{+0.04}_{-0.02}$ & $1.46^{+0.05}_{-0.05}$   
               & $5.10^{+0.04}_{-0.04}$ & $11.00^{+0.10}_{-0.09}$   
               & $2.17^{+0.04}_{-0.04}$ & $21.00^{+0.10}_{-0.09}$   \\
         &  W & $2.18^{+0.05}_{-0.03}$ & $0.817^{+0.012}_{-0.020}$   
               & $1.039^{+0.000}_{-0.001}$ & $0.222^{+0.020}_{-0.012}$   
               & $2.53^{+0.04}_{-0.02}$ & $1.40^{+0.05}_{-0.04}$   
               & $5.08^{+0.04}_{-0.04}$ & $11.04^{+0.09}_{-0.09}$   
               & $2.20^{+0.04}_{-0.04}$ & $20.91^{+0.09}_{-0.09}$   \\
\enddata

\tablecomments{A machine-readable version of Table \ref{tab:struc} is
published in its entirety in the
electronic edition of the {\it Astrophysical Journal Supplement Series};
only a portion is shown here, for guidance regarding its form and content.
Key to columns:
\hfill\break
{\bf Column (1)}---Cluster name.
\hfill\break
{\bf Column (2)}---Model fit.
\hfill\break
{\bf Column (3)}---Logarithm of cluster tidal radius in pc: $\log\,r_t=
c+\log\,r_0$. By construction, $r_t=\infty$ for power-law models.
\hfill\break
{\bf Column (4)}---Logarithm of projected core (half-power) radius at which
$I(R_c)=I(0)/2$, in pc.
\hfill\break
{\bf Column (5)}---Logarithm of projected half-light (effective radius) in pc.
Half of the total cluster luminosity is emitted from within $R_h$.
\hfill\break
{\bf Column (6)}---Logarithm of ratio $R_h/R_c$, a measure of cluster
concentration.
\hfill\break
{\bf Column (7)}---Logarithm of central luminosity surface density in $V$-band
$L_\odot\,{\rm pc}^{-2}$.
\hfill\break
{\bf Column (8)}---Logarithm of central luminosity volume density in $V$-band
$L_\odot\,{\rm pc}^{-3}$.
\hfill\break
{\bf Column (9)}---Total $V$-band luminosity of fitted model.
\hfill\break
{\bf Column (10)}---Apparent $V$-band magnitude of model cluster, corrected
for extinction: $V_{\rm tot}=4.85-2.5\,\log\,(L_{\rm tot}/L_\odot)
+5\,log(D/10\,{\rm kpc})$.
\hfill\break
{\bf Column (11)}---Logarithm of $V$-band luminosity surface density averaged
inside the half-light radius: $I_h\equiv L_{\rm tot}/2\pi R_h^2$.
\hfill\break
{\bf Column (12)}---$V$-band surface brightness corresponding to average
projected luminosity density in Column (11): $\langle\mu_V\rangle_h\equiv
26.422-2.5\,\log\,(I_h/L_\odot\,{\rm pc}^{-2})$.
}

\end{deluxetable*}

\clearpage
\end{landscape}

\begin{landscape}
\begin{deluxetable*}{lrcrrrrrrrrr}
\tabletypesize{\scriptsize}
\tablewidth{0pt}
\tablecaption{Derived Dynamical Parameters \label{tab:dyn}}              
\tablecolumns{12}
\tablehead{
\colhead{Cluster}                    & \colhead{$\Upsilon_V^{\rm pop}$} &
\colhead{mod}                        & \colhead{$\log\,M_{\rm tot}$}    &
\colhead{$\log\,E_b$}                & \colhead{$\log\,\Sigma_0$}       &
\colhead{$\log\,\rho_0$}             & \colhead{$\log\,\Sigma_h$}       &
\colhead{$\sigma_{{\rm p},0}$}       & \colhead{$v_{{\rm esc},0}$}      &
\colhead{$\log\,t_{\rm rh}$}         &
\colhead{$\log\,f_0$}                \\
\colhead{} & \colhead{[$M_\odot\,L_\odot^{-1}$]} & \colhead{} &
\colhead{[$M_\odot$]}                & \colhead{[erg]} &
\colhead{[$M_\odot\,{\rm pc}^{-2}$]} & \colhead{[$M_\odot\,{\rm pc}^{-3}$]} &
\colhead{[$M_\odot\,{\rm pc}^{-2}$]} & \colhead{[${\rm km\,s}^{-1}$]} &
\colhead{[${\rm km\,s}^{-1}$]} & \colhead{[yr]} & \colhead{}     \\
\colhead{(1)}  & \colhead{(2)}  & \colhead{(3)}  & \colhead{(4)}  &
\colhead{(5)}  & \colhead{(6)}  & \colhead{(7)}  & \colhead{(8)}  &
\colhead{(9)}  & \colhead{(10)} & \colhead{(11)} & \colhead{(12)}
}
\startdata
 LMC-HODGE11 & $1.893^{+0.160}_{-0.160}$ &  K   
          & $5.17^{+0.07}_{-0.06}$ & $49.91^{+0.11}_{-0.11}$   
          & $3.13^{+0.05}_{-0.05}$ & $2.31^{+0.06}_{-0.07}$   
          & $2.69^{+0.07}_{-0.10}$ & $3.63^{+0.22}_{-0.21}$   
          & $14.14^{+0.87}_{-0.83}$ & $9.40^{+0.12}_{-0.08}$   
          & $-0.605^{+0.062}_{-0.064}$   \\
          &           &  W   
          & $5.22^{+0.10}_{-0.07}$ & $49.92^{+0.11}_{-0.11}$   
          & $3.12^{+0.05}_{-0.05}$ & $2.29^{+0.06}_{-0.06}$   
          & $2.62^{+0.10}_{-0.19}$ & $3.63^{+0.22}_{-0.21}$   
          & $14.17^{+0.89}_{-0.82}$ & $9.50^{+0.25}_{-0.12}$   
          & $-0.629^{+0.067}_{-0.060}$   \\
          &           & PL   
          & $5.31^{+0.13}_{-0.10}$ & $49.95^{+0.11}_{-0.11}$   
          & $3.12^{+0.05}_{-0.05}$ & $2.29^{+0.07}_{-0.07}$   
          & $2.43^{+0.19}_{-0.40}$ & $3.64^{+0.22}_{-0.21}$   
          & $14.27^{+0.89}_{-0.85}$ & $9.75^{+0.45}_{-0.25}$   
          & $-0.642^{+0.070}_{-0.074}$   \\
   SMC-KRON3 & $1.129^{+0.113}_{-0.113}$ &  K   
          & $5.15^{+0.06}_{-0.06}$ & $49.65^{+0.12}_{-0.12}$   
          & $2.61^{+0.06}_{-0.06}$ & $1.51^{+0.07}_{-0.07}$   
          & $2.22^{+0.06}_{-0.06}$ & $2.74^{+0.19}_{-0.18}$   
          & $10.51^{+0.71}_{-0.67}$ & $9.72^{+0.03}_{-0.03}$   
          & $-1.045^{+0.059}_{-0.052}$   \\
          &           &  W   
          & $5.13^{+0.06}_{-0.06}$ & $49.63^{+0.12}_{-0.12}$   
          & $2.59^{+0.06}_{-0.06}$ & $1.46^{+0.07}_{-0.06}$   
          & $2.26^{+0.06}_{-0.06}$ & $2.78^{+0.19}_{-0.18}$   
          & $10.42^{+0.71}_{-0.66}$ & $9.67^{+0.03}_{-0.03}$   
          & $-1.137^{+0.056}_{-0.041}$   \\
\enddata

\tablecomments{A machine-readable version of Table \ref{tab:dyn} is
published in its entirety in the
electronic edition of the {\it Astrophysical Journal Supplement Series};
only a portion is shown here, for guidance regarding its form and content.
Key to columns:
\hfill\break
{\bf Column (1)}---Cluster name.
\hfill\break
{\bf Column (2)}---Population-synthesis $V$-band mass-to-light ratio, from
Column (6) of Table \ref{tab:poptable}.
\hfill\break
{\bf Column (3)}---Model fit.
\hfill\break
{\bf Column (4)}---Total cluster mass from extrapolation of fitted model
density profile.
\hfill\break
{\bf Column (5)}---Total cluster binding energy.
\hfill\break
{\bf Column (6)}---Logarithm of central mass surface density.
\hfill\break
{\bf Column (7)}---Logarithm of central mass volume density.
\hfill\break
{\bf Column (8)}---Logarithm of mass surface density averaged over the
projected half-light/half-mass radius:
$\Sigma_h\equiv M_{\rm tot}/2\pi R_h^2$,
\hfill\break
{\bf Column (9)}---Predicted line-of-sight velocity dispersion at cluster
center ($\sigma_{{\rm p},0}^2 \propto \Upsilon_V^{\rm pop}$).
\hfill\break
{\bf Column (10)}---Predicted velocity of escape (to infinity) at cluster
center ($v_{{\rm esc},0}^2 \propto \Upsilon_V^{\rm pop}$).
\hfill\break
{\bf Column (11)}---Median two-body relaxation timescale, evaluated at
the half-light radius; see text.
\hfill\break
{\bf Column (12)}---Central phase-space density
$f_0\equiv\rho_0/(2\pi \sigma_c^2)^{3/2}$, where $\sigma_c$ is the
unprojected, one-dimensional velocity dispersion at $r=0$ in the model.
Units are $M_\odot\,{\rm pc}^{-3}\,{\rm km}^{-3}\,{\rm s}^3$, in which
case the core relaxation time is given roughly by
$\log\,(t_{\rm rc}/{\rm yr}) \simeq 8.28-\log\,f_0$.
}

\end{deluxetable*}

\clearpage
\end{landscape}

\LongTables
\begin{landscape}
\begin{deluxetable}{lrrrcrrrrrrr}
\tabletypesize{\scriptsize}
\tablewidth{0pt}
\tablecaption{Population vs.~Dynamical Mass-to-Light Ratios
              \label{tab:popvdyn}}
\tablecolumns{12}
\tablehead{
\multicolumn{6}{c}{}              &
\multicolumn{2}{c}{King fits}     &
\multicolumn{2}{c}{Wilson fits}   &
\multicolumn{2}{c}{Power-Law fits}        \\
\cline{7-8}   \cline{9-10}   \cline{11-12}   \\
\colhead{Cluster}                 & \colhead{$\Upsilon_V^{\rm pop}$}        &
\colhead{$\sigma_{\rm p, obs}$}   & \colhead{$R_{\rm ap}$}            &
\colhead{Ref.}                    &
\colhead{$\sigma_{\rm p}(R=0)$}   &
\colhead{$\Upsilon_V^{\rm dyn}$}  & \colhead{$\Delta\,(\log\,\Upsilon_V)$}  &
\colhead{$\Upsilon_V^{\rm dyn}$}  & \colhead{$\Delta\,(\log\,\Upsilon_V)$}  &
\colhead{$\Upsilon_V^{\rm dyn}$}  & \colhead{$\Delta\,(\log\,\Upsilon_V)$}  \\
\colhead{}                          & \colhead{[$M_\odot\,L_\odot^{-1}$]} &
\colhead{[${\rm km\,s}^{-1}$]}      & \colhead{[sec]}                     &
\colhead{}                          & \colhead{[${\rm km\,s}^{-1}$]}      &
\colhead{[$M_\odot\,L_\odot^{-1}$]} & \colhead{}                          &
\colhead{[$M_\odot\,L_\odot^{-1}$]} & \colhead{}                          &
\colhead{[$M_\odot\,L_\odot^{-1}$]} & \colhead{}                          \\
\colhead{(1)}  & \colhead{(2)}  & \colhead{(3)}  & \colhead{(4)}  &
\colhead{(5)}  & \colhead{(6)}  & \colhead{(7)}  & \colhead{(8)}  &
\colhead{(9)}  & \colhead{(10)} & \colhead{(11)} & \colhead{(12)}
}
\startdata
LMC-NGC1466 & $1.88\pm0.03$ & $3.80^{+2.80}_{-3.80}$ &   2.8 &  1
          & $3.81^{+2.81}_{-3.81}$   
          & $1.15^{+2.48}_{-1.15}$ & $-0.213^{+0.506}_{-99}$   
          & $1.14^{+2.45}_{-1.14}$ & $-0.217^{+0.509}_{-99}$   
          & $1.13^{+2.44}_{-1.13}$ & $-0.221^{+0.512}_{-99}$ \\
LMC-NGC1754 & $1.88\pm0.15$ & $7.80^{+3.80}_{-2.20}$ &   2.8 &  1
          & $7.91^{+3.88}_{-2.24}$   
          & $3.79^{+5.21}_{-2.04}$ & $0.304^{+0.413}_{-0.369}$   
          & $3.41^{+4.74}_{-1.80}$ & $0.258^{+0.440}_{-0.360}$   
          & $3.48^{+4.80}_{-1.82}$ & $0.267^{+0.435}_{-0.355}$ \\
LMC-NGC1786 & $1.87\pm0.16$ & $8.30^{+1.50}_{-1.50}$ &   2.8 &  1
          & $8.36^{+1.52}_{-1.51}$   
          & $1.78^{+0.89}_{-0.68}$ & $-0.023^{+0.214}_{-0.244}$   
          & $1.80^{+0.86}_{-0.66}$ & $-0.017^{+0.212}_{-0.234}$   
          & $1.85^{+0.85}_{-0.66}$ & $-0.005^{+0.208}_{-0.227}$ \\
       ~~ &           ~~ & $9.90^{+3.00}_{-3.00}$ &   1.1 &   2
          & $9.91^{+3.01}_{-3.00}$   
          & $2.50^{+2.07}_{-1.38}$ & $0.125^{+0.300}_{-0.385}$   
          & $2.53^{+2.03}_{-1.37}$ & $0.130^{+0.298}_{-0.375}$   
          & $2.60^{+2.02}_{-1.39}$ & $0.143^{+0.292}_{-0.367}$ \\
       ~~ &           ~~ &           ~~ &           ~~ &    ave.
          & $9.17^{+0.74}_{-0.81}$   
          & $2.14^{+0.36}_{-0.36}$ & $0.058^{+0.106}_{-0.116}$   
          & $2.16^{+0.36}_{-0.36}$ & $0.063^{+0.106}_{-0.115}$   
          & $2.22^{+0.37}_{-0.37}$ & $0.075^{+0.106}_{-0.115}$   \\
LMC-NGC1835 & $1.87\pm0.16$ & $10.40^{+1.40}_{-1.40}$ &   2.8 &  1
          & $10.52^{+1.43}_{-1.42}$   
          & $1.42^{+0.50}_{-0.41}$ & $-0.119^{+0.170}_{-0.181}$   
          & $1.38^{+0.50}_{-0.39}$ & $-0.132^{+0.173}_{-0.179}$   
          & $1.42^{+0.43}_{-0.41}$ & $-0.120^{+0.170}_{-0.182}$ \\
\enddata

\tablecomments{The electronic edition of {\it Astrophysical Journal
Supplement Series} also includes a machine-readable version of Table
\ref{tab:popvdyn}.
Key to columns:
\hfill\break
{\bf Column (1)}---Cluster name.
\hfill\break
{\bf Column (2)}---Population-synthesis $V$-band mass-to-light ratio, from
Column (6) of Table \ref{tab:poptable}.
\hfill\break
{\bf Column (3)}---Observed stellar velocity dispersion from literature.
\hfill\break
{\bf Column (4)}---Radius of effective circular aperture corresponding to
velocity-dispersion measurement in Column (3).
\hfill\break
{\bf Column (5)}---Source of $\sigma_{\rm p, obs}$ and $R_{\rm ap}$ estimates
(see below).
\hfill\break
{\bf Column (6)}---Aperture-corrected line-of-sight velocity dispersion at
cluster center (mean of separate results using King, Wilson, and power-law
structural models).
\hfill\break
{\bf Column (7)}---Dynamical mass-to-light ratio calculated within King-model
fit to cluster surface-brightness profile.
\hfill\break
{\bf Column (8)}---King-model $\Delta(\log\,\Upsilon_V)\equiv
\log\,(\Upsilon_V^{\rm dyn}/\Upsilon_V^{\rm pop})$.
\hfill\break
{\bf Column (9)}---Dynamical mass-to-light ratio calculated within
Wilson-model fit to cluster surface-brightness profile.
\hfill\break
{\bf Column (10)}---Wilson-model $\Delta(\log\,\Upsilon_V)\equiv
\log\,(\Upsilon_V^{\rm dyn}/\Upsilon_V^{\rm pop})$.
\hfill\break
{\bf Column (11)}---Dynamical mass-to-light ratio calculated within
power-law model fit to cluster surface-brightness profile.
\hfill\break
{\bf Column (12)}---Power-law model $\Delta(\log\,\Upsilon_V)\equiv
\log\,(\Upsilon_V^{\rm dyn}/\Upsilon_V^{\rm pop})$.
}

\tablerefs{
(1)--\citet{dmm97};
(2)--\citet{mwf91};
(3)--\citet{fis93};
(4)--\citet{fis92};
(5)--\citet{lup89};
(6)--\citet{fea80};
(7)--\citet{dmm92};
(8)--\citet{pry93}.
}

\end{deluxetable}

\clearpage
\end{landscape}

\begin{deluxetable*}{lccrrr}
\tabletypesize{\scriptsize}
\tablewidth{0pt}
\tablecaption{Galactocentric Radii and $\kappa$-Space Parameters
\label{tab:kappa}}
\tablecolumns{10}
\tablehead{
\colhead{Cluster}             &
\colhead{$R_{\rm gc}$ [kpc]}  &
\colhead{model}               &
\colhead{$\kappa_1$}    & \colhead{$\kappa_2$}   &\colhead{$\kappa_3$}  \\
\colhead{(1)}  & \colhead{(2)}  & \colhead{(3)}  &
\colhead{(4)}  & \colhead{(5)}  & \colhead{(6)}
}
\startdata
 LMC-HODGE11 &   3.71 &  K & $-0.734^{+0.052}_{-0.045}$ 
             & $3.305^{+0.074}_{-0.107}$ & $0.500^{+0.022}_{-0.021}$   \\
             &        &  W & $-0.694^{+0.096}_{-0.053}$ 
             & $3.225^{+0.106}_{-0.212}$ & $0.507^{+0.030}_{-0.022}$   \\
             &        & PL & $-0.593^{+0.181}_{-0.100}$ 
             & $3.020^{+0.217}_{-0.433}$ & $0.535^{+0.075}_{-0.032}$   \\
   SMC-KRON3 &   2.16 &  K & $-0.749^{+0.040}_{-0.041}$ 
             & $2.919^{+0.049}_{-0.050}$ & $0.370^{+0.025}_{-0.025}$   \\
             &        &  W & $-0.759^{+0.041}_{-0.041}$ 
             & $2.962^{+0.049}_{-0.049}$ & $0.373^{+0.025}_{-0.025}$   \\
             &        & PL & $-0.764^{+0.041}_{-0.040}$ 
             & $2.989^{+0.050}_{-0.049}$ & $0.376^{+0.025}_{-0.025}$   \\
     FORNAX1 &   1.59 &  K & $-1.264^{+0.034}_{-0.034}$ 
             & $1.749^{+0.042}_{-0.042}$ & $0.517^{+0.022}_{-0.022}$   \\
             &        &  W & $-1.259^{+0.034}_{-0.034}$ 
             & $1.757^{+0.041}_{-0.041}$ & $0.518^{+0.021}_{-0.021}$   \\
\enddata

\tablecomments{A machine-readable version of Table \ref{tab:kappa} is
published in its entirety in the
electronic edition of the {\it Astrophysical Journal Supplement Series};
only a portion is shown here, for guidance regarding its form and content.
Key to columns:
\hfill\break
{\bf Column (1)}---Cluster name.
\hfill\break
{\bf Column (2)}---Distance from center of parent galaxy, in kpc. For LMC,
SMC, and Fornax-dwarf clusters, $R_{\rm gc}$ is a projected distance (with
galaxy centers given in eq.~[\ref{eq:centers}] of the text); for Milky Way
globulars, $R_{\rm gc}$ is the three-dimensional Galactocentric distance,
taken from \citet{har96}.
\hfill\break
{\bf Column (3)}---Model fit.
\hfill\break
{\bf Column (4)}---
$\kappa_1\equiv (\log\,\sigma_{{\rm p},0}^2 + \log\,R_h)/\sqrt{2}$,
for $\sigma_{{\rm p},0}$ from Table \ref{tab:dyn} in km~s$^{-1}$ and
$R_h$ from Table \ref{tab:struc} in kpc.
\hfill\break
{\bf Column (5)}---
$\kappa_2\equiv (\log\,\sigma_{{\rm p},0}^2 + 2\,\log\,\Sigma_h -
\log\,R_h)/\sqrt{6}$,
for $\sigma_{{\rm p},0}$ from Table \ref{tab:dyn} in km~s$^{-1}$,
$\Sigma_h$ from Table \ref{tab:dyn} in $M_\odot\,{\rm pc}^{-2}$, and
$R_h$ from Table \ref{tab:struc} in kpc.
\hfill\break
{\bf Column (6)}---
$\kappa_3\equiv (\log\,\sigma_{{\rm p},0}^2 - \log\,\Sigma_h -
\log\,R_h)/\sqrt{3}$,
for $\sigma_{{\rm p},0}$ from Table \ref{tab:dyn} in km~s$^{-1}$,
$\Sigma_h$ from Table \ref{tab:dyn} in $M_\odot\,{\rm pc}^{-2}$, and
$R_h$ from Table \ref{tab:struc} in kpc.
}
\end{deluxetable*}

\clearpage

\setcounter{figure}{11}

\begin{figure*}
\epsscale{1.00}
\plotone{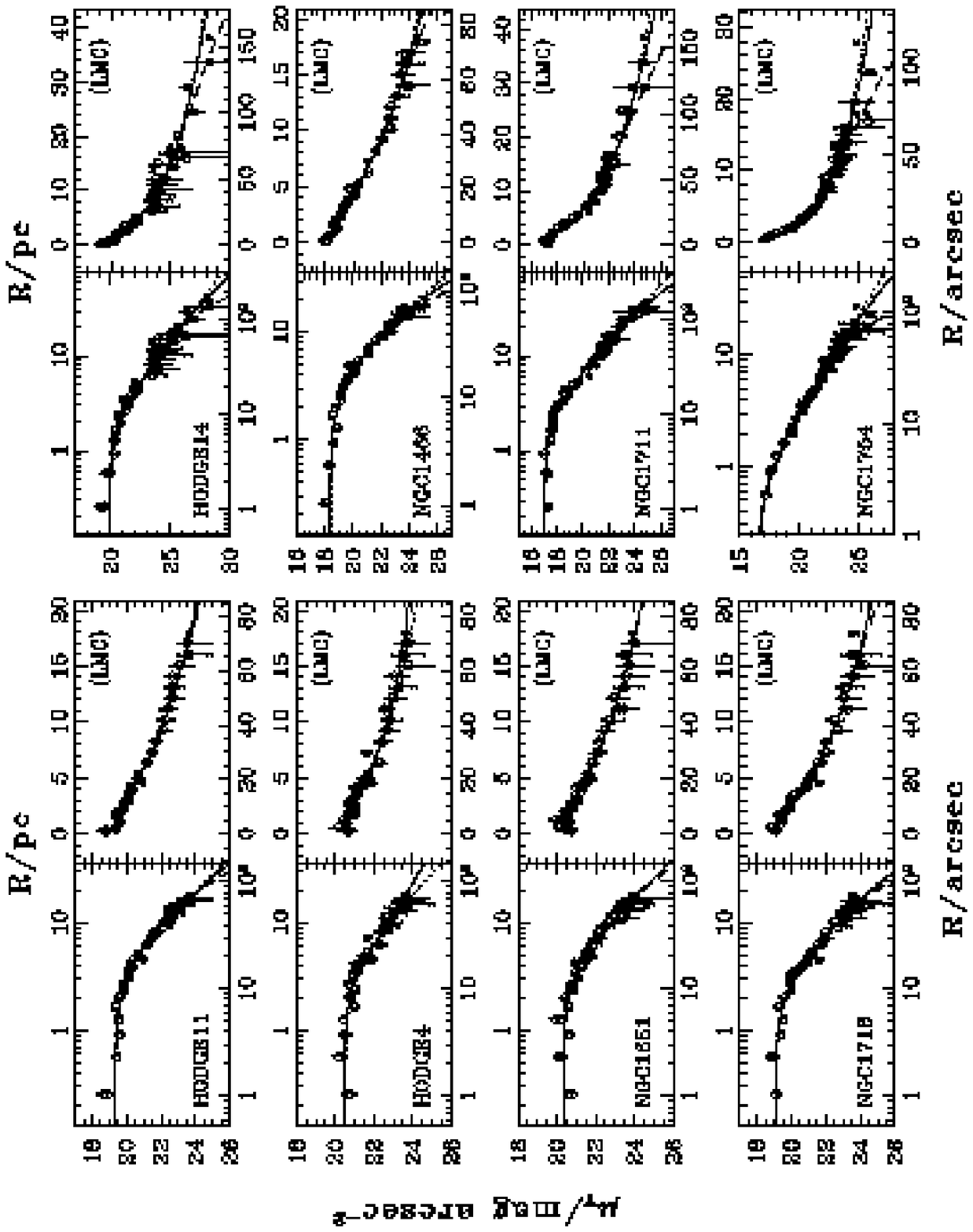}
\caption{\label{fig:sbfits}
Fits of \citet{king66} models, isotropic \citet{wil75} models, and power-law
models (eq.~[\ref{eq:power}]) to the surface-brightness profiles of 53 LMC
clusters and 10 SMC clusters spanning a wide range of ages;
5 globular clusters in the Fornax dwarf spheroidal; and 84 Galactic globular
clusters besides $\omega$ Centauri. In all panels, {\it solid} lines
correspond to the Wilson-model fits; {\it dashed} lines, to King-model fits;
and {\it dotted} lines, to power-law fits. All SB data are plotted with
corrections for zeropoint changes and $V$-band extinction {\it included}
(Columns 2 and 3 of Table \ref{tab:basic}). Radial scales are in arcsec along
the lower horizontal axes of all panels, and in pc along the upper
horizontal axes. Numerical details of every fit are given in Table
\ref{tab:basic}, and derived cluster parameters are in Tables \ref{tab:struc}
and \ref{tab:dyn}.
}
\end{figure*}

\clearpage

\begin{figure*}
\figurenum{\ref{fig:sbfits} [continued]}
\epsscale{1.00}
\plotone{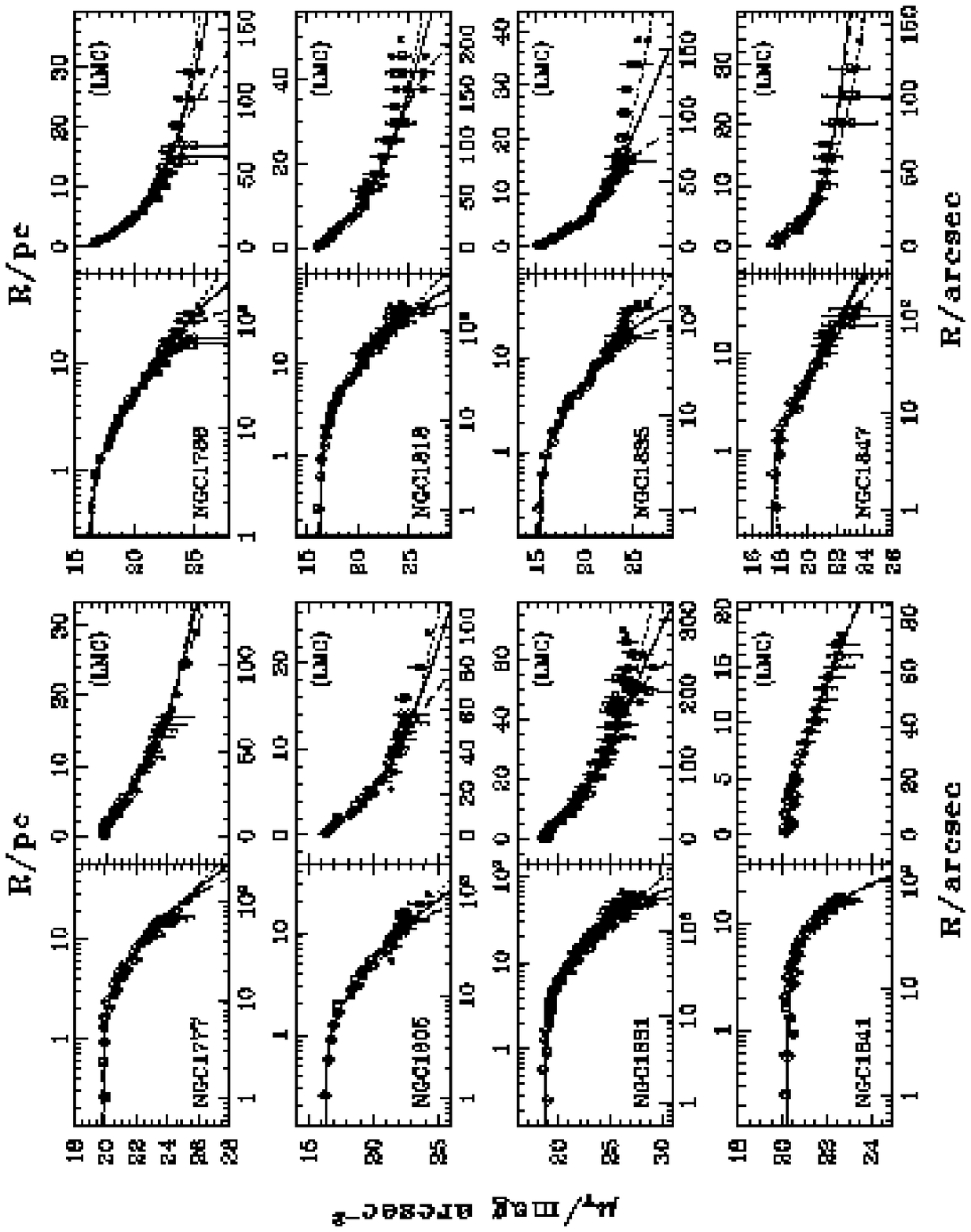}
\caption{}
\end{figure*}

\clearpage

\begin{figure*}
\figurenum{\ref{fig:sbfits} [continued]}
\epsscale{1.00}
\plotone{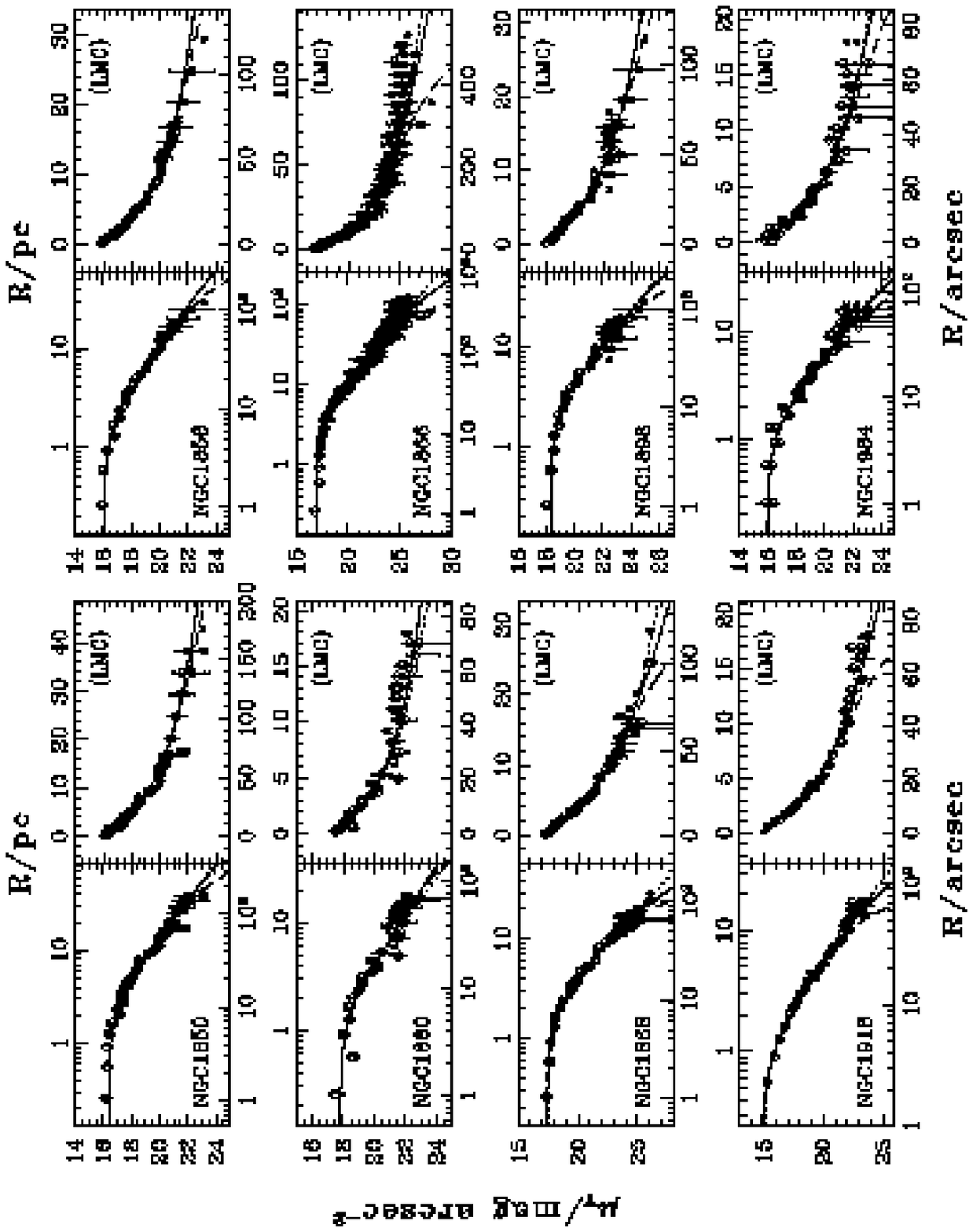}
\caption{}
\end{figure*}

\clearpage

\begin{figure*}
\figurenum{\ref{fig:sbfits} [continued]}
\epsscale{1.00}
\plotone{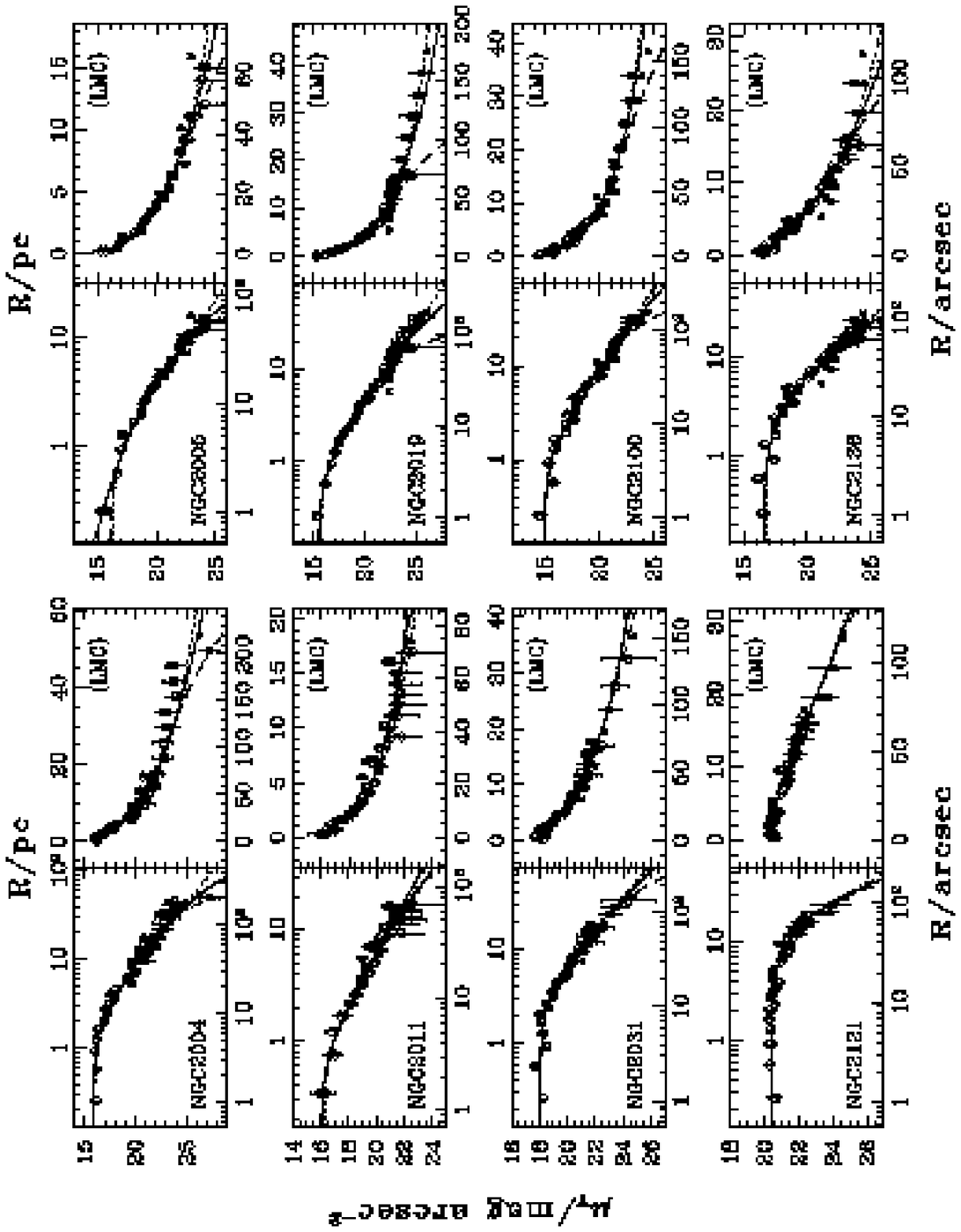}
\caption{}
\end{figure*}

\clearpage

\begin{figure*}
\figurenum{\ref{fig:sbfits} [continued]}
\epsscale{1.00}
\plotone{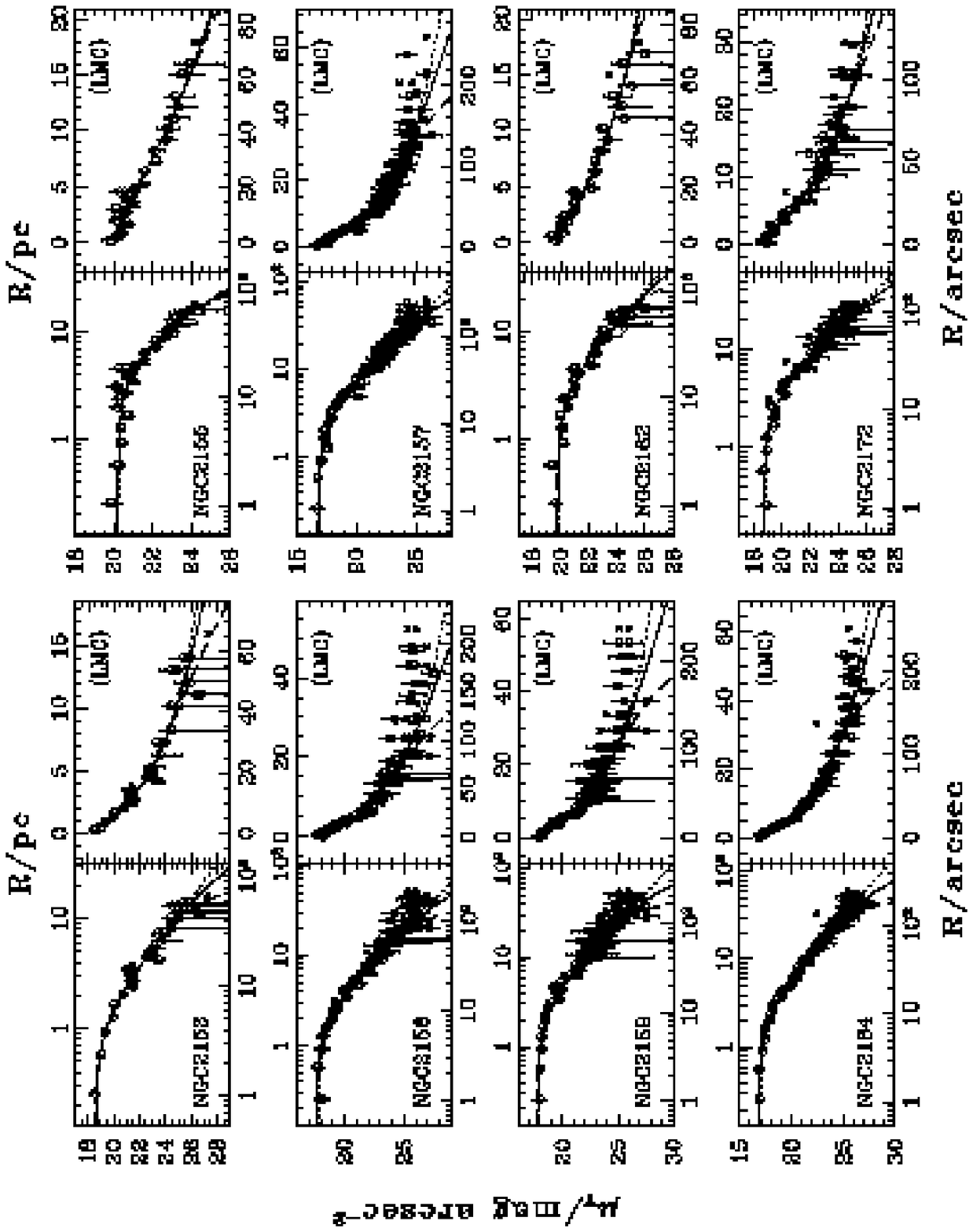}
\caption{}
\end{figure*}

\clearpage

\begin{figure*}
\figurenum{\ref{fig:sbfits} [continued]}
\epsscale{1.00}
\plotone{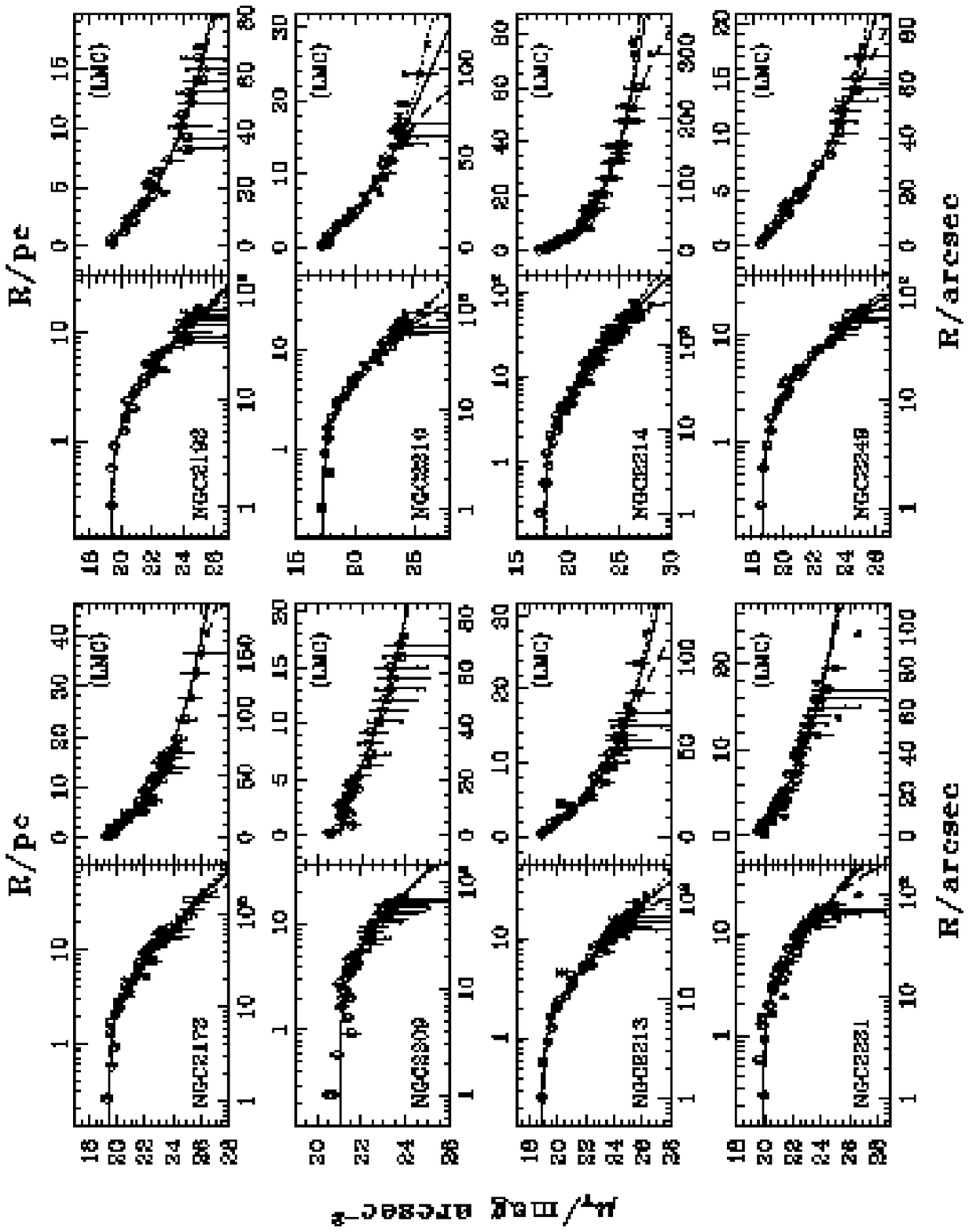}
\caption{}
\end{figure*}

\clearpage

\begin{figure*}
\figurenum{\ref{fig:sbfits} [continued]}
\epsscale{1.00}
\plotone{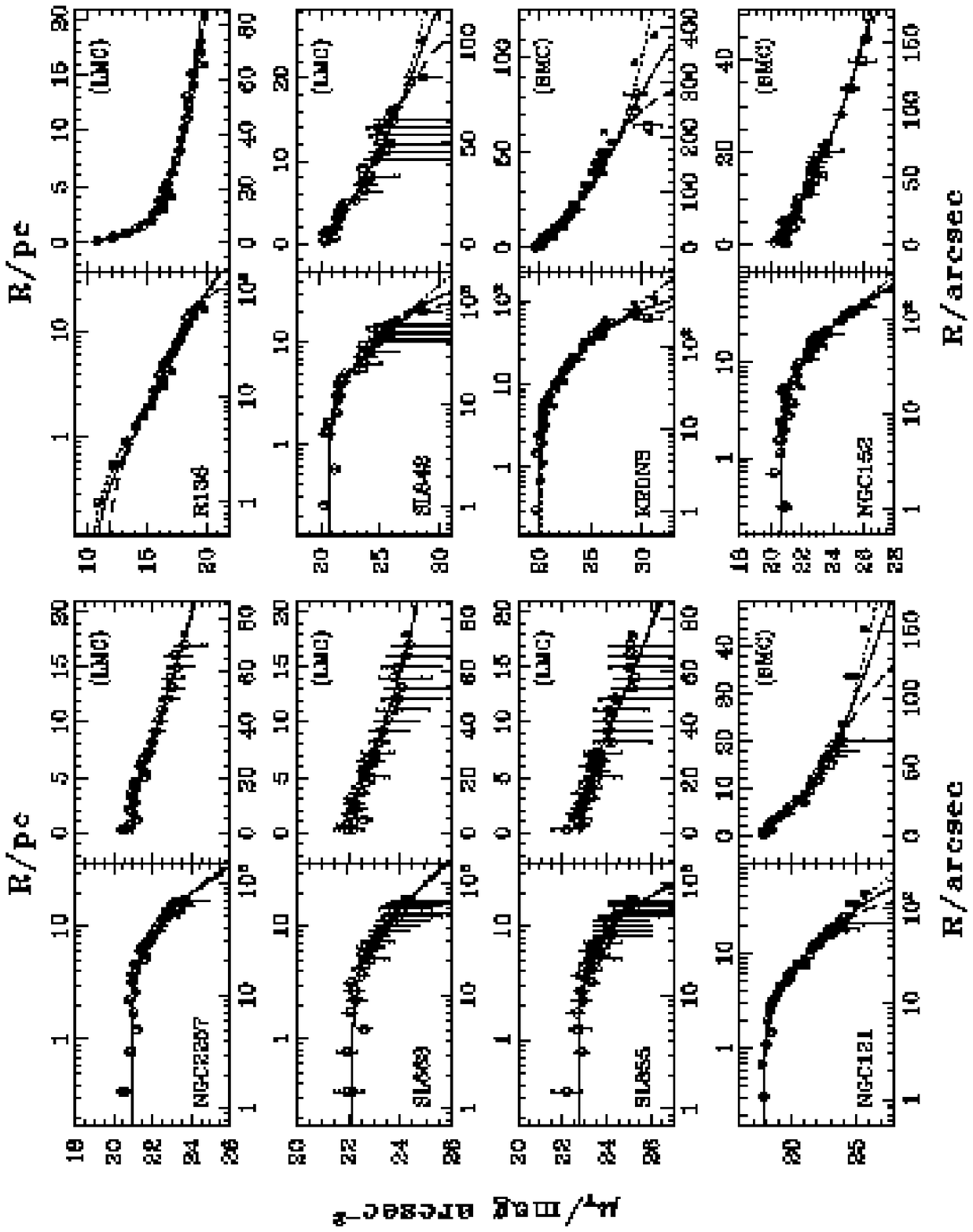}
\caption{}
\end{figure*}

\clearpage

\begin{figure*}
\figurenum{\ref{fig:sbfits} [continued]}
\epsscale{1.00}
\plotone{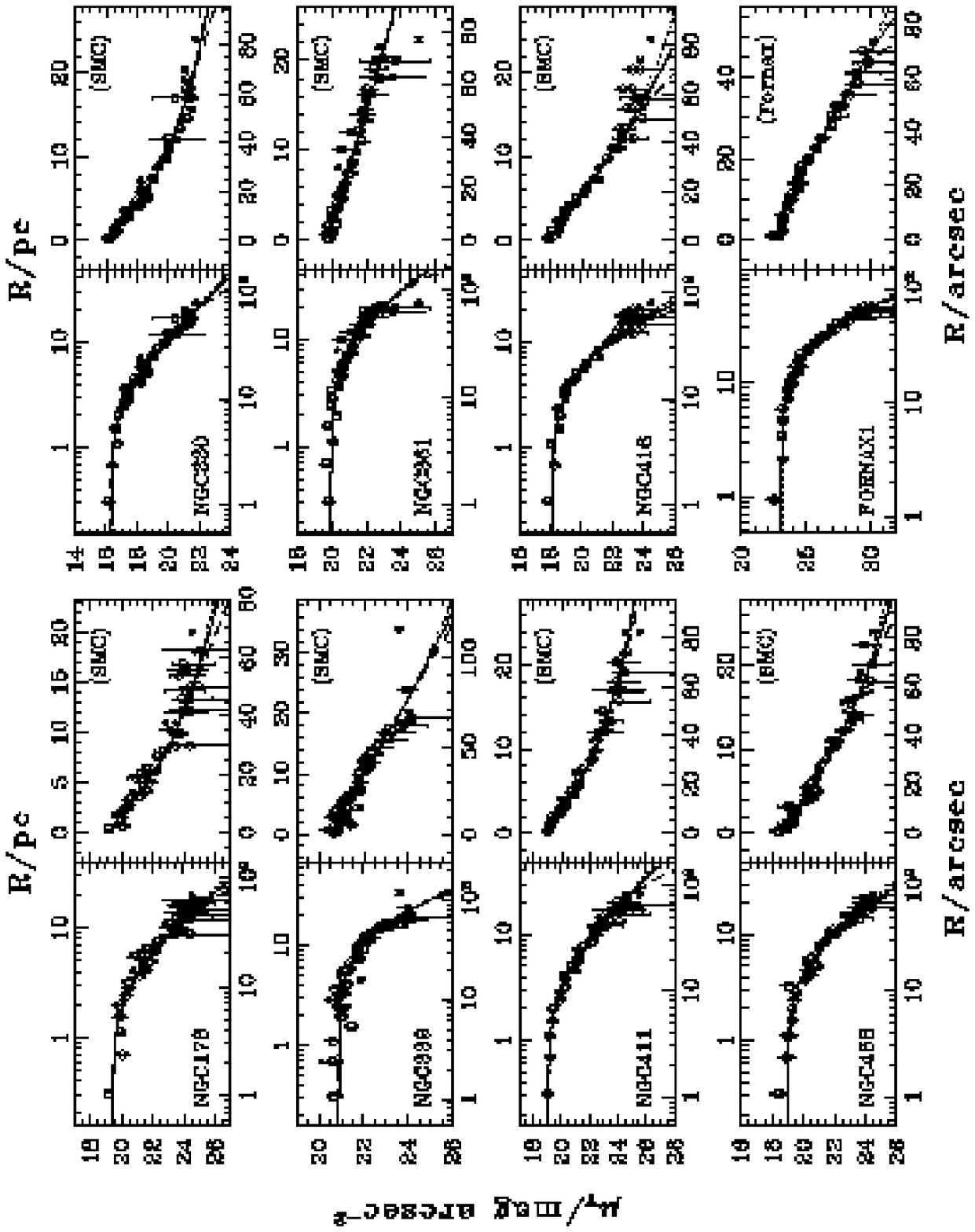}
\caption{}
\end{figure*}

\clearpage

\begin{figure*}
\figurenum{\ref{fig:sbfits} [continued]}
\epsscale{1.00}
\plotone{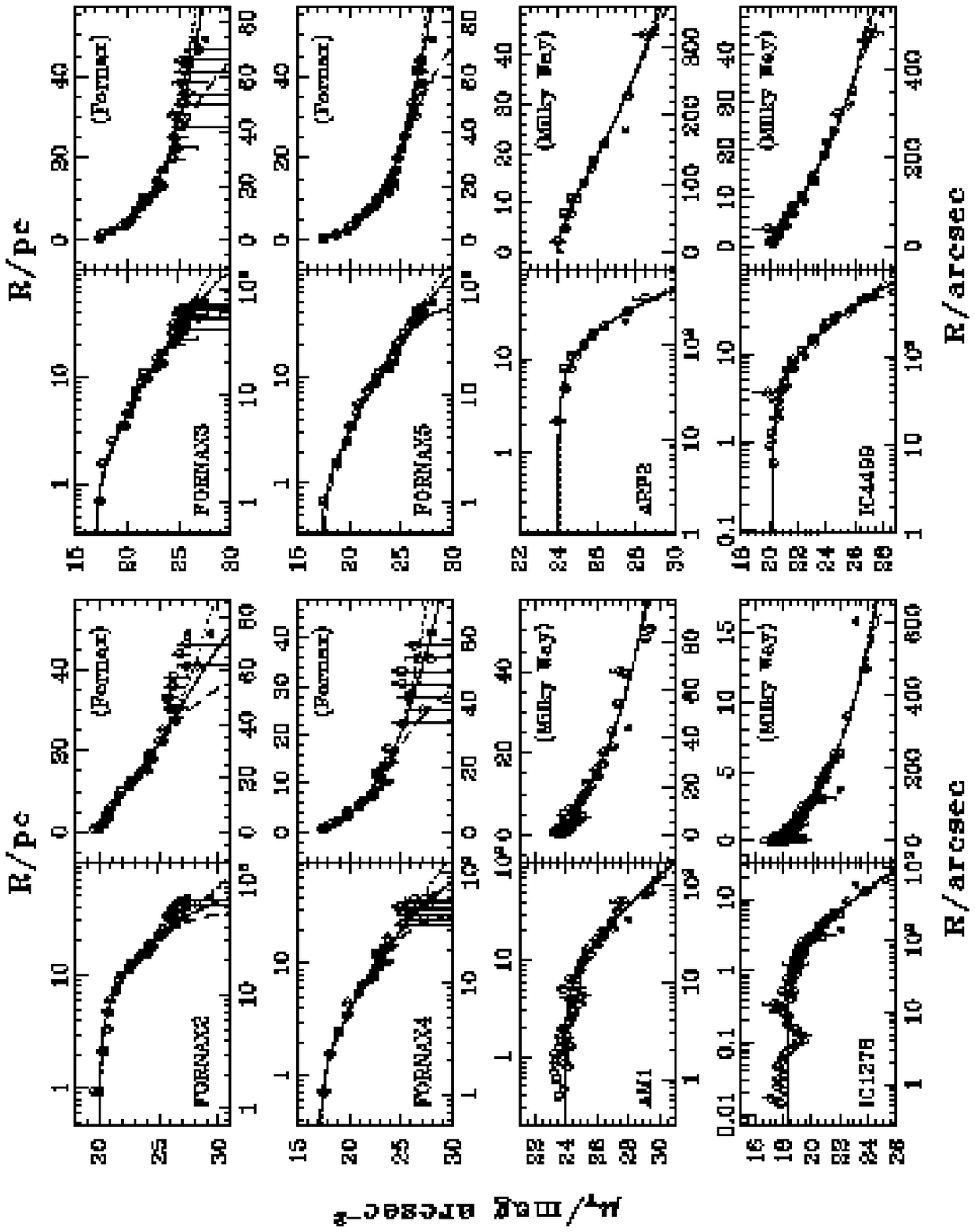}
\caption{}
\end{figure*}

\clearpage

\begin{figure*}
\figurenum{\ref{fig:sbfits} [continued]}
\epsscale{1.00}
\plotone{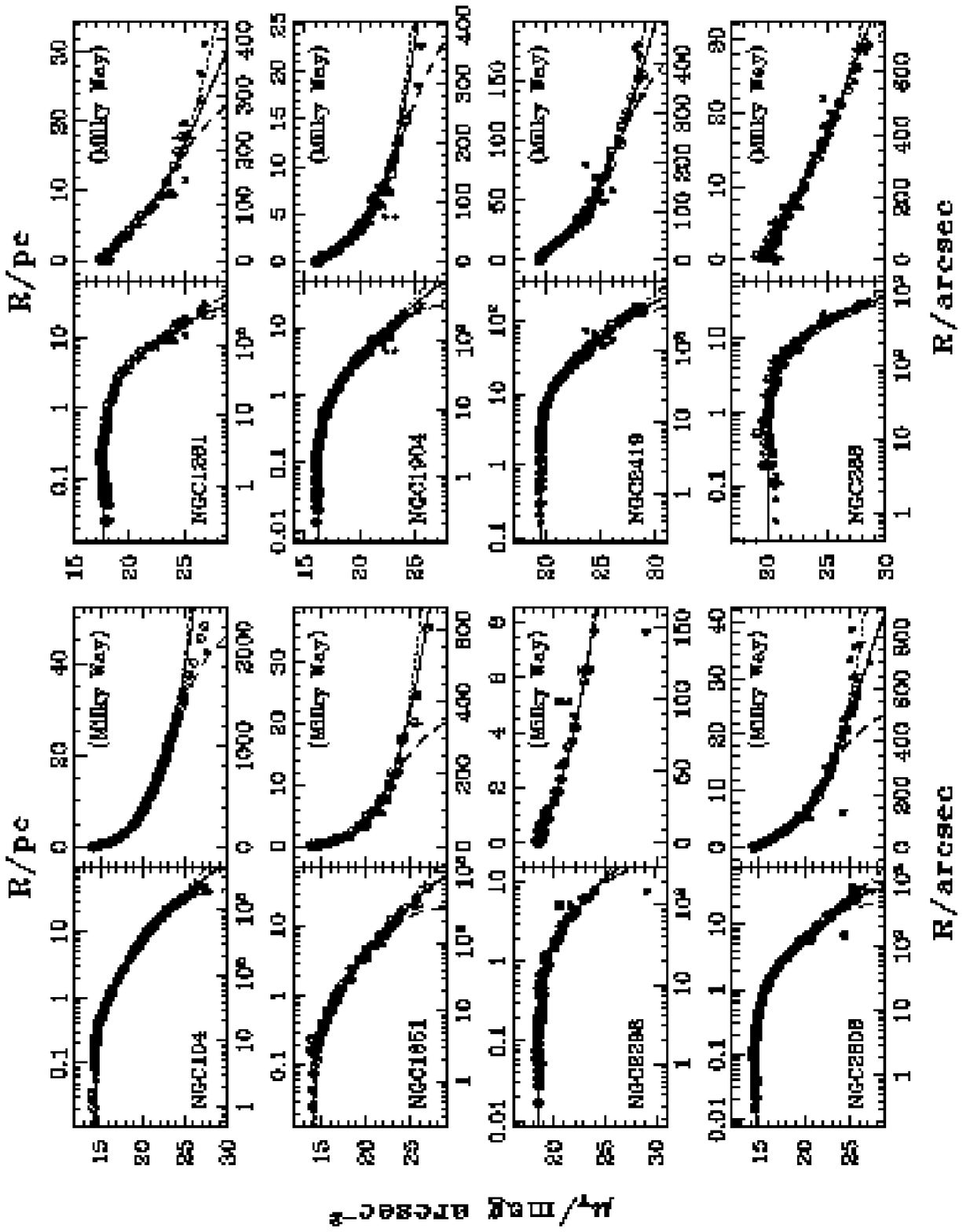}
\caption{}
\end{figure*}

\clearpage

\begin{figure*}
\figurenum{\ref{fig:sbfits} [continued]}
\epsscale{1.00}
\plotone{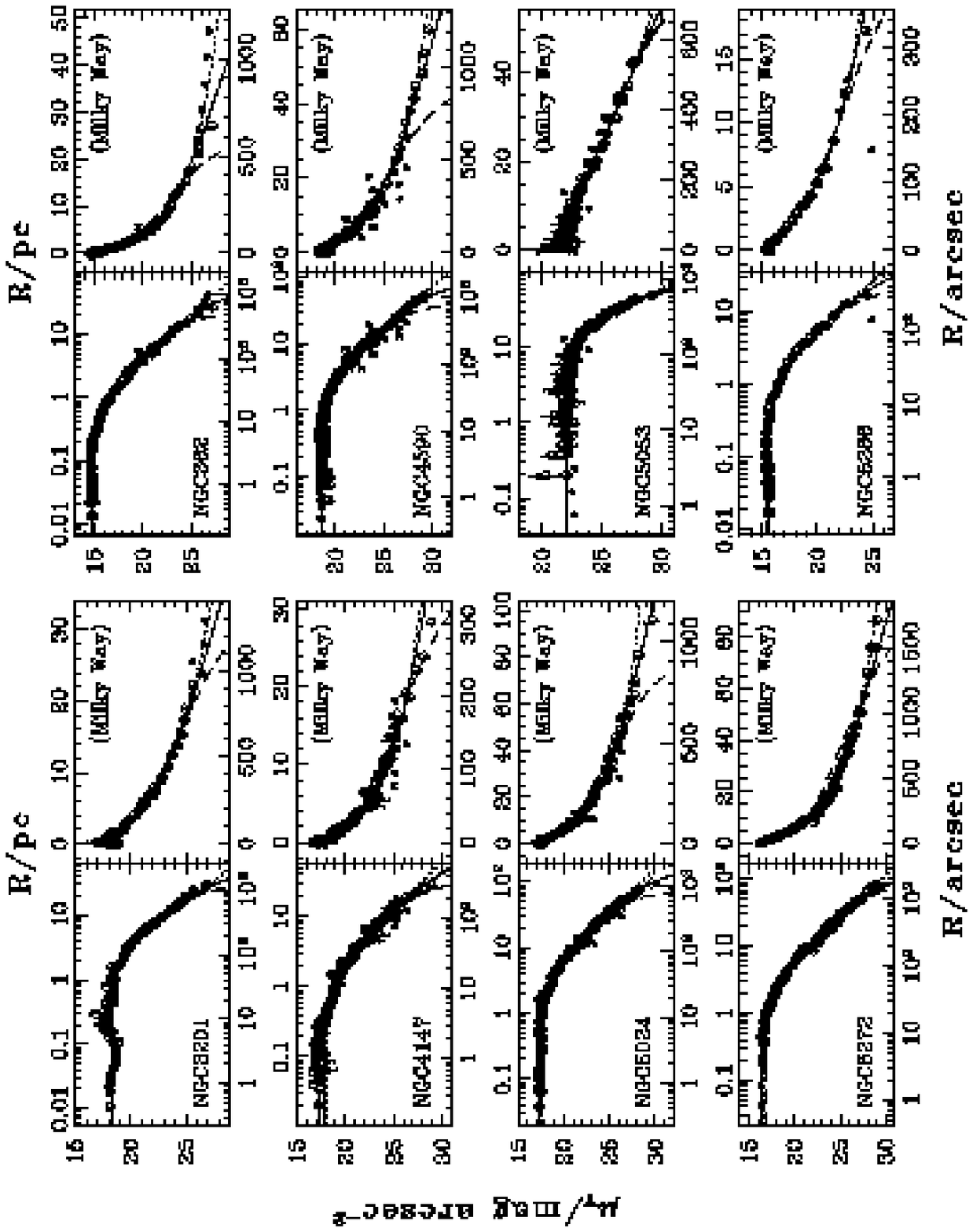}
\caption{}
\end{figure*}

\clearpage

\begin{figure*}
\figurenum{\ref{fig:sbfits} [continued]}
\epsscale{1.00}
\plotone{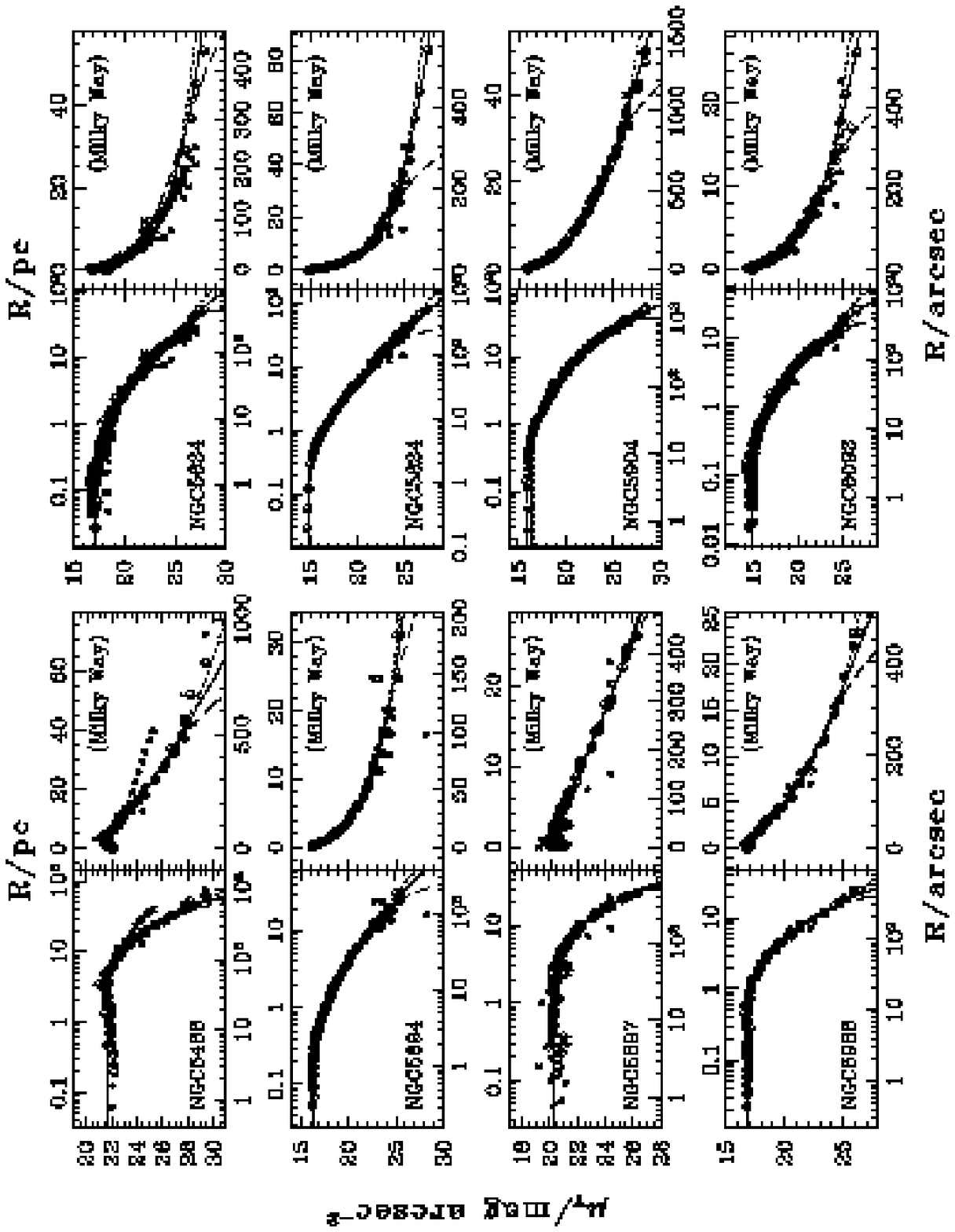}
\caption{}
\end{figure*}

\clearpage

\begin{figure*}
\figurenum{\ref{fig:sbfits} [continued]}
\epsscale{1.00}
\plotone{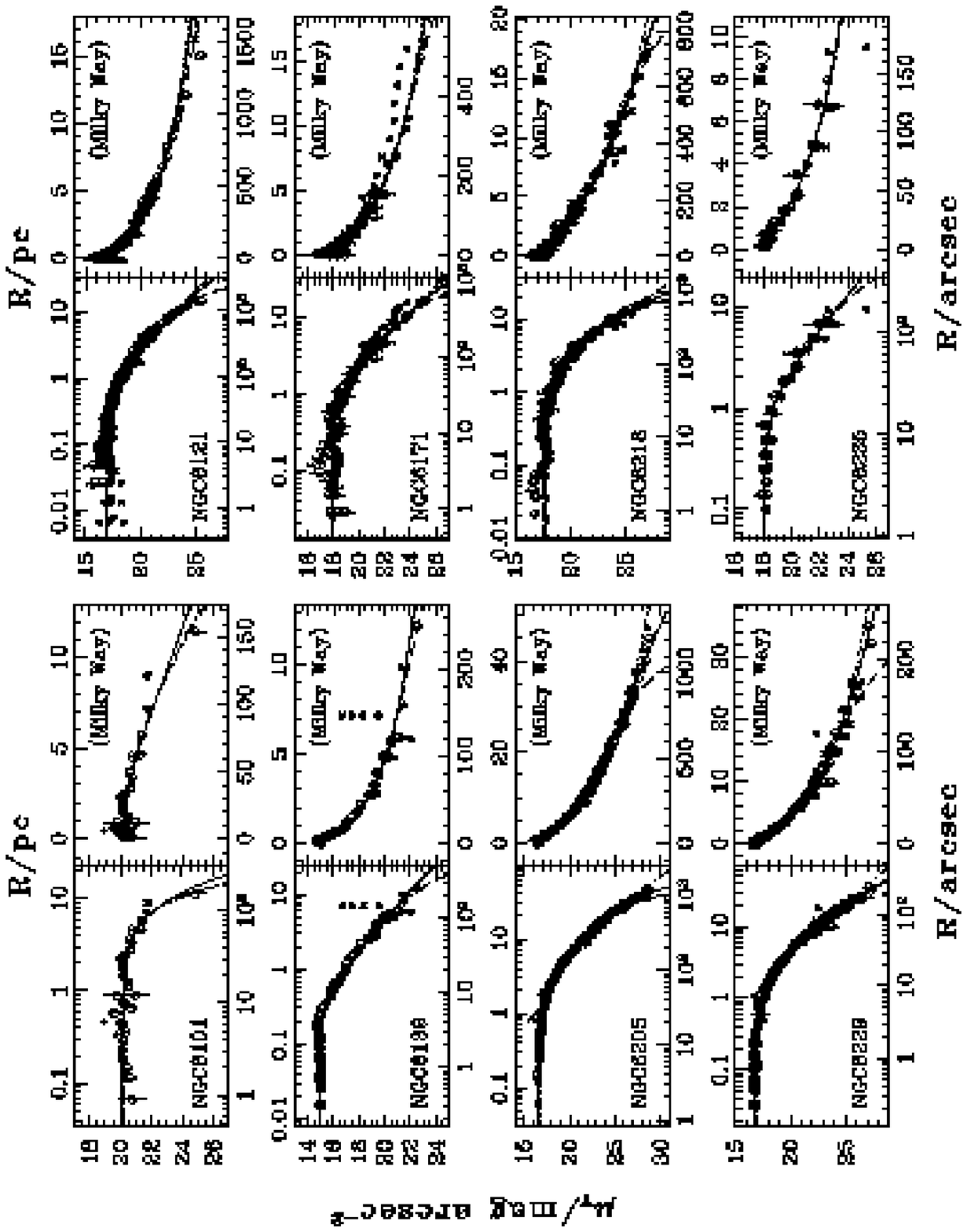}
\caption{}
\end{figure*}

\clearpage

\begin{figure*}
\figurenum{\ref{fig:sbfits} [continued]}
\epsscale{1.00}
\plotone{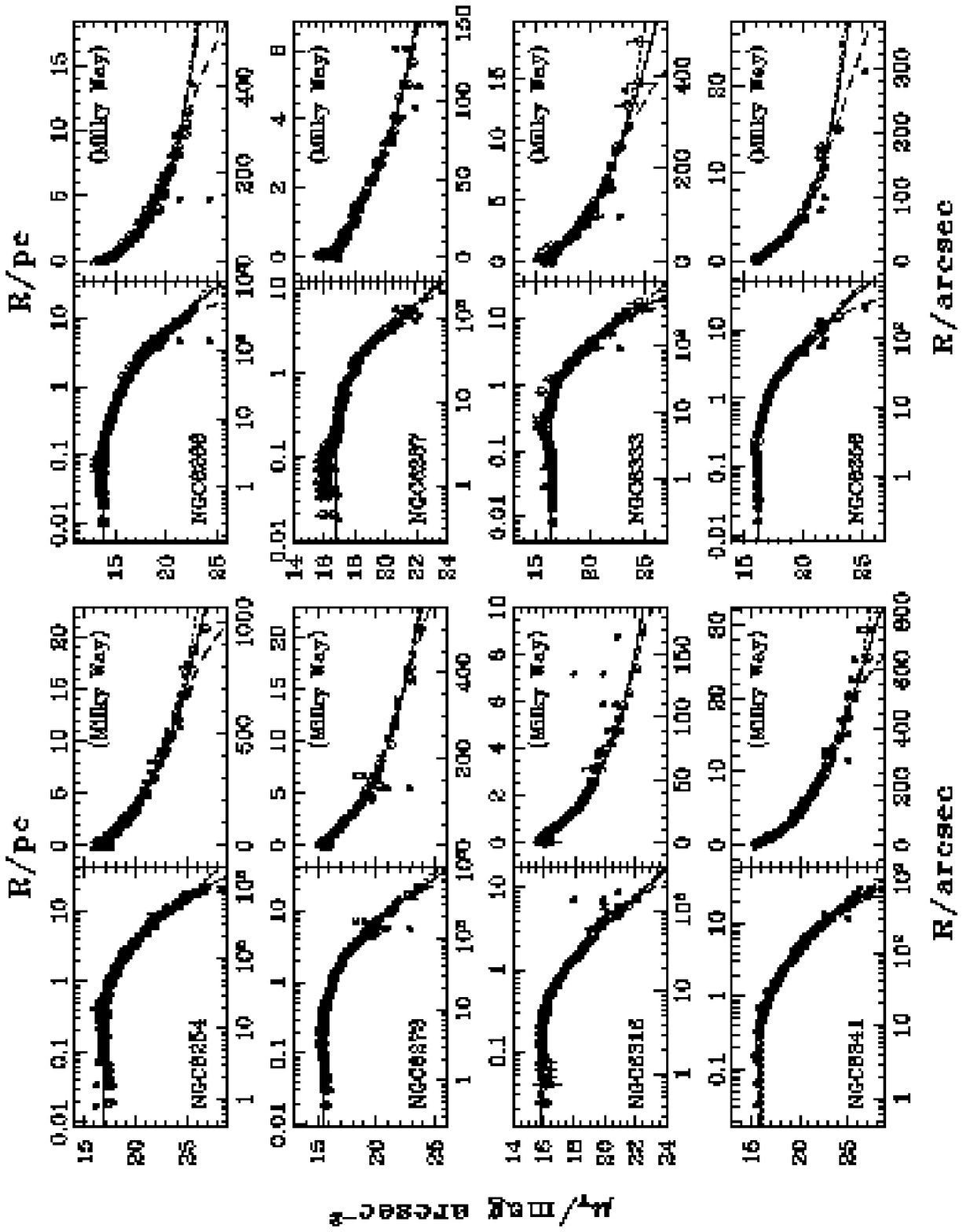}
\caption{}
\end{figure*}

\clearpage

\begin{figure*}
\figurenum{\ref{fig:sbfits} [continued]}
\epsscale{1.00}
\plotone{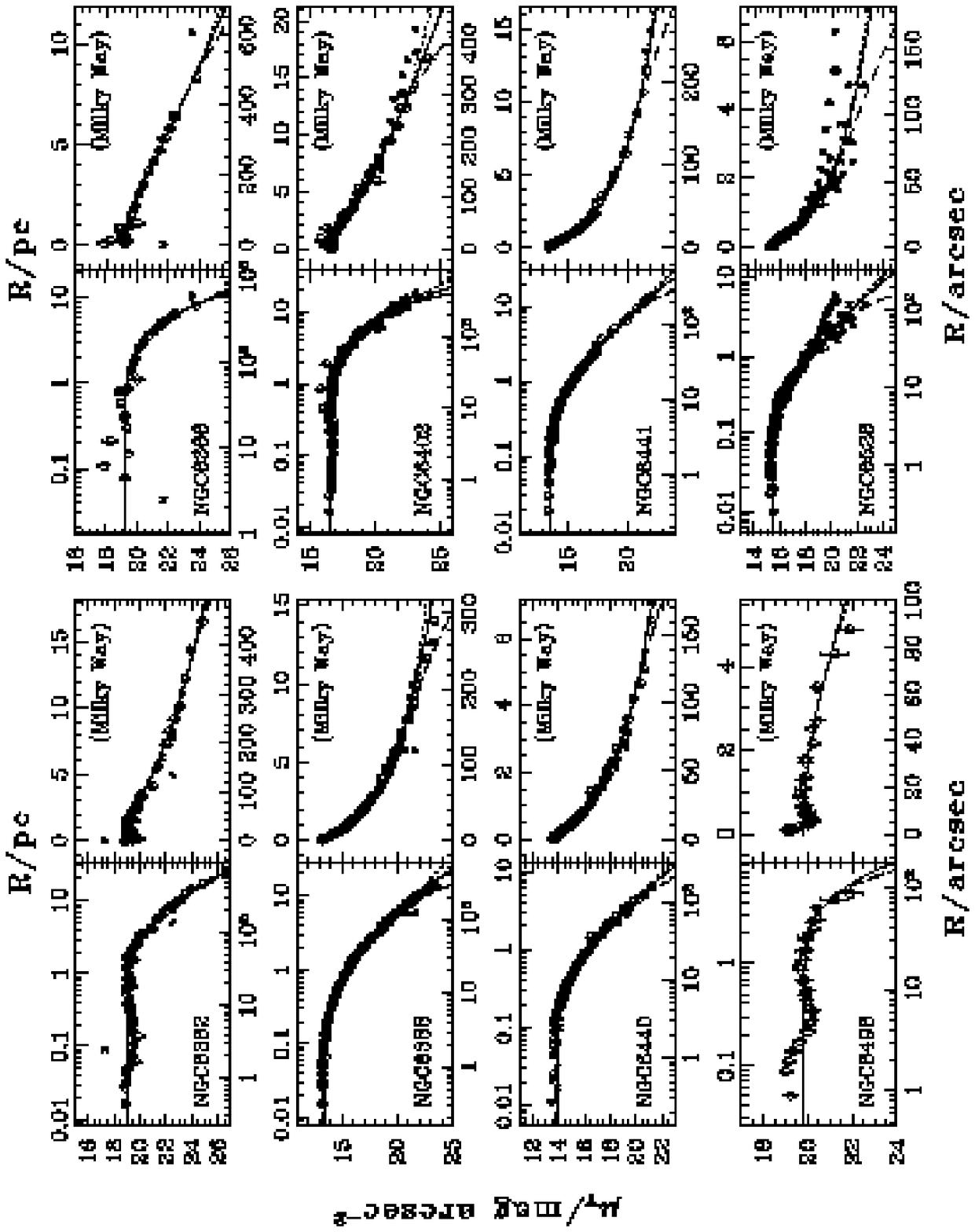}
\caption{}
\end{figure*}

\clearpage

\begin{figure*}
\figurenum{\ref{fig:sbfits} [continued]}
\epsscale{1.00}
\plotone{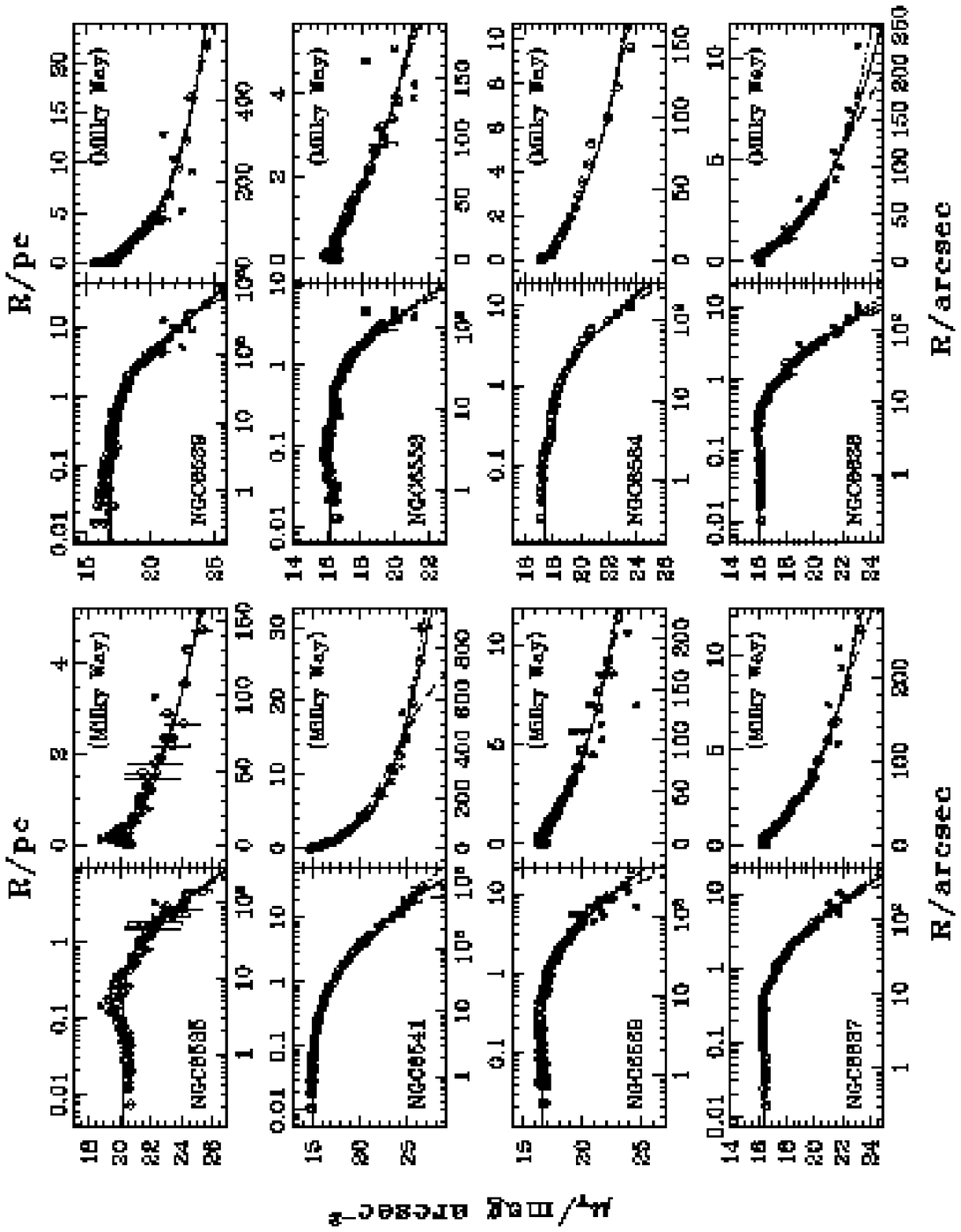}
\caption{}
\end{figure*}

\clearpage

\begin{figure*}
\figurenum{\ref{fig:sbfits} [continued]}
\epsscale{1.00}
\plotone{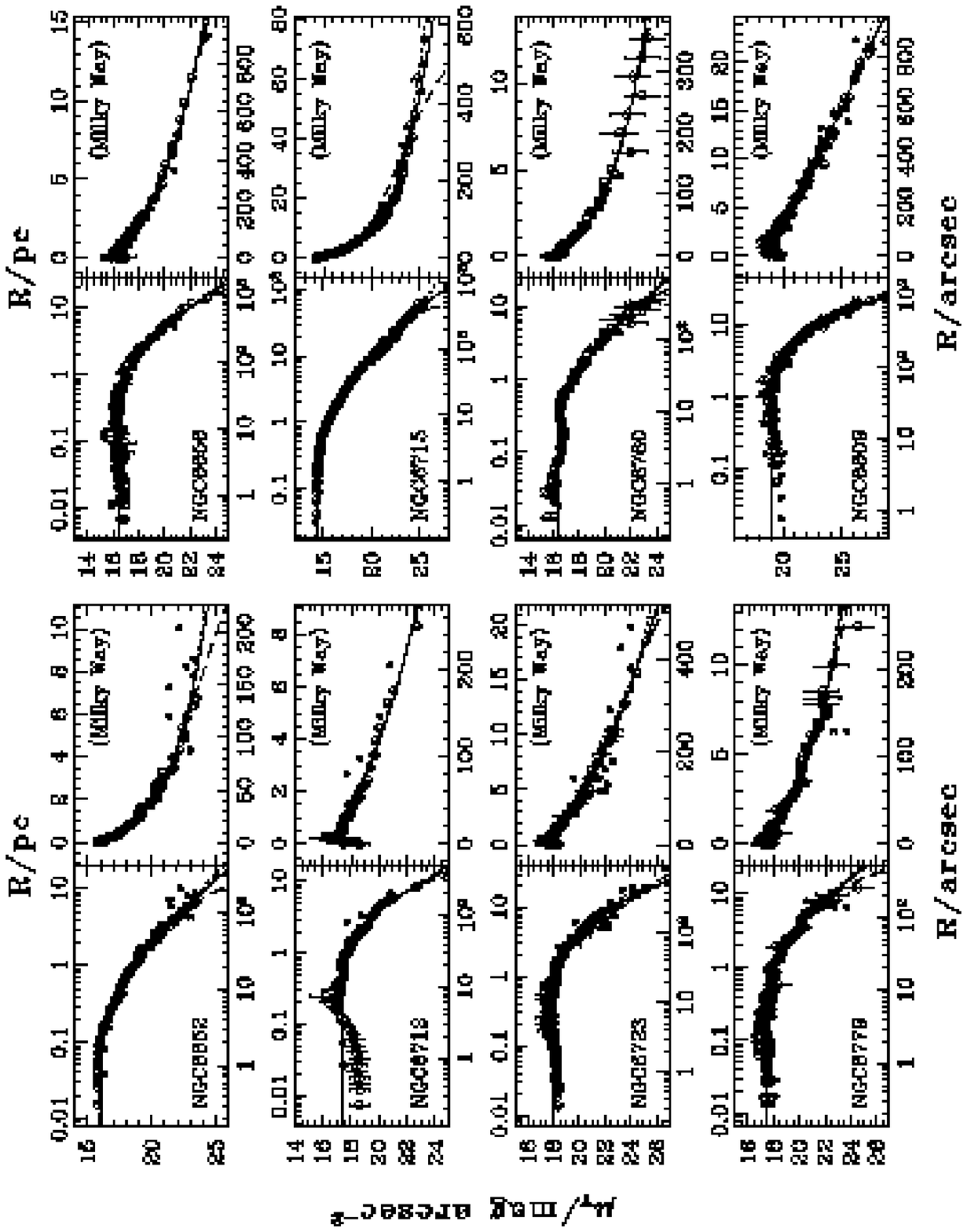}
\caption{}
\end{figure*}

\clearpage

\begin{figure*}
\figurenum{\ref{fig:sbfits} [continued]}
\epsscale{1.00}
\plotone{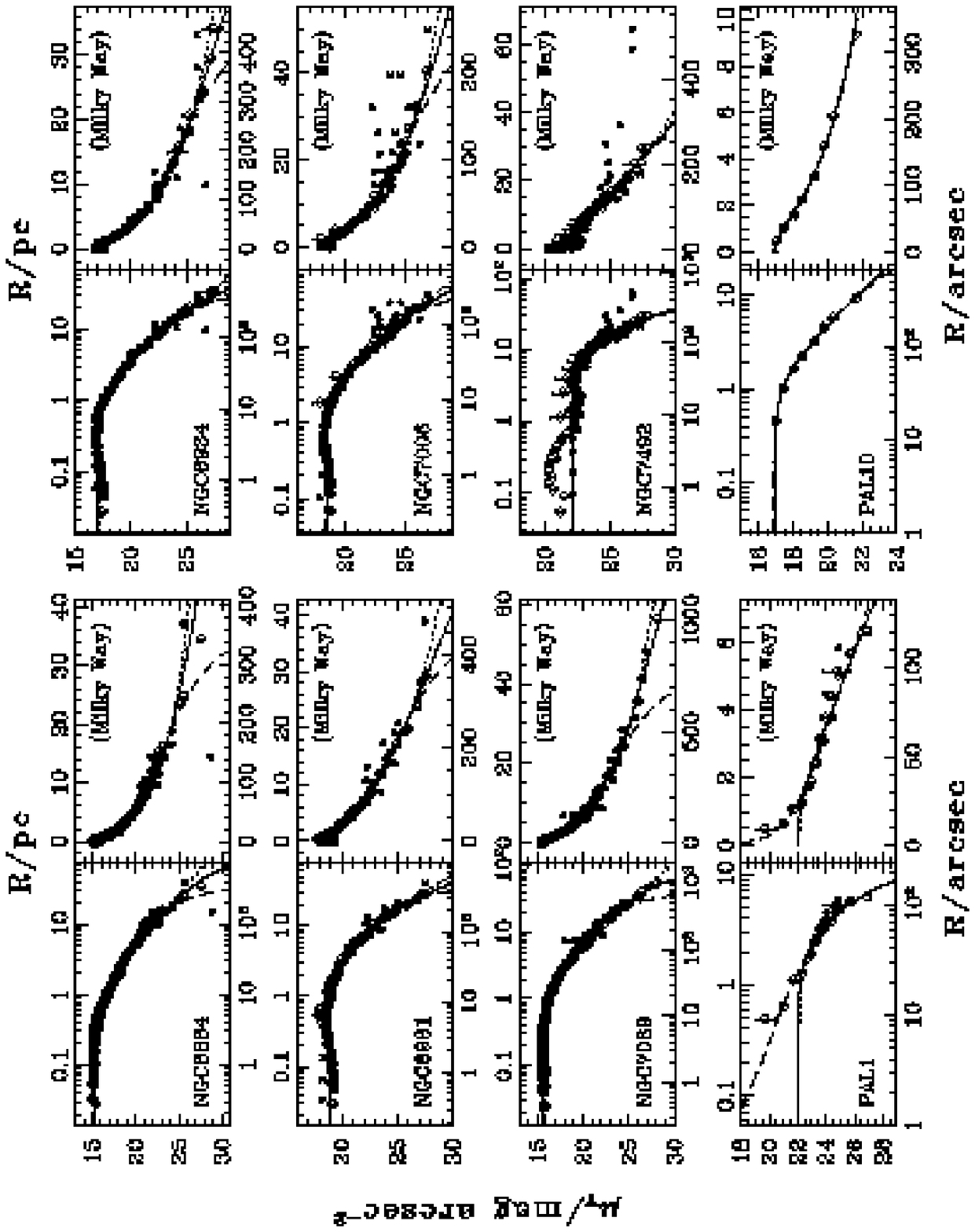}
\caption{}
\end{figure*}

\clearpage

\begin{figure*}
\figurenum{\ref{fig:sbfits} [continued]}
\epsscale{1.00}
\plotone{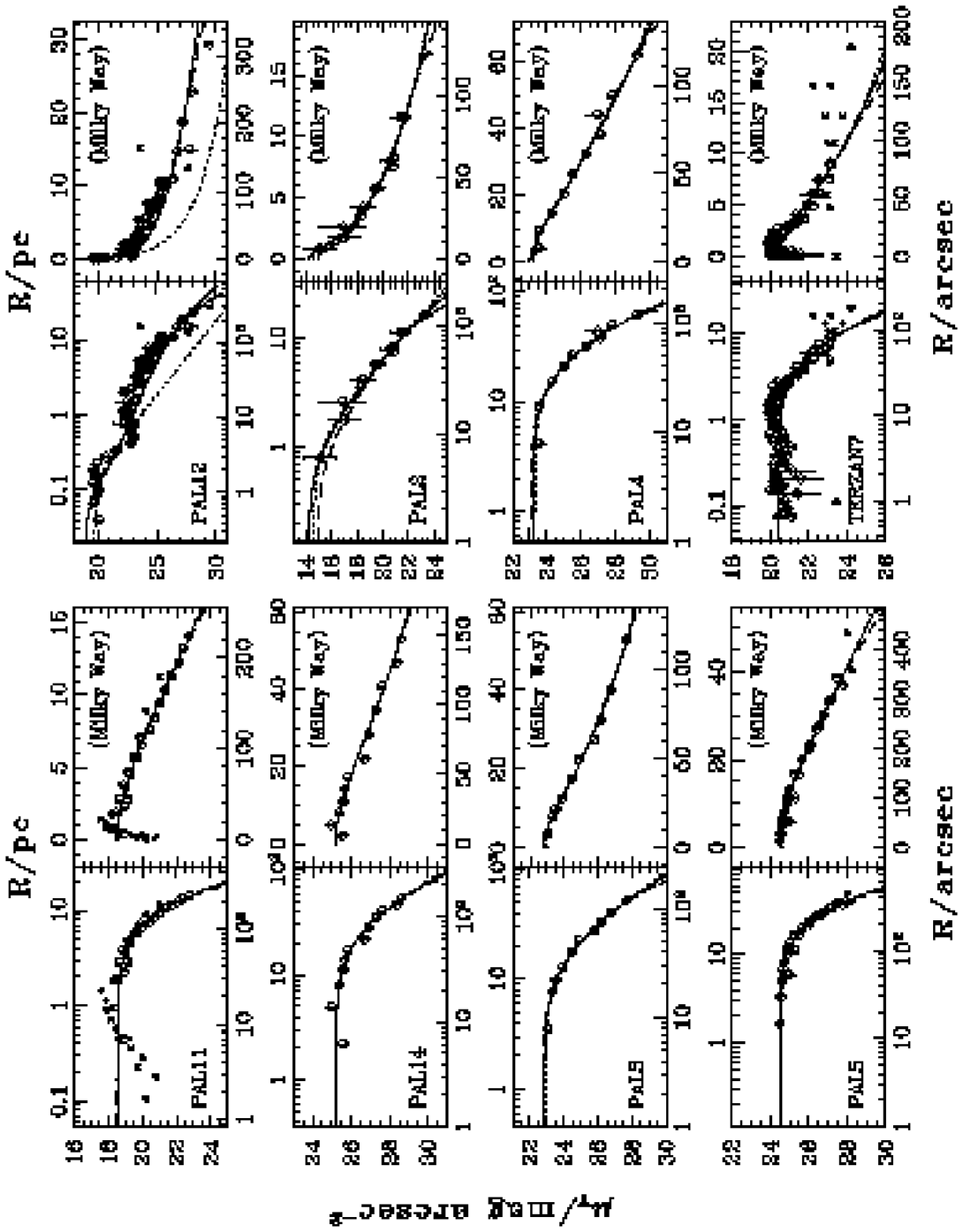}
\caption{}
\end{figure*}

\end{document}